\documentclass[aps,prd,preprintnumbers,superscriptaddress,showpacs,nofootinbib,11pt,floatfix]{revtex4}
\usepackage{graphicx}% Include figure files
\usepackage{dcolumn}% Align table columns on decimal point
\usepackage{longtable}
\usepackage{bm}% bold math
\usepackage{amsmath}
\usepackage{amssymb}

\def\beq{\begin{equation}}
\def\eeq{\end{equation}}
\def\[{\left[}
\def\]{\right]}

\def\gsim{\lower.7ex\hbox{$\;\stackrel{\textstyle>}{\sim}\;$}}
\def\lsim{\lower.7ex\hbox{$\;\stackrel{\textstyle<}{\sim}\;$}}

\begin{document}
\begin{flushright}
SHEP-06-35\\
{DCP-07-04}\\
{LPT-ORSAY-07-98}\\
\today
\end{flushright}
%\draft
\title{Charged Higgs boson phenomenology \\
in Supersymmetric models with Higgs triplets}
\author{J. L. D\'{\i}az-Cruz}
\email{jldiaz@fcfm.buap.mx} \affiliation{Fac. de Cs.
F\'{\i}sico-Matem\'aticas, BUAP. Apdo. Postal 1364, C.P. 72000
Puebla, Pue., M\'exico and Dual C-P Institute of High Energy Physics,  M\'exico.}
\author{J. Hern\' andez--S\' anchez}
\email{jaimeh@uaeh.edu.mx} \affiliation{Centro de Investigaci\'on en Matem\'aticas, UAEH,
Carr. Pachuca-Tulancingo Km. 4.5, C.P. 42184, Pachuca, Hgo., M\'exico and Dual C-P Institute of High Energy Physics.}
\author{S. Moretti}
\email{stefano@hep.phys.soton.ac.uk} \affiliation{ School of  Physics and Astronomy,
University of Southampton, Highfield, Southampton SO17 1BJ, UK\\
and\\
Laboratoire de Physique Th\'eorique, U. Paris-Sud and CNRS, F-91405 Orsay,
France.}
\author{A. Rosado}
\email{rosado@sirio.ifuap.buap.mx} \affiliation{ Instituto de
F\'{\i}sica, BUAP. Apdo. Postal J-48, C.P. 72570 Puebla, Pue.,
M\'exico.}
\date{\today}
\begin{abstract}
We present a detailed study of the Higgs sector  within an
extension of the Minimal Supersymmetric
Standard Model that includes one Complex Higgs Triplet
(MSSM+1CHT). The model spectrum includes three singly charged Higgs
bosons as well as three CP-even (or scalar) and two CP-odd
(or pseudoscalar) neutral Higgs bosons.
We present an approximated calculation of the one-loop radiative
corrections to the neutral CP-even Higgs masses ($m_{H_i^0}$) and
the couplings $H_i^0 Z^0 Z^0$ ($i=1$, 2, $3$), which determine the
magnitude of the Higgs-strahlung processes  $e^+ e^-\to Z^0 H^0_i$.
Limits from LEP2 are then considered, in order to obtain bounds on
the neutral Higgs sector. Further, we also include the experimental
limits from LEP2 on $e^+e^-\to H^+H^-$ and those on BR($t \to b \,
H^+$) from Tevatron, to derive bounds  on the mass of the two
lightest charged Higgs bosons ($H_1^{\pm}$ and $H_2^{\pm}$).
Concerning the latter, we find some cases, where $m_{H_1^{\pm}}
\simeq 90$ GeV, that are not excluded by any experimental bound,
even for large values of $\tan\beta$, so that they should be looked
for at the Large Hadron Collider (LHC).
\end{abstract}
\maketitle

\newpage

%%%%%%%%%%%%%%%%%%%%%%%%%%%%%%%%%%%%%%%%%%%%%%%%%%%%%%%%%%%%%%%%%%%%%%%
%\narrowtext
%%%%%%%%%%%%%%%%%%%%%%%%%%%%%%%%%%%%%%%%%%%%%%%%%%%%%%%%%%%%%%%%%%%%%%%
\section{Introduction}
%%%%%%%%%%%%%%%%%%%%%%%%%%%%%%%%%%%%%%%%%%%%%%%%%%%%%%%%%%%%%%%%%%%%

The Higgs spectrum of many well motivated extensions of the Standard
Model (SM) often include charged Higgs bosons whose detection at
future colliders would constitute a clear evidence of a Higgs sector
beyond that of the Standard Model (SM) \cite{stanmod}. In particular, the
2-Higgs-Doublet-Model (2HDM, hereafter, of Type II), in both its
Supersymmetric
(SUSY) and non-SUSY versions \cite{kanehunt}, has been extensively
studied as a prototype of a Higgs sector that includes one charged
Higgs boson pair ($H^\pm$), whose detection is expected to take
place at the LHC \cite{LHCrev}. However, a
definitive test of the mechanism of Electro-Weak Symmetry Breaking (EWSB)
will require further studies aiming at pinning down the underlying
complete Higgs spectrum. In particular, probing the properties of
charged Higgs bosons could help to find out whether they are indeed
associated with a
 weakly-interacting theory, as in
the case of the most popular SUSY extension of the
SM, the so-called Minimal Supersymmetric Standard Model (MSSM)
\cite{susyhix}, or with a strongly-interacting scenario, like the
ones discussed recently \cite{stronghix}. Ultimately, while many
analyses in this direction can be carried out at the LHC, it will be
a future International Linear Collider (ILC) saying the definite
word about which mechanism leads to mass generation. Notice that
 these tests should also allow one to
probe the symmetries of the Higgs potential and to determine
whether the charged Higgs bosons belong to a weak doublet or to some
larger multiplet. Among the latter, in particular, Higgs triplets
have been considered \cite{Rizzo:1990uu}, mainly to search for
possible manifestations of an explicit breaking of the custodial
$SU(2)_c$ symmetry, which keeps Veltman's so-called `rho parameter' close to
one, i.e. $\rho\simeq 1$. Motivations to discuss Higgs triplets can
also be drawn from models of neutrino masses \cite{cppaper} as well
as scenarios with extra spacial dimensions \cite{ourexdims}. Though
most of the work has been  within non-SUSY models \cite{vegunwud},
there have also been studies of SUSY scenarios with complex Higgs
triplets, such as in \cite{Espinosa:1991wt}, where some
phenomenological aspects of the Higgs sector were explored.
Subsequent work in this model has been done in
\cite{Felix-Beltran:2002tb}.

Decays of charged Higgs bosons have been studied in the literature
\cite{Hdecays},
including the radiative modes $W^{\pm}\gamma, W^{\pm}Z^0$
\cite{hcdecay}, mostly within the context of the 2HDM or its MSSM
incarnation (including into SUSY particles \cite{hctoSUSY}), but also
for the effective Lagrangian extension of the
2HDM defined in \cite{ourpaper} and more recently within an extension of the
MSSM with one Complex Higgs Triplet (MSSM+1CHT)
\cite{Barradas-Guevara:2004qi}. All these activities are particularly
relevant especially in view of the fact that
charged Higgs boson decays can be exploited to determine key parameters of
the underlying Higgs sector \cite{Ketevi}. Charged Higgs boson production at
hadron colliders was studied long ago \cite{ldcysampay} and, more
recently, systematic calculations of production processes at the
upcoming LHC have been presented
\cite{newhcprod}, including some higher order effects in QCD and
SUSY QCD  \cite{newhcprod-HO}. Current bounds on the mass of the
charged Higgs bosons have been obtained at Tevatron, by studying the
top decay $t \to b \, H^+$, which already eliminates some regions of
the 2HDM and MSSM
parameter spaces \cite{Abulencia:2005jd}, whereas LEP2 gives a model
independent bound of $m_{H^{\pm}} > 80$ GeV \cite{lepbounds,partdat}.

In this paper we present a detailed study of the spectrum
and discuss the phenomenology of the Higgs sector of the MSSM+1CHT model, i.e.,
the scenario that includes one complex Higgs triplet in addition to
the usual MSSM Higgs content, namely two Higgs doublets. Our main focus
will eventually be on the production and decay phenomenology of the
charged Higgs states of the model.

This
article is organized as follows. In section II, we discuss the Higgs
sector of this model, in particular, we present the charged Higgs
boson spectrum and the inclusion of an estimated calculation of the
one-loop radiative corrections for the CP-even neutral Higgs sector.
In this section, we also present a study of the couplings
$H_i^0 Z^0 Z^0$, which are modified by radiative corrections. Then, in
section III, we derive the expressions for the vertex $H^{\pm} f f'$
(where $f$ and $f'$ are generic fermions with cumulative
electromagnetic charge $\pm1$) and we calculate the decay $t \to
H^+_i \, b$ in the framework of the MSSM+1CHT model, also presenting
numerical results for the most relevant
charged Higgs Branching Ratios (BR's)\footnote{In the framework of
the MSSM+1CHT  the three charged Higgs states are denoted by $H^\pm_i$ with
the convention: $m_{H^\pm_1}<m_{H^\pm_2}<m_{H^\pm_3}$.}. (A comparison
with latest bounds from Tevatron Run2 is also given therein.) A
discussion of the main production mechanism at the LHC is presented
in section IV. LHC event rates are given in section V.
Finally, we summarize and conclude in section VI.

%%%%%%%%%%%%%%%%%%%%%%%%%%%%%%%%%%%%%%%%%%%%%%%%%%%%%%%%%%%%%%%%%%%%%%%
\section{The charged Higgs spectrum  in a SUSY model with an
additional complex Higgs triplet}
%%%%%%%%%%%%%%%%%%%%%%%%%%%%%%%%%%%%%%%%%%%%%%%%%%%%%%%%%%%%%%%%%%%%%%%

The SUSY model with two doublets and a complex Higgs triplet
(MSSM+1CHT) of \cite{Espinosa:1991wt} is one of
the simplest extensions of the
MSSM that allows one to study phenomenological consequences of
an explicit breaking of the custodial $SU(2)_c$ symmetry
\cite{Espinosa:1991wt,Felix-Beltran:2002tb}. In the reminder of this section, we recap its
main theoretical features.
%%%%%%%%%%%%%%%%%%%%%%%%%%%%%%%%%%%%%%%%%%%%%%%%%%%%%%%%%%%%%%%%%%%%%%%%%%
\subsection{The Higgs potential of the model}
%%%%%%%%%%%%%%%%%%%%%%%%%%%%%%%%%%%%%%%%%%%%%%%%%%%%%%%%%%%%%%%%%%%%%%%%%
The MSSM+1CHT model includes two Higgs doublets and a complex Higgs
triplet given by
\begin{equation}
\Phi_1 = \left(
\begin{array}{c}
{\phi_1}^0 \\
{\phi_1}^-
\end{array}
\right) \,\,\, , \,\,\, \Phi_2= \left(
\begin{array}{c}
{\phi_2}^+ \\
{\phi_2}^0
\end{array}
\right) \,\,\, , \,\,\, \sum = \left(
\begin{array}{cc}
\sqrt{\frac{1}{2}} \xi^0 & - \xi_2^+ \\
\xi_1^- & - \sqrt{\frac{1}{2}} \xi^0
\end{array}
\right) \,\,\, .
\end{equation}
The Higgs triplet, of zero hypercharge, is described in terms of a
$2 \times 2$ matrix representation: $\xi^0$ is the complex neutral
field and $\xi_1^-, \, \xi_2^+$ denote the charged fields.  The
most general gauge invariant and renormalizable Superpotential that
can be written for the Higgs Superfields $\Phi _{1,2}$ and $\Sigma $
is given by:
\begin{equation}
W=\lambda \Phi _{1}\cdot \Sigma \Phi _{2}+\mu _{1}\Phi _{1}\cdot
\Phi _{2}+\mu _{2}\mbox{Tr}(\Sigma ^{2})\,\,\,,
\end{equation}
where we have used the notation $\Phi _{1}\cdot \Phi _{2}\equiv
\epsilon _{ab}\Phi _{1}^{a}\Phi _{2}^{b}$. The resulting scalar
potential involving only the Higgs fields is thus written as
$$
V=V_{SB}+V_{F}+V_{D}\,\,\,,
$$
\noindent where $V_{SB}$ denotes the most general
soft-Supersymmetry breaking potential, which is given by
\begin{eqnarray}
V_{SB} & = & m_1^2 |\Phi_1|^2+ m_2^2|\Phi_2|^2+m_3^2 Tr(\Sigma^\dag \Sigma)
\nonumber \\
&+& [A\lambda \Phi_1\cdot \Sigma \Phi_2+ B_1 \mu_1 \Phi_1 \cdot \Phi_2+B_2 \mu_2 Tr(\Sigma^2)+h.c. ],
\end{eqnarray}
$V_F$ is the SUSY potential from F-terms
\begin{eqnarray}
V_{F} & = &  \bigg| \mu_1 \phi_2^0+\lambda \bigg( \phi_2^+ \xi^-_1-\frac{1}{\sqrt{2}} \phi_2^0 \xi^0 \bigg) \bigg|^2+
\bigg| \mu_1 \phi_1^0+\lambda \bigg( \phi_1^- \xi^+_2-\frac{1}{\sqrt{2}} \phi_1^0 \xi^0 \bigg) \bigg|^2  \nonumber \\
&+& \bigg| \mu_1 \phi_2^++\lambda \bigg( \frac{1}{\sqrt{2}}\phi_2^+ \xi^0- \phi_2^0 \xi_2^+ \bigg) \bigg|^2+
\bigg| \mu_1 \phi_1^-+\lambda \bigg( \frac{1}{\sqrt{2}}\phi_1^- \xi^0- \phi_1^0 \xi^-_1 \bigg) \bigg|^2 \nonumber \\
&+& \bigg|2 \mu_2 \xi^0-\frac{\lambda }{\sqrt{2}} \bigg( \phi_1^0 \phi^0_2+\phi_1^- \phi^+_2 \bigg) \bigg|^2+
\bigg| \lambda  \phi_1^0 \phi^+_2-2\mu_2 \xi_2^+ \bigg|^2+\bigg| \lambda  \phi_1^- \phi^0_2-2\mu_2 \xi_1^- \bigg|^2,
\end{eqnarray}
$V_D$ is the SUSY potential from D-terms
\begin{eqnarray}
V_{D} & = & \frac{g^2}{8}\bigg[ |\phi_1^0|^2-|\phi_1^-|^2+ |\phi^+_2|^2-|\phi_2^0|^2 +2 |\xi^+_2|^2 -2|\xi_1^-|^2 \bigg]^2
\nonumber \\ &+& \frac{g^{'2}}{8}\bigg[ |\phi_1^0|^2+|\phi_1^-|^2- |\phi^+_2|^2-|\phi_2^0|^2  \bigg]^2 \nonumber \\
&+& \frac{g^2}{8} \bigg[ \phi_1^{0*} \phi_1^- + \phi^{+*}_2 \phi_2^0 + \sqrt{2} ( \xi^+_2 + \xi_1^- ) \xi^{0*}
+h.c. \bigg]^2 \nonumber \\
&-& \frac{g^2}{8} \bigg[ \phi_1^{-*} \phi_1^0 + \phi^{0*}_2 \phi_2^+ + \sqrt{2} ( \xi^+_2 - \xi_1^- ) \xi^{0*}
-h.c. \bigg]^2.
\end{eqnarray}
In turn, the full scalar potential can be split into its neutral
and charged parts, {i.e.}, $V = V_{\rm{charged}} + V_{\rm{neutral}}$ \cite{Espinosa:1991wt,Felix-Beltran:2002tb}.

Besides the Supersymmetry-breaking mass terms, $m_i^2$ ($i =
1,\,2,\,3$), the potential depends on the parameters $\lambda,\,
\mu_1,\,\mu_2$, that appear in Eq. (2), as well as the trilinear and bilinear
terms, $A$ and $B_i$, respectively. For simplicity, we will assume that there is no CP
violation in the Higgs sector and thus all the parameters and
Vacuum Expectations Values (VEVs)
 are assumed to be real. In the charged Higgs sector with the basis of states
$(\phi_2^{+}, \, \phi_1^{-*}, \, \xi_2^{+}, \,\xi_2^{-*} )$
one has a $4 \times 4$ squared-matrix mass $(M_{\pm}^2)_{ij}$, $i=j=1, \, ...4$.
For the CP-odd Higgs sector with the basis of states
 $(\frac{1}{\sqrt{2}} Im \phi_1^{0}, \, \frac{1}{\sqrt{2}} Im  \phi_2^{0}, \, \frac{1}{\sqrt{2}} Im \xi^{0} )$
one obtains a  $3 \times 3$ squared-matrix mass $(M_{P}^2)_{ij}$, $i=j=1,
\, 2, \, 3$.
For the CP-even Higgs sector with the basis of states
 $(\frac{1}{\sqrt{2}} Re \phi_1^{0}, \, \frac{1}{\sqrt{2}} Re  \phi_2^{0}, \, \frac{1}{\sqrt{2}} Re \xi^{0} )$
one also has a  $3 \times 3$ squared-matrix mass $(M_{S}^2)_{ij}$,
$i=j=1, \, 2, \, 3$.
The explicit expression of the resulting Higgs potential is given in Refs.~\cite{Espinosa:1991wt,Felix-Beltran:2002tb}.

%%%%%%%%%%%%%%%%%%%%%%%%%%%%%%%%%%%%%%%%%%%%%%%%%%%%%%%%%%%%%
\subsection{Parameters of the model and definition of scenarios}
%%%%%%%%%%%%%%%%%%%%%%%%%%%%%%%%%%%%%%%%%%%%%%%%%%%%%%%%%%%%%

We can combine the VEVs of the doublet Higgs fields through the
relation $v_{D}^{2} \equiv v_{1}^{2}+v_{2}^{2}$ and define
$\tan\beta \equiv {v_{2}}/{v_{1}}$. Furthermore, the parameters
$v_{D}$, $v_{T}$, $m_W^2$ and $m_Z^2$ are related as follows:
\begin{equation}
\begin{array}{cl}
m_{W}^{2}= & \frac{1}{2}g^{2}(v_{D}^{2}+4v_{T}^{2}), \nonumber \\
m_{Z}^{2}= &
{\displaystyle{\frac{{\frac{1}{2}g^{2}v_{D}^{2}}}{{\mbox{cos}
^{2}{\theta }_{W}}}}}\,\, ,
\end{array}
\end{equation}
which implies that the $\rho$-parameter is different from 1 at the
tree level, namely,
\begin{equation}
\rho \equiv \frac{M_{W}^{2}}{M_{Z}^{2}\mbox{cos}^{2}\theta _{W}}
=1+4R^{2},\,\,\,\,\,\,R\equiv \frac{v_{T}}{v_{D}}\,\,\,.
\end{equation}
The bound on $R$ is obtained from the $\rho $ parameter measurement, which
presently lies in the range 0.9993--1.0006, from the global fit reported
in
Refs.~\cite{partdat,Rizzo:1990uu}. Thus, one has $R \leq 0.012$ and
$v_{T}\leq 3$ GeV. We have taken into account this bound in our
numerical analyses.

Thus, the Higgs sector of this model depends of the following
parameters: (i) the gauge-related parameters ($g$, $g'$, $v$, $R$,
$\tan\beta$); (ii) the Yukawa couplings ($\lambda $, $\mu_1$, $\mu_2$)
and (iii) the soft Supersymmetry-breaking parameters ($A$, $B_1$,
$B_2$). The gauge-related parameters can be replaced by the
quantities ($G_F$, $\alpha$, $m_W$, $\rho$, $\tan\beta$). For the
numerical analysis to be realized in the remainder of this paper, we
must make sure  that the following theoretical conditions of the
MSSM+1CHT are satisfied: (a) the global stability condition of the
potential; (b) the necessary condition for having a global minimum
and (c) the positivity of the mass eigenvalues of the full spectrum
of charged, pseudoscalar and scalar Higgs bosons
\cite{Espinosa:1991wt}.

The parameter space analyzed here for the MSSM+1CHT is the same that
was considered before in the specialized literature
\cite{Espinosa:1991wt,Barradas-Guevara:2004qi}, namely, we consider
characteristic values  below their perturbative limits. Small and large
values of $\tan\beta$ are both considered. Typical cases for $A$,
$B_1$, $B_2$, $\mu_1$ and $\mu_2$ are used to define the following
scenarios. \vskip0.5cm \noindent{\bf Scenario A.} It is defined by
considering $B_1=\mu_1=0$, $B_2=-A$ and $\mu_2=100$ GeV while for
$\lambda$ we shall consider the values $\lambda=0.1,\,0.5,\,1.0$. In
this scenario it happens that the additional Higgs triplet plays a
significant role in EWSB. \vskip0.5cm \noindent{\bf Scenario B.}
This scenario is defined by choosing: $B_2=\mu_2=0$, $B_1=-A$, while
for $\lambda$ we shall consider again the values
$\lambda=0.1,\,0.5,\,1.0$. Most results will take  $\mu_1=200$ GeV,
though other values (such as $\mu_1=400, 700$ GeV) will also be
considered.  Here, the effects of the additional Higgs triplet are
smaller, hence the behaviour of the model is similar to that of the
MSSM.

%A third scenario, where both Doublet and Triplet effects are
%competing could be defined, but we did not show results in order to
%keep under control the number of graphs to be shown.

%%%%%%%%%%%%%%%%%%%%%%%%%%%%%%%%%%%%%%%%%%%%%%%%%%%%%%%%%%%%%
\subsection{One-loop radiative corrections to the CP-even Higgs bosons masses in the MSSM+1CHT}
%%%%%%%%%%%%%%%%%%%%%%%%%%%%%%%%%%%%%%%%%%%%%%%%%%%%%%%%%%%%%

In some cases, within the Scenarios A and B, we will show that a very
light CP-even Higgs boson appears at the tree level, with a mass around
10 GeV, which can even be as small as  ${\cal O}(0.1)$ GeV. However, it
is known that, in the MSSM, the inclusion of radiative corrections
from top and stop loops can alter significantly the (lightest) neutral CP-even
Higgs mass. Thus, we can expect that similar effects will appear
here and, furthermore, one also needs to consider in the MSSM+1CHT a
possible large
correction from  Higgs-chargino loops, which  could lift the corresponding
Higgs mass above current experimental bounds. This
means that one needs to include all such radiative corrections in order
to avoid misleading conclusions. We are also interested in
discussing the neutral Higgs bosons masses here because of their
possible appearance in charged Higgs boson decays. As it will be
shown later, this effect is important for large regions of the MSSM+1CHT
parameter
space.

The radiative corrections to Supersymmetric Higgs boson masses can
be evaluated using the effective potential technique
\cite{Ellis-Ridolfi}, which at one-loop reads:
\begin{eqnarray}
V_1(Q)=V_0 (Q) +\Delta V_1(Q), \\
\Delta V_1(Q)=\frac{1}{64 \pi^2} Str M^4 (log\frac{M^2}{Q^2}-\frac{3}{2}),
\end{eqnarray}
where $V_0(Q)$ is the tree-level potential evaluated with couplings
renormalized at some scale $Q$, $Str$ denotes the conventional
Supertrace and $M^2$ is the mass matrix for the CP-even sector. As
discussed in \cite{Ellis-Ridolfi}, the radiatively-corrected Higgs
mass-squared matrix is given by the matrix of the second derivatives
of  $V_1$ with respect to the Higgs fields, which is written as a
function of their self-energies. In the MSSM, we know that the most
important contributions to the Higgs self-energies at the one-loop
level come from the diagrams with the top quark (and its scalar
partner) circulating in the loop, due to the large top Yukawa
coupling. However, for very large values of $\tan \beta$, the
bottom-sbottom contributions can become non-negligible. Therefore,
for our settings, the dominant contributions to the Supertrace in
the MSSM+1CHT are due to the top-stop and bottom-sbottom loops.
 Within this approximation, it happens that the squared-mass matrix of the
 CP-even Higgs bosons only gets corrected along its $(1,1)$ and $(2,2)$ elements, given as follows:
\begin{eqnarray}
(\Delta M_S^2)_{1,1} = \frac{3 }{8 \pi^2 } \lambda^2_b m_b^2 log
\frac{m_{\tilde{b}}^4}{m_b^4},  \label{b-sb}\nonumber \\
(\Delta M_S^2)_{2,2} = \frac{3 }{8 \pi^2 } \lambda^2_t m_t^2 log \frac{m_{\tilde{t}}^4}{m_t^4},  \label{t-st}
\end{eqnarray}
where $\lambda_{t,b} $ are the Yukawa couplings and the D-terms are
omitted.
%Possible effects from stop and sbottom mixing are also neglected.\\
In short, in the MSSM+1CHT, the radiative corrections to the Higgs boson
masses must include the dominant
contribution from the top-stop and bottom-sbottom systems. For this, it is enough to suitably modify the elements
$(\Delta M_S^2)_{1,1}$, $(\Delta M_S^2)_{2,2}$  to the squared-mass matrix $M_S^2$ of the CP-even Higgs boson.
Furthermore, as intimated already, we must also evaluate the
contribution from the
fermionic partner of the Higgs Superfields, which includes the
Higgs-Higgsino triplets, because there is a potentially large
effect emerging in the calculation of the squared-mass matrix of the
CP-even Higgs bosons when the parameter $\lambda$ is large.
Similarly to the top-stop and bottom-sbottom corrections, we estimate that
the
correction from the Higgs-Higgsino only  modifies the element
$(M_S^2)_{3,3}$
\begin{eqnarray}
(\Delta M_S^2)_{3,3} =\frac{3 }{8 \pi^2 } \lambda^2 m_{\chi^\pm}^2  log \frac{m_{\chi^\pm}^4}{m_{H^\pm}^4}, \label{h-h}
\end{eqnarray}
where $\lambda$ is the Yukawa coupling that appears in the
Superpotential of the Higgs Superfields and -- within our
approximation -- we take $m_{\chi^\pm}$ and $m_{H^\pm}$ as the
mass scales of the lightest charginos and charged Higgs bosons,
respectively, i.e., $m_{\chi^\pm}\simeq m_{\chi^\pm_1}$ and
$m_{H^\pm}\simeq m_{H^\pm_1}$. D-terms are omitted, as well as
possible effects from stop, sbottom and Higgsino mixing. Previous
studies of Higgs mass bounds of this model were considered by J.
R. Espinosa and M. Quir\'os \cite{espqui2}, who concluded that the
lightest Higgs boson of the model satisfy the bound,
\begin{eqnarray}
m_{H_1^0} \lsim m_Z \sqrt{\cos^2(2 \beta) + 1/2 (\lambda^2
v^2/m_Z^2) \sin^2(2 \beta)}.
\end{eqnarray}
Thus, for values of $\lambda$ that are consistent with
perturbativity, wich then implies a bound of the order $m_{H_1^0}
\lsim 155$ GeV. Throughout this paper we take values of $\lambda$
that do not saturate this bound.
 A more complete calculation of the
radiative corrections at one-loop level for this model is in
progress \cite{future-work}.

The main consequence of these radiative corrections is that the lightest CP-even Higgs mass can be enhanced
at such levels that it makes it
possible to pass current experimental  bounds from LEP2.
Besides, the radiative corrections affect mainly the neutral Higgs bosons sector, in particular the
production of the neutral scalar Higgs in $e^+ e^-$ collisions, which is
the
Higgs-strahlung processes $e^+ e^- \to H_i^0 Z^0$, whose  cross sections
can be expressed in terms  of the
SM Higgs boson (herein denoted by $\phi_{SM}^0$)
production formula and the Higgs-$Z^0Z^0$ coupling, as follows
\cite{table-LEP}:
\begin{eqnarray}
\sigma_{H_i^0 Z} = R^2_{H_i^0 Z^0 Z^0} \sigma_{H_i^0 Z}^{SM}, \nonumber \\
R^2_{H_i^0 Z^0 Z^0}  = \frac{g^2_{H_i^0 Z^0 Z^0}}{g^2_{\phi_{SM}^0 Z^0
Z^0} },
\end{eqnarray}
where $g^2_{H_i^0 Z^0 Z^0}$ is the coupling $H_i^0 Z^0 Z^0 $  in the
MSSM+1CHT
and $g^2_{\phi_{SM}^0 Z^0 Z^0}$ is the SM coupling
$\phi_{SM}^0  Z^0 Z^0$,
which obey the relation
\begin{eqnarray}
\sum_{i=1}^{3} g^2_{H_i^0 Z^0 Z^0} = g^2_{\phi_{SM}^0 Z^0 Z^0}.
\end{eqnarray}

In particular, for our model the factor  $R^2_{H_i^0 Z^0 Z^0}$  is given by:
\begin{eqnarray}
R^2_{H_i^0 Z^0 Z^0} = (V_{1i}^S  c_\beta +V_{2 i}^S s_\beta)^2,
\end{eqnarray}
where $V^S_{ij}$ denote the $ij$-elements of the rotation matrix for the
CP-even neutral sector,
which relates the physical states $H_i^0$ and the real part of the fields $\phi_1^0$, $\phi_2^0$, $\xi^0$ in the following way:
\begin{eqnarray}
 \left(
\begin{array}{c}
\frac{1}{\sqrt{2}}\phi_1^0 \\
\frac{1}{\sqrt{2}}\phi_2^0 \\
\frac{1}{\sqrt{2}}\xi^0 \\
\end{array}
\right) = \left(
\begin{array}{ccc}
V_{11}^S & V_{12}^S & V_{13}^S \\
V_{21}^S & V_{22}^S & V_{23}^S \\
V_{31}^S & V_{32}^S & V_{33}^S \\
\end{array}
\right) \left(
\begin{array}{c}
H_1^0 \\
H_2^0 \\
H_3^0 \\
\end{array}
\right),
\end{eqnarray}
where the $V^S_{ij}$ are modified by the one-loop radiative corrections
to the CP-even sector of our model. For our
numerical analysis of the Higgs mass spectrum in the MSSM+1CHT  we
consider the experimental limits on the charged Higgs mass from LEP2
and apply it to the lightest charged Higgs state $H_1^\pm $
\cite{lepbounds,partdat}. The bounds on the neutral Higgs bosons
$H_1^0$, $H_2^0$ are expressed in terms of the LEP2 bounds  for
$R_{H_i^0 Z^0 Z^0}^2$ \cite{table-LEP}. We will show that this excludes
large regions of the parameter space of the MSSM+1CHT model. This is
summarized in Tables \ref{tab:1}--\ref{tab:4}. Herein,
we define as
the ``marginal regions'' those cases that
almost pass LEP2 bounds on the neutral Higgs, i.e.,  when $m_{H_{1,2}^0}
\sim 110 $ GeV and/or $R_{H_{1,2}^0 Z^0 Z^0}^2$  are not consistent with
experimental bounds but for which we expect that the complete calculation
of the
one-loop radiative corrections to the mass of the neutral Higgs boson in
question could enhance its mass, thereby allowing it to eventually pass
said
experimental limits.

%%%%%%%%%%%%%%%%%%%%%%%%%%%%%%%%%%%%%%%%%%%%%%%%%%%%%%%%%%%%%
\subsection{Higgs masses: numerical results}
%%%%%%%%%%%%%%%%%%%%%%%%%%%%%%%%%%%%%%%%%%%%%%%%%%%%%%%%%%%%%

\begin{figure}
\centering
\includegraphics[width=5in]{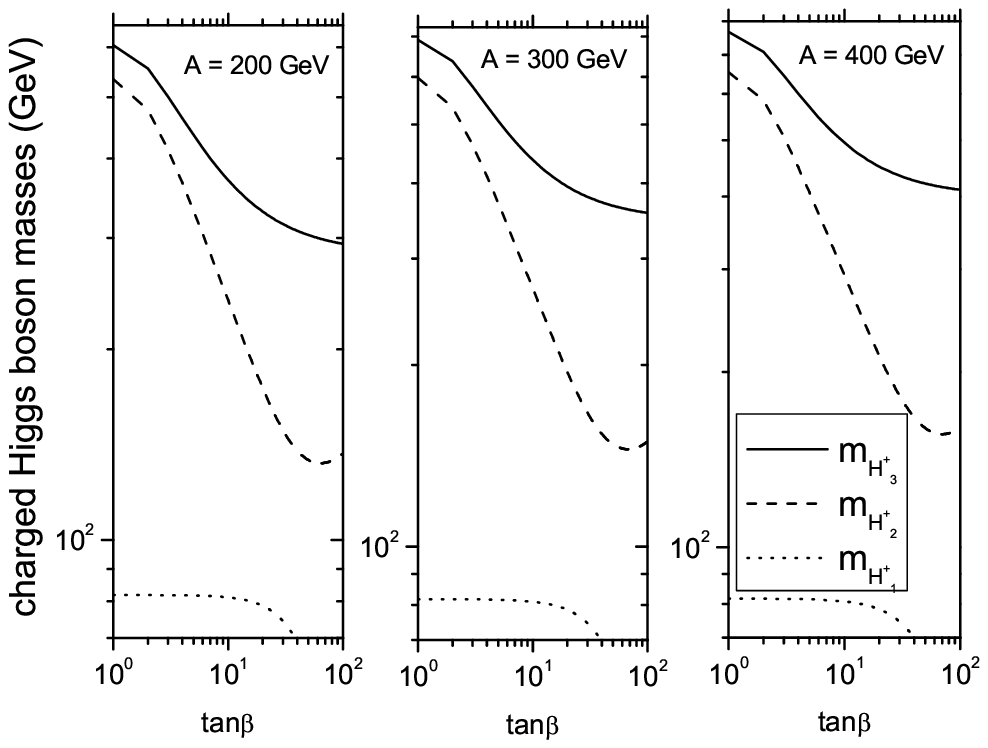}
\caption{Mass spectrum of the charged Higgs bosons, taking
$\mu_{2}=100$ GeV and $\lambda = 0.1$, for: $A=200$ GeV  (left),
$A=300$ GeV (center), $ A=400$ GeV (right).} \label{fig:ec1}
\end{figure}

\begin{figure}
\centering
\includegraphics[width=5in]{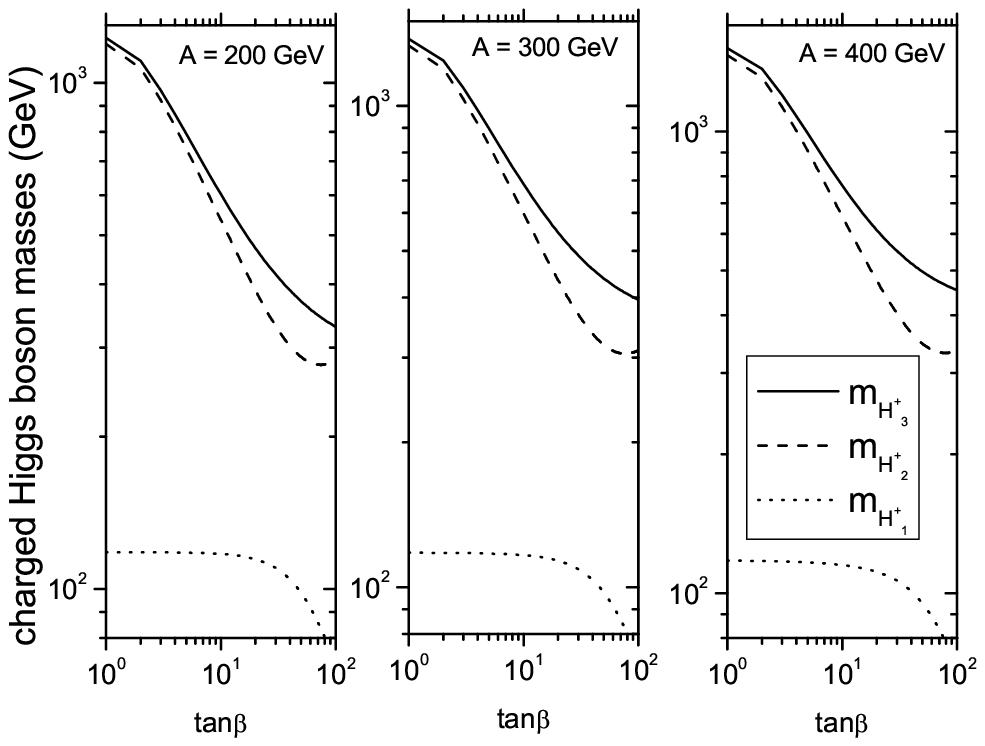}
\caption{ Same as in Figure \ref{fig:ec1}, but taking $\lambda =
0.5$.} \label{fig:ec12}
\end{figure}

\begin{figure}
\centering
\includegraphics[width=5in]{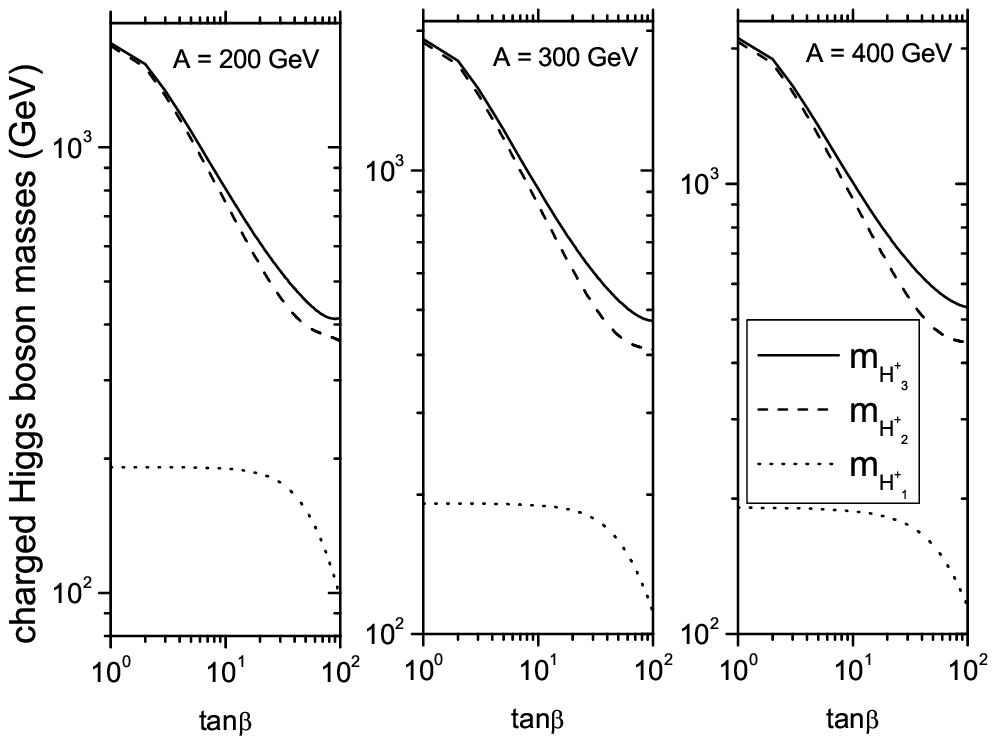}
\caption{ Same as in Figure \ref{fig:ec1}, but taking $\lambda = 1$.}
\label{fig:ec13}
\end{figure}

\begin{figure}
\centering
\includegraphics[width=5in]{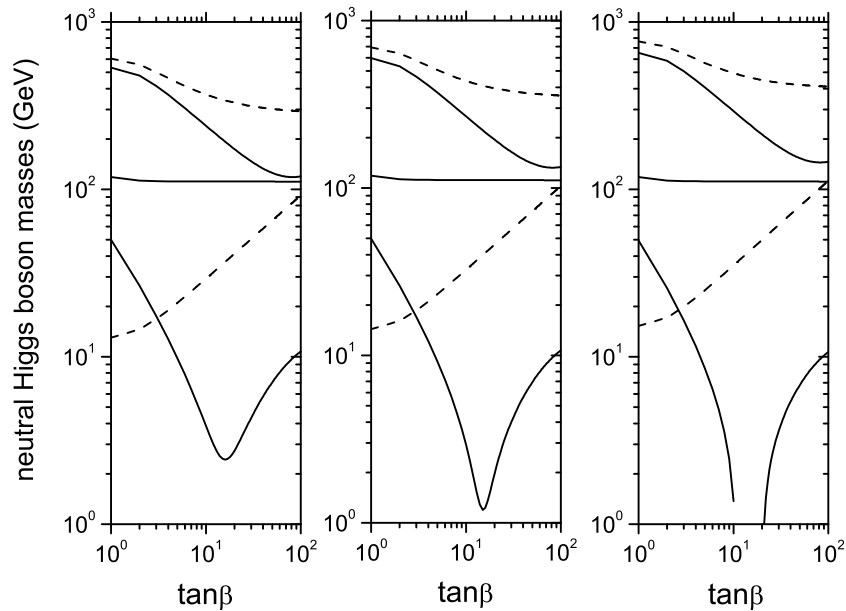}
\caption{Mass spectrum of the neutral Higgs bosons, taking
$\mu_{2}=100$ GeV and $\lambda = 0.1$, for: $A=200$ GeV  (left),
$A=300$ GeV (center), $ A=400$ GeV (right). The solid lines
correspond to scalar and the dashed lines to pseudoscalar
eigenstates.} \label{fig:en1}
\end{figure}

\begin{figure}
\centering
\includegraphics[width=5in]{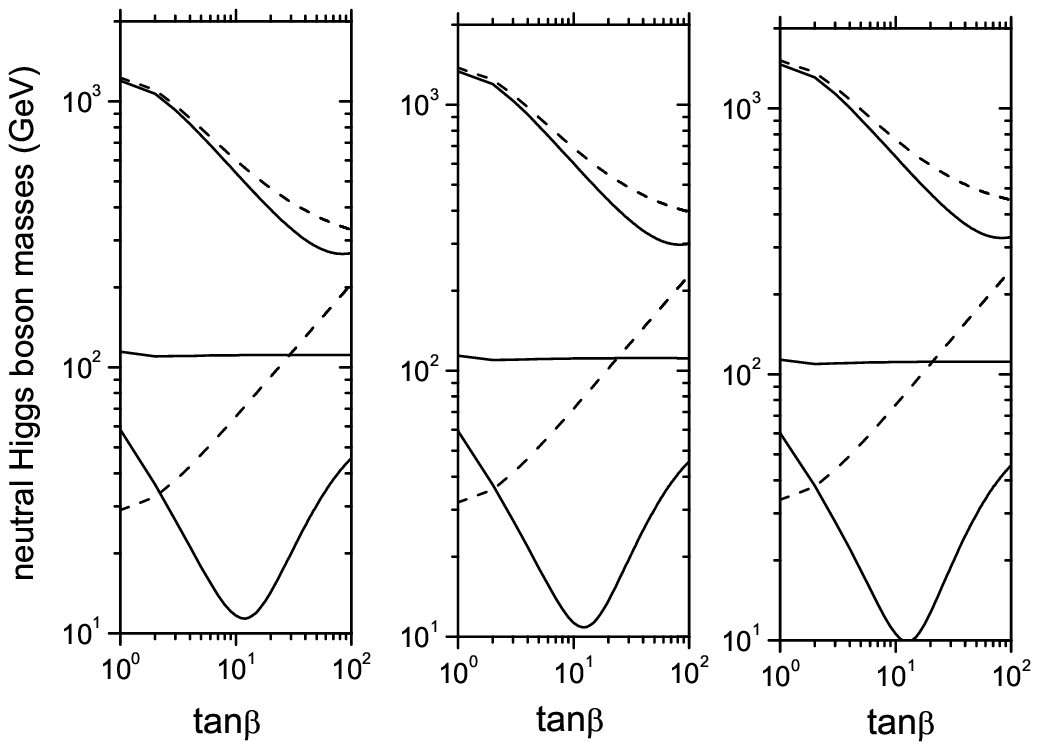}
\caption{ Same as in Figure \ref{fig:en1}, but taking $\lambda = 0.5$.}
\label{fig:en12}
\end{figure}

\begin{figure}
\centering
\includegraphics[width=5in]{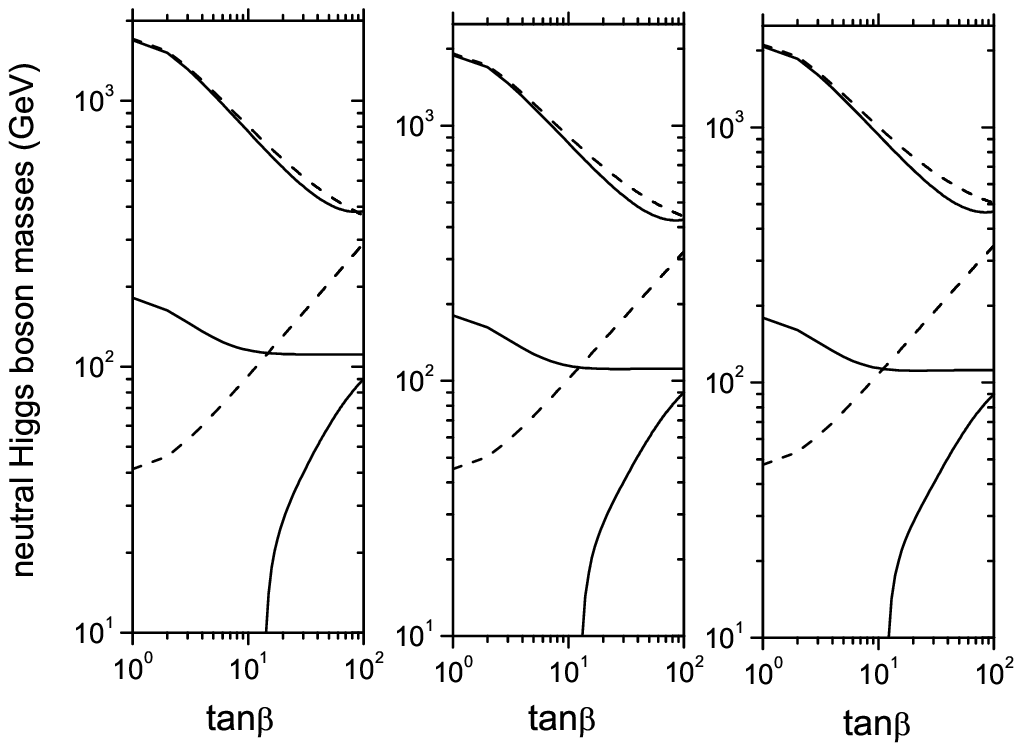}
\caption{ Same as in Figure \ref{fig:en1}, but taking $\lambda = 1$.}
\label{fig:en13}
\end{figure}

Let us consider first Scenario A.  Figures \ref{fig:ec1},
\ref{fig:ec12} and
 \ref{fig:ec13}(\ref{fig:en1}, \ref{fig:en12} and \ref{fig:en13}) show
the results for charged(neutral) Higgs bosons masses
 as a function   of $\tan\beta$, in the range $1 \leq \tan\beta  \leq 100$,
for the cases $\lambda = 0.1,\,0.5,\,1.0$, while taking
$A=200,\,300,\,400$ GeV, respectively. Throughout this paper we
shall assume that the numerical values for stop and sbottom
masses, taken at the electroweak scale, are degenerated. The above
results for charged Higgs massed is based on the tree-level
analysis. Similarly, the coming results for the pseudoscalar
masses is also based on the tree-level formulae. However, the
masses of the neutral CP-even Higgs bosons is based in the
previous discussion of one-loop radiative corrections to the Higgs
masses. For the stop, sbottom and chargino masses we take as input
the value $m = 1$ TeV. In Figure \ref{fig:ec1} we present the
charged Higgs boson masses for $\lambda = 0.1$. We can see that
the lightest charged Higgs boson has a mass $ m_{H^\pm_1} \lsim
m_{W^{\pm}}$, which is not below the theoretical limit that one
obtains in the MSSM. Similarly, Figure \ref{fig:ec12} shows the
charged Higgs boson masses  for the case $\lambda = 0.5$, and
again we have that $m_{H^{\pm}_1} \lsim m_{W^{\pm}}$ is possible
but only for large $\tan\beta$. Furthermore, here it is possible
for both $H_1^{\pm}$ and $H_2^{\pm}$ to be lighter than the top
quark. Figure \ref{fig:ec13} shows the charged Higgs boson masses
for the case $\lambda = 1$: now the lightest charged Higgs boson
has a mass in the range $100$ GeV $ < m_{H^{\pm}_1} < 200$ GeV.

Figure \ref{fig:en1} shows the neutral Higgs spectrum for the case
$\lambda = 0.1$, and we notice the presence of a light CP-even
Higgs boson with 11 GeV $<m_{H_1^0}< 50$ GeV, especially for low
values of  $\tan \beta ~(\leq  5)$, that at first sight it would
seem excluded by the LEP2 experimental limits. In fact, when one
compares the results for $R_{H_1^0 Z^0 Z^0}^2$ obtained for this
model , which measures  the strength of the Higgs-strahlung
process, with the LEP2 bounds  \cite{table-LEP}, which require it
to be less than 0.01, we conclude that this scenario is indeed
excluded,  as it is summarized in our Table \ref{tab:1}. We assume
that the lightest neutral Higgs boson decays predominantly into $b
\bar{b}$ mode. Similarly, Figure \ref{fig:en12} considers the case
$\lambda =0.5$, and again we find 11 GeV $<m_{H_1^0}< 50$ GeV for
$1 \leq \tan \beta \leq 100$. However, we find that, for $\tan
\beta \leq 77$, $R_{H_1^0 Z^0 Z^0}^2 $ is within the range allowed
by LEP2.
 There is also a region where  111 GeV $<m_{H_2^0}< 114$ GeV, which we
identify as marginal.
Finally, Figure \ref{fig:en13} corresponds to the case $\lambda =
1.0$, and we find that  14 GeV $<m_{H_1^0}< 89$ GeV,  for $15 \leq
\tan \beta  $ and, although  $R_{H_1^0 Z^0 Z^0}^2 <0.01 $, again we find
that this is a marginal region because 111 GeV $<m_{H_2^0}< 114$
GeV.

As a lesson from these figures, for the case
$\lambda=0.5$,  we find that the LEP2 limit on the charged Higgs mass
allows cases where $m_{H^{\pm}_1} \lsim m_{W^{\pm}}$, while the
neutral Higgs bosons (chiefly $H^0_1$)
satisfy the experimental limits of LEP2.
However, the case $\lambda =0.1$  is not a  favorable scenario,
because $R_{H_1^0 Z^0 Z^0}$ does not satisfy the experimental bounds.
In contrast, for $\lambda=1.0$, the charged Higgs boson masses are
significantly heavier.  A
complete list of bounds for all cases  considered within Scenario A
is shown in Table \ref{tab:1}.
%%%%%%%%%%%%%%%%%%%%%%%%%%%%%%%
% Table 1
%%%%%%%%%%%%%%%%%%%%%%%%%%%%%%%%%
\squeezetable
\begin{table*}[htdp]
\caption{\label{tab:1} Analysis of  $R^2_{H_i^0 Z^0 Z^0}$  consistent with
LEP. We consider experimental limits allowed by LEP2 for charged and
neutral Higgs bosons,  for Scenario A with $A=200, \, 300,
\, 400$  GeV and $\mu_2 =100$ GeV.  }
%\begin{center}
\begin{tabular}{|c|c|c|c|c|}
\hline\hline
$\lambda = 0.1$ & $ \tan \beta \leq 5$   &
\begin{tabular}{c}
$m_{H_1^\pm} \approx 81$ GeV \\
11 GeV $< m_{H_1^0} < 50 $ GeV \\
111 GeV $< m_{H_2^0} <118$ GeV
\end{tabular} &
\begin{tabular}{c}
$0.15 < R_{H_1^0 Z^0 Z^0}^2 < 0.8$ \\
$R_{H_2^0 Z^0 Z^0}^2 < 0.8 $
\end{tabular}
& Excluded by $R_{H_1^0 Z^0 Z^0}^2 $ \\ \hline
$\lambda = 0.5$ & $ \tan \beta \leq 77$  &
\begin{tabular}{c}
79.8 GeV $ < m_{H_1^\pm} <118 $ GeV  \\
12 GeV $ < m_{H_1^0} < 50$ GeV  \\
111 GeV $< m_{H_2^0}<114 $ GeV
\end{tabular}  &
\begin{tabular}{c}
$0.002 < R_{H_1^0 Z^0 Z^0}^2 < 0.2$ \\
$0.9 < R_{H_2^0 Z^0 Z^0} ^2 $
\end{tabular}  &
\begin{tabular}{c}
Allowed by $R_{H_1^0 Z^0 Z^0}^2$,  \\
but marginal for $R_{H_2^0 Z^0 Z^0}^2$
\end{tabular} \\ \hline
$\lambda = 1$ & 15 $ \leq \tan \beta $  &
\begin{tabular}{c}
89 GeV $ < m_{H_1^\pm} <187 $ GeV  \\
14 GeV $ < m_{H_1^0} < 89$ GeV  \\
111 GeV $< m_{H_2^0}<114 $ GeV
\end{tabular}  &
\begin{tabular}{c}
$ R_{H_1^0 Z^0 Z^0}^2 < 0.01$ \\
$0.9 < R_{H_2^0 Z^0 Z^0}^2  $
\end{tabular}  &
\begin{tabular}{c}
Allowed by $R_{H_1^0 Z^0 Z^0}^2$,   \\
but marginal for $R_{H_2^0 Z^0 Z^0}^2$
\end{tabular}
\\  \hline\hline
\end{tabular}
%\end{center}
\label{default1}
\end{table*}
%%%%%%%%%%%%%%%%%%%%%%%%%%%%%%%%

\begin{figure}
\centering
\includegraphics[width=5in]{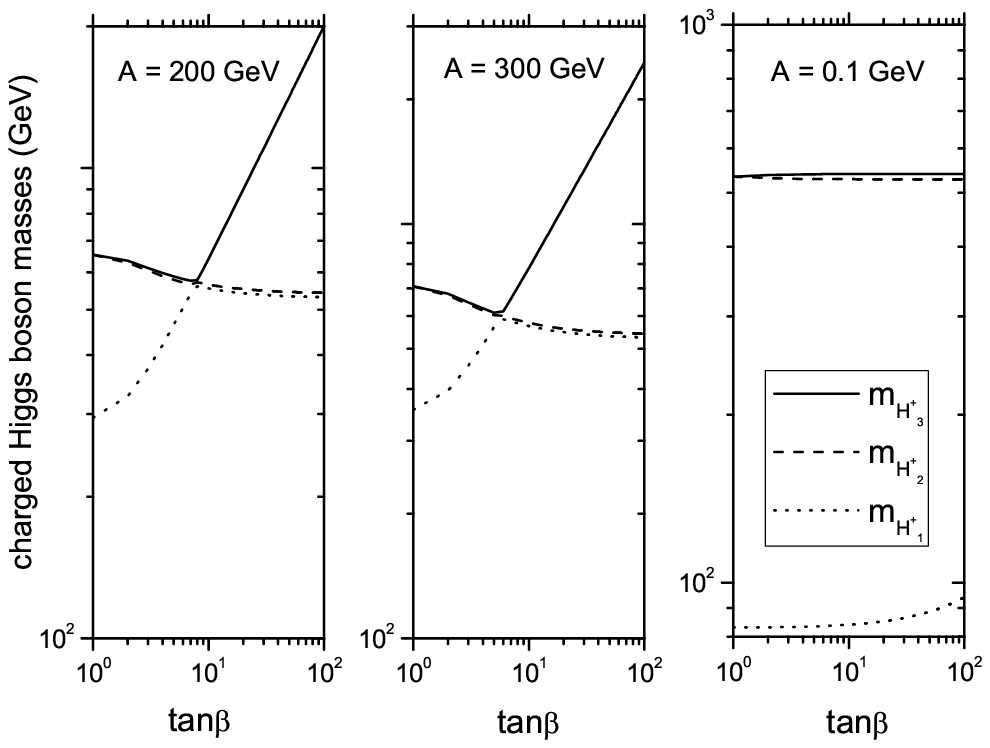}
\caption{ Mass spectrum of the charged Higgs bosons, taking
$\mu_{1}=200$ GeV and $\lambda = 0.1$, for: $A=200$ GeV  (left),
$A=300$ GeV (center), $ A=0.1$ GeV (right).} \label{fig:ec2}
\end{figure}

\begin{figure}
\centering
\includegraphics[width=5in]{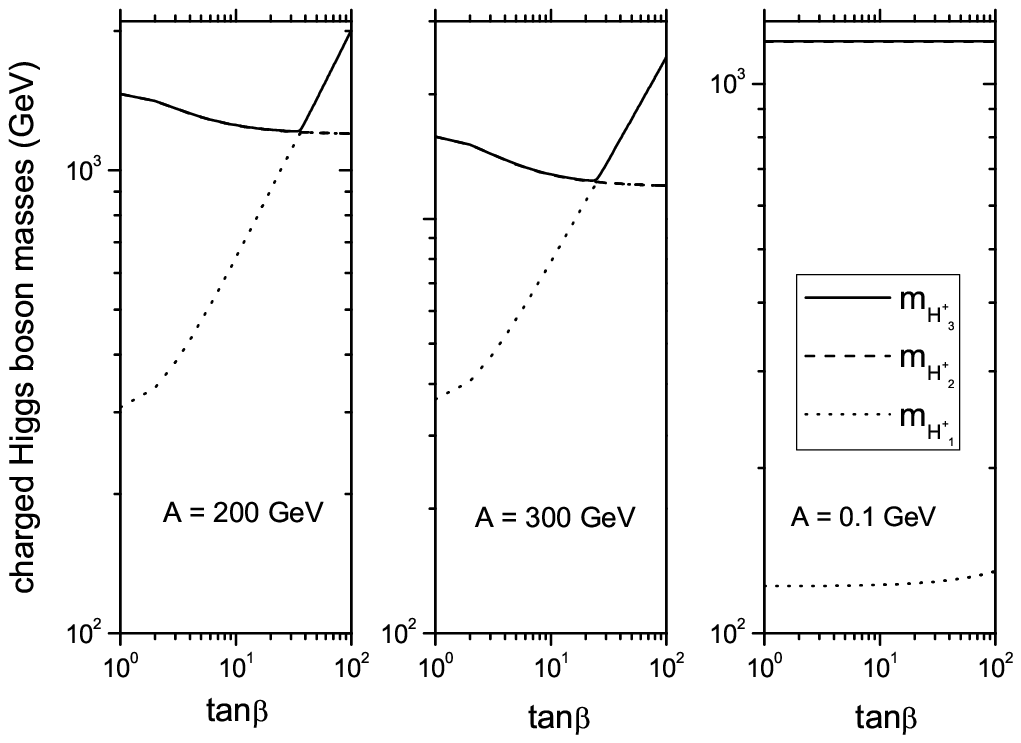}
\caption{ Same as in Figure \ref{fig:ec2}, but taking $\lambda =
0.5$.} \label{fig:ec22}
\end{figure}

\begin{figure}
\centering
\includegraphics[width=5in]{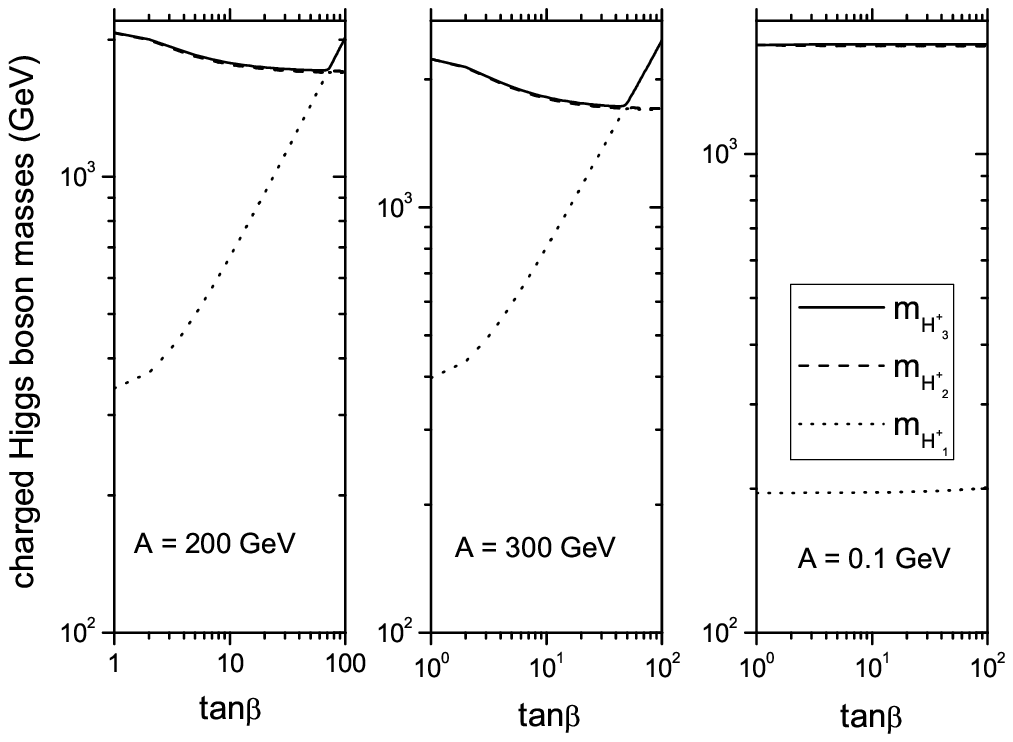}
\caption{Same as in Figure \ref{fig:ec2}, but taking $\lambda = 1$.}
\label{fig:ec23}
\end{figure}

\begin{figure}
\centering
\includegraphics[width=5in]{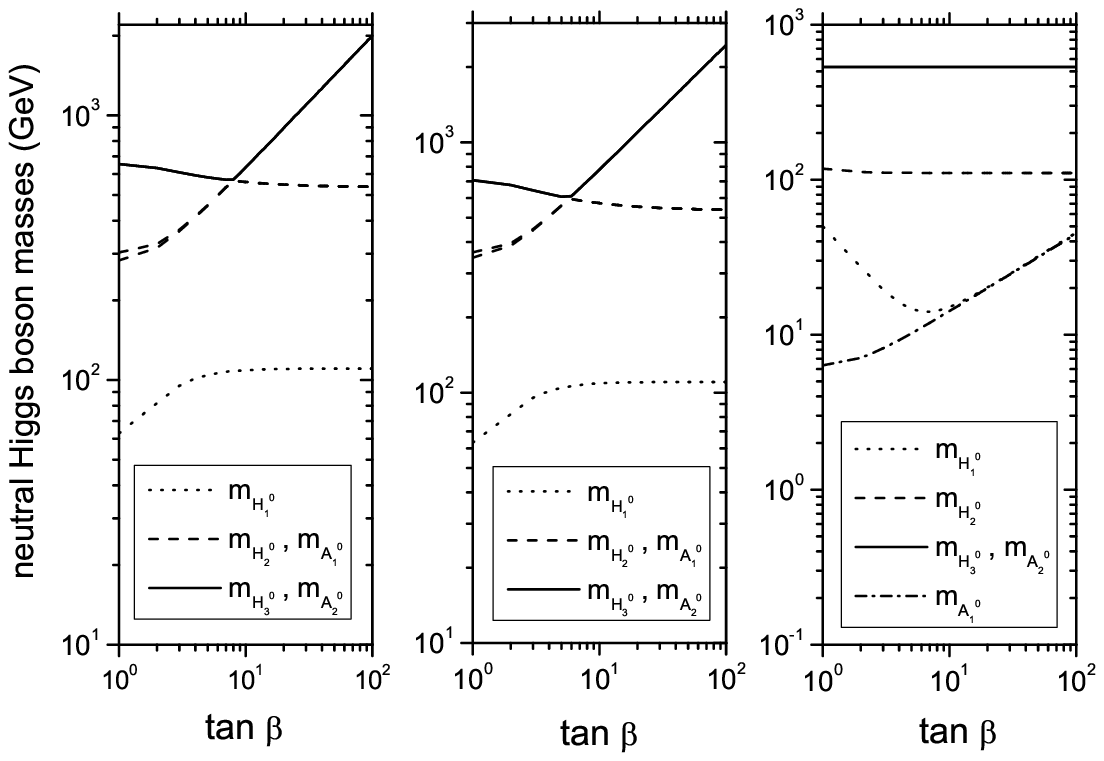}
\caption{ Mass spectrum of the neutral Higgs bosons, taking
$\mu_{1}=200$ GeV and $\lambda = 0.1$, for: $A=200$ GeV  (left),
$A=300$ GeV (center), $ A=0.1$ GeV (right).} \label{fig:en2}
\end{figure}

\begin{figure}
\centering
\includegraphics[width=5in]{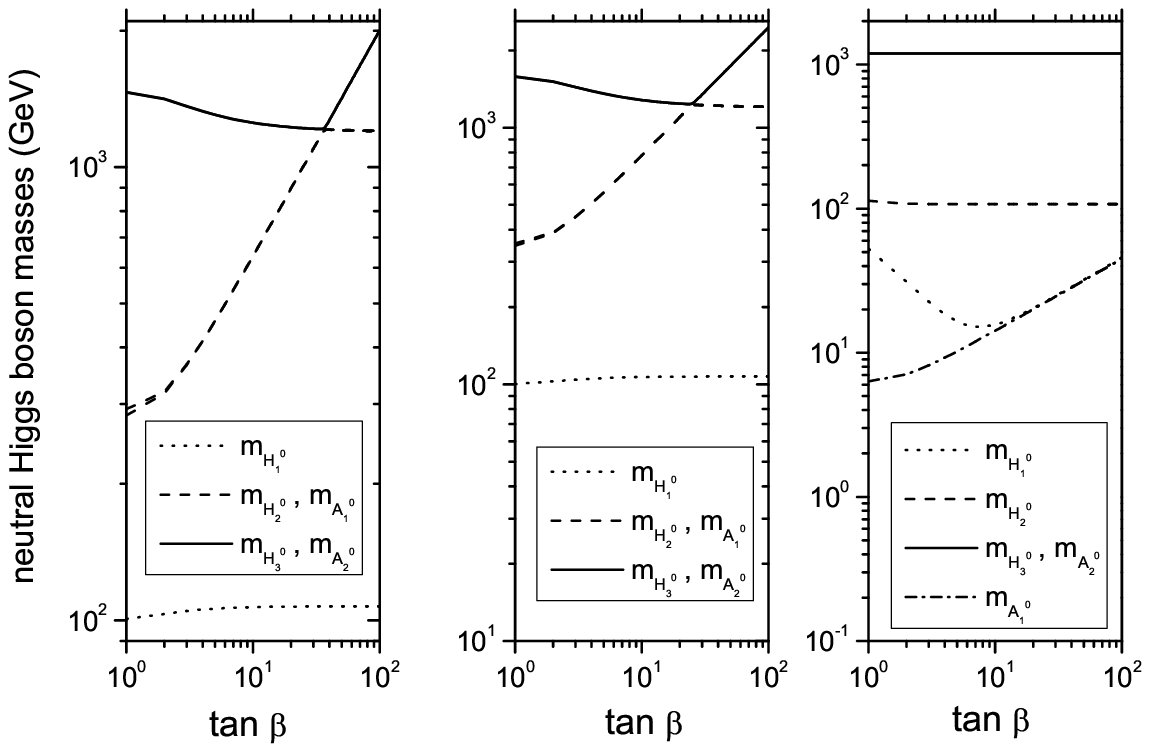}
\caption{ Same as in Figure \ref{fig:en2}, but taking $\lambda = 0.5$.}
\label{fig:en22}
\end{figure}

\begin{figure} \centering
\includegraphics[width=5in]{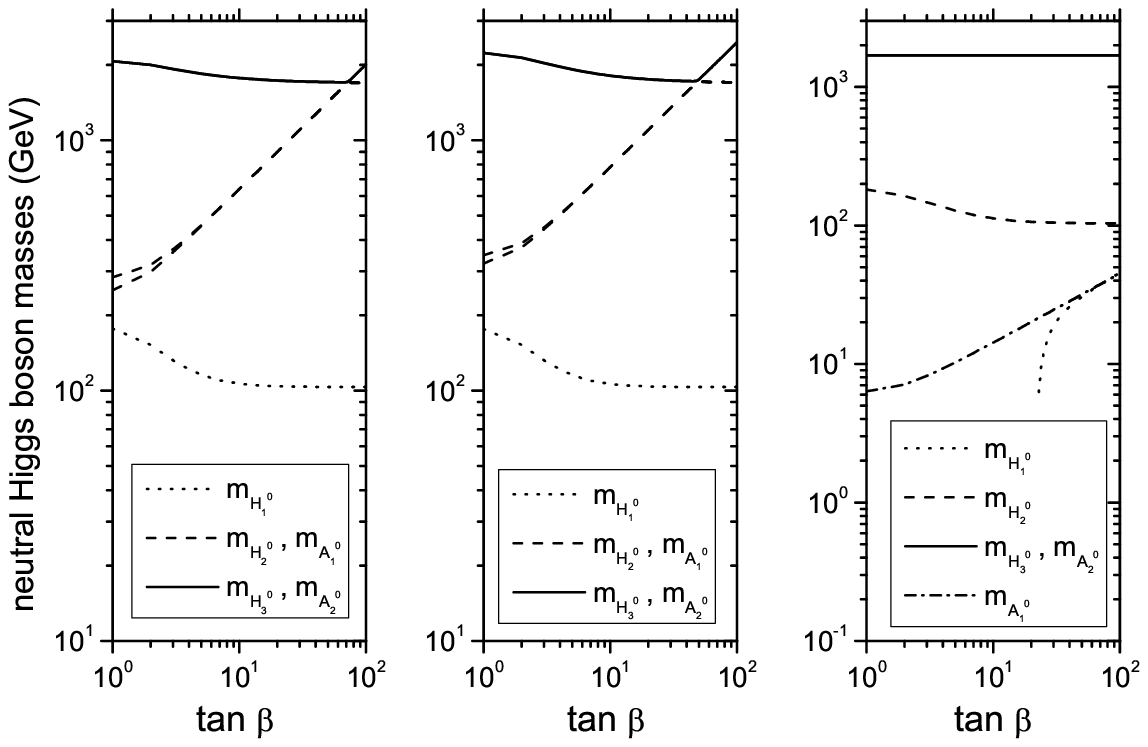}
\caption{Same as in Figure
\ref{fig:en2}, but taking $\lambda = 1$.} \label{fig:en23}
\end{figure}

For Scenario B, Figures \ref{fig:ec2}, \ref{fig:ec22} and
\ref{fig:ec23}(\ref{fig:en2}, \ref{fig:en22} and \ref{fig:en23})
show the charged(neutral) Higgs bosons masses, as a function  of
$\tan\beta$ in the range $1 \leq \tan\beta \leq 100$, and for the
cases $\lambda = 0.1,\,0.5,\,1.0$, taking $A=200,\,300,\,0.1$ GeV,
respectively. The lowest value ($A=0.1$) is designed in order to get
charged Higgs masses below the top mass. Let us comment first the
results found for the charged Higgs mass in the case $\lambda =
0.1$, that appear in Figure \ref{fig:ec2}. We can see that the
lightest charged Higgs boson has a mass above $ 300$ GeV for
$A=200,\,300$ GeV, while even for $A=0.1$ GeV, it has a mass above $
m_{W^{\pm}}$, but it is still lighter than the top quark. Similarly,
Figure \ref{fig:ec22} shows the charged Higgs boson masses for the
case $\lambda = 0.5$.
 We find that, for $A=200,\,300$ GeV,  $m_{H_1^\pm} > 300$ GeV, while, for $A=0.1$ GeV,
the mass is still in the range $100$ GeV $ < m_{H^{\pm}_1} < m_t$. In
turn, Figure \ref{fig:ec23} shows the charged Higgs boson masses for
$\lambda = 1$. Now, we have that the lightest charged Higgs boson is
heavier than the top quark, even for $A=0.1$ GeV.
%%%%%%%%%%%%%%%%%%%%%%%%%%%%%%%
% Table 2
%%%%%%%%%%%%%%%%%%%%%%%%%%%%%%%%%
\squeezetable
\begin{table*}[htdp]
\caption{\label{tab:2} Analysis of  $R^2_{H_i^0 Z^0 Z^0}$  consistent with
LEP. We consider experimental limits by LEP2 for charged and neutral
Higgs bosons,  for Scenario B with $A =200, \, 300$  GeV and
$\mu_1=200$ GeV.}
%\begin{center}
\begin{tabular}{|c|c|c|c|c|}
\hline\hline
$\lambda = 0.1$ & $10 \leq \tan \beta \leq 100$   &
\begin{tabular}{c}
294 GeV $< m_{H_1^\pm} < 532$ GeV \\
 $ m_{H_1^0} \approx 110 $ GeV \\
300 GeV $< m_{H_2^0} <538$ GeV
\end{tabular} &
\begin{tabular}{c}
$0.99 < R_{H_1^0 Z^0 Z^0}^2 $ \\
$R_{H_2^0 Z^0 Z^0}^2 < 0.001 $
\end{tabular}
& Marginally allowed by $R_{H_1^0 Z^0 Z^0}^2$  \\ \hline $\lambda = 0.5$
& 1 $ \leq \tan \beta \leq 100$  &
\begin{tabular}{c}
300 GeV $ < m_{H_1^\pm} <1200$ GeV  \\
100 GeV $ < m_{H_1^0} < 107$ GeV  \\
290 GeV $< m_{H_2^0}<1200 $ GeV
\end{tabular}  &
\begin{tabular}{c}
$0.99 < R_{H_1^0 Z^0 Z^0}^2 $ \\
$R_{H_2^0 Z^0 Z^0}^2 <0.001  $
\end{tabular}  &
\begin{tabular}{c}
Marginally allowed by $R_{H_1^0 Z^0 Z^0}^2$
%\\ for $H_2^0 Z^0 Z^0$
\end{tabular} \\ \hline
$\lambda = 1$ & $ 1\leq \tan \beta \leq 100 $  &
\begin{tabular}{c}
340 GeV $ < m_{H_1^\pm} <1690 $ GeV  \\
104 GeV $ < m_{H_1^0} < 176$ GeV  \\
252 GeV $< m_{H_2^0}<1700 $ GeV
\end{tabular}  &
\begin{tabular}{c}
$0.99< R_{H_1^0 Z^0 Z^0}^2 $ \\
$R_{H_2^0 Z^0 Z^0}^2 <0.001  $
\end{tabular}  &
\begin{tabular}{c}
Allowed by $R_{H_1^0 Z^0 Z^0}^2$
% \\ marginal region for $H_2^0 Z^0 Z^0$
\end{tabular}
\\  \hline\hline
\end{tabular}
%\end{center}
\label{default2}
\end{table*}

Let us now discuss the neutral Higgs spectrum. Figure \ref{fig:en2}
shows the case $\lambda = 0.1$ for $A=200$, 300 GeV, where one finds
that 60 GeV $<m_{H_1^0}<110$ GeV, for $1 \leq \tan \beta \leq  100$,
but the region allowed by $R_{H_1^0 Z^0 Z^0}^2$ corresponds to $10 \leq
\tan \beta$, while the parameter area corresponding to  $m_{H_1^0} \sim
110$ GeV is of marginal type. Then, the case $A=0.1$ GeV gives
neutral Higgs masses within the range 14 GeV $<m_{H_1^0}<50$ GeV,
$m_{H_2^0} \sim 110$ GeV, for which one finds that
$R_{H_1^0 Z^0 Z^0}^2<0.01$. In this case we have a marginally allowed
region. Figure \ref{fig:en22} corresponds to the
case $\lambda =0.5$, for $A=200$, 300 GeV, and now we have 100 GeV
$<m_{H_1^0}< 107$ GeV for  $1 \leq \tan \beta \leq 100$. However, we
find that $0.9 <R_{H_1^0 Z^0 Z^0}^2 $, so that this region is marginal for
$m_{H_1^0}$. For the case $A=0.1$ GeV we have 16 GeV $<m_{H_1^0}<50$
GeV, $m_{H_2^0} \sim 107$ GeV, but also $R_{H_1^0 Z^0 Z^0}^2<0.01$, as we
can see in Table \ref{tab:3}. We consider that this is a
marginally allowed region by LEP2. Furthermore, Figure  \ref{fig:en23}
includes the case $\lambda = 1$, for $A=200$, 300 GeV, and we find
104 GeV $<m_{H_1^0}< 176$ GeV, with $0.9<R_{H_1^0 Z^0 Z^0}^2  $, again
 a marginal region allowed by LEP2, namely  when $m_{H_1^0}<
115$ GeV and $10\leq \tan \beta $.  Finally, in the case $A=0.1$
GeV, we have 14 GeV $<m_{H_1^0}<46$ GeV, $m_{H_2^0} \sim 104$ GeV,
and $R_{H_1^0 Z^0 Z^0}^2<0.01$, as we can see in Table \ref{tab:3}. We
also identify this as a possible marginal region for LEP2
\cite{table-LEP}.

As a lesson from this second set of figures,
 we can state that, for the cases $\lambda =0.1$, 0.5, 1.0, which
leave the range $\tan \beta \geq 25$ and  $A=0.1$ as acceptable, we
have 84 GeV $ < m_{H^{\pm}} < 200$ GeV, while  the neutral Higgs
bosons lay in the mass range disallowed by the experimental limits
of LEP2. However, in the cases with $A=200$, 300 GeV, we obtain masses of
the charged Higgs boson heavier than the top quark. A complete list
of bounds for all cases considered within Scenario B is shown in
Tables \ref{tab:2} and \ref{tab:3}.

%%%%%%%%%%%%%%%%%%%%%%%%%%%%%%%
% Table 3
%%%%%%%%%%%%%%%%%%%%%%%%%%%%%%%%%
\squeezetable
\begin{table*}[htdp]
\caption{\label{tab:3} Same analysis of Table \ref{tab:2}, but taking $A =0.1$  GeV. }
%\begin{center}
\begin{tabular}{|c|c|c|c|c|}
\hline\hline
$\lambda = 0.1$ & $12 \leq \tan \beta \leq 100$   &
\begin{tabular}{c}
84 GeV $< m_{H_1^\pm} < 95$ GeV \\
14 GeV $< m_{H_1^0} < 50 $ GeV \\
$ m_{H_2^0} \approx 110$ GeV
\end{tabular} &
\begin{tabular}{c}
$R_{H_1^0 Z^0 Z^0} ^2< 0.01 $ \\
$ 0.99<R_{H_2^0 Z^0 Z^0}^2  $
\end{tabular}
& \begin{tabular}{c}
Allowed for $R_{H_1^0 Z^0 Z^0}^2$  \\
Marginal region for $R_{H_2^0 Z^0 Z^0}^2$
\end{tabular} \\ \hline
$\lambda = 0.5$ & 4 $ \leq \tan \beta \leq 100$  &
\begin{tabular}{c}
121 GeV $ < m_{H_1^\pm} <129$ GeV  \\
16 GeV $ < m_{H_1^0} < 50$ GeV  \\
$m_{H_2^0}\approx 107 $ GeV
\end{tabular}  &
\begin{tabular}{c}
$ R_{H_1^0 Z^0 Z^0}^2<0.01 $ \\
$0.98<R_{H_2^0 Z^0 Z^0}^2   $
\end{tabular}  &
\begin{tabular}{c}
Allowed for $R_{H_1^0 Z^0 Z^0}^2$  \\
Marginal region for $R_{H_2^0 Z^0 Z^0}^2$
\end{tabular} \\ \hline
$\lambda = 1$ & $ 27\leq \tan \beta \leq 100 $  &
\begin{tabular}{c}
197 GeV $ < m_{H_1^\pm} <200 $ GeV  \\
14 GeV $ < m_{H_1^0} < 46$ GeV  \\
103 GeV $< m_{H_2^0}<105 $ GeV
\end{tabular}  &
\begin{tabular}{c}
$R_{H_1^0 Z^0 Z^0}^2 <0.01$ \\
$0.99< R_{H_2^0 Z^0 Z^0}^2  $
\end{tabular}  &
\begin{tabular}{c}
Allowed  for $R_{H_1^0 Z^0 Z^0}^2$
\\ Marginal region for $R_{H_2^0 Z^0 Z^0}^2$
\end{tabular}
\\  \hline\hline
\end{tabular}
%\end{center}
\label{default3}
\end{table*}
%%%%%%%%%%%%%%%%%%%%%%%%%%%%%%%
% Table 4
%%%%%%%%%%%%%%%%%%%%%%%%%%%%%%%%%
\squeezetable
\begin{table*}[htdp]
\caption{\label{tab:4} Same analysis of Table \ref{tab:2}, but taking  $A =0$  GeV, $\mu_1=200, \, 400, \, 700$ GeV
and
$\lambda = 0.5$. }
%\begin{center}
\begin{tabular}{|c|c|c|c|c|}
\hline\hline
$\lambda = 0.5$ & $1< \tan \beta < 6$   &
\begin{tabular}{c}
121 GeV $< m_{H_1^\pm} < 130$ GeV \\
10 GeV $< m_{H_1^0} < 50 $ GeV \\
 97 GeV$ <m_{H_2^0} < 113$ GeV
\end{tabular} &
\begin{tabular}{c}
$0.006<R_{H_1^0 Z^0 Z^0}^2 < 0.2 $ \\
$ 0.76<R_{H_2^0 Z^0 Z^0}^2  $
\end{tabular}
& \begin{tabular}{c}
Allowed for $R_{H_1^0 Z^0 Z^0}^2$  \\
Marginal region for $R_{H_2^0 Z^0 Z^0}^2$
\end{tabular} \\ \hline\hline
\end{tabular}
%\end{center}
\label{default4}
\end{table*}

\begin{figure}
\centering
\includegraphics[width=5in]{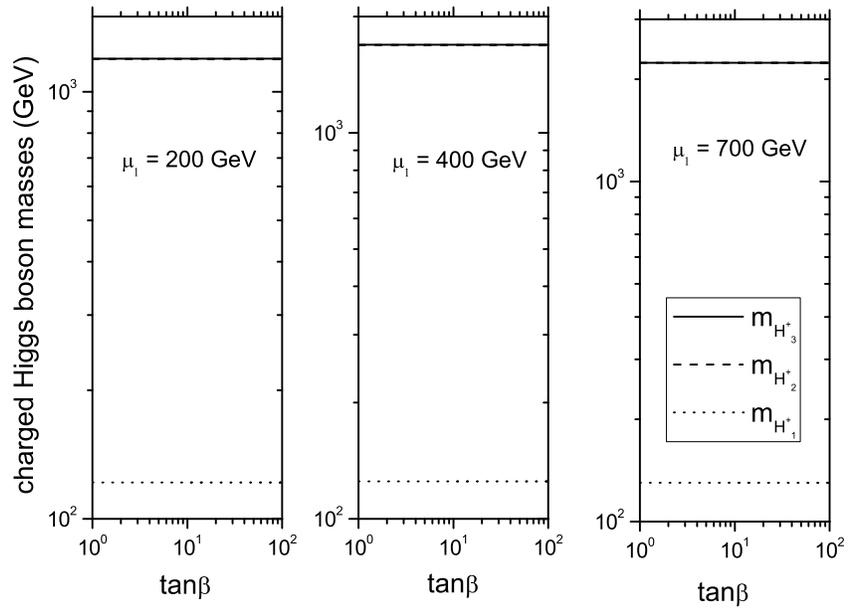}
\caption{ Mass spectrum of the charged Higgs bosons, taking $A=0$
GeV and $\lambda = 0.5$, for: $\mu_1=200$ GeV  (left), $\mu_1=400$
GeV (center), $ \mu_1=700$ GeV (right).} \label{fig:ec3}
\end{figure}

\begin{figure}
\centering
\includegraphics[width=5in]{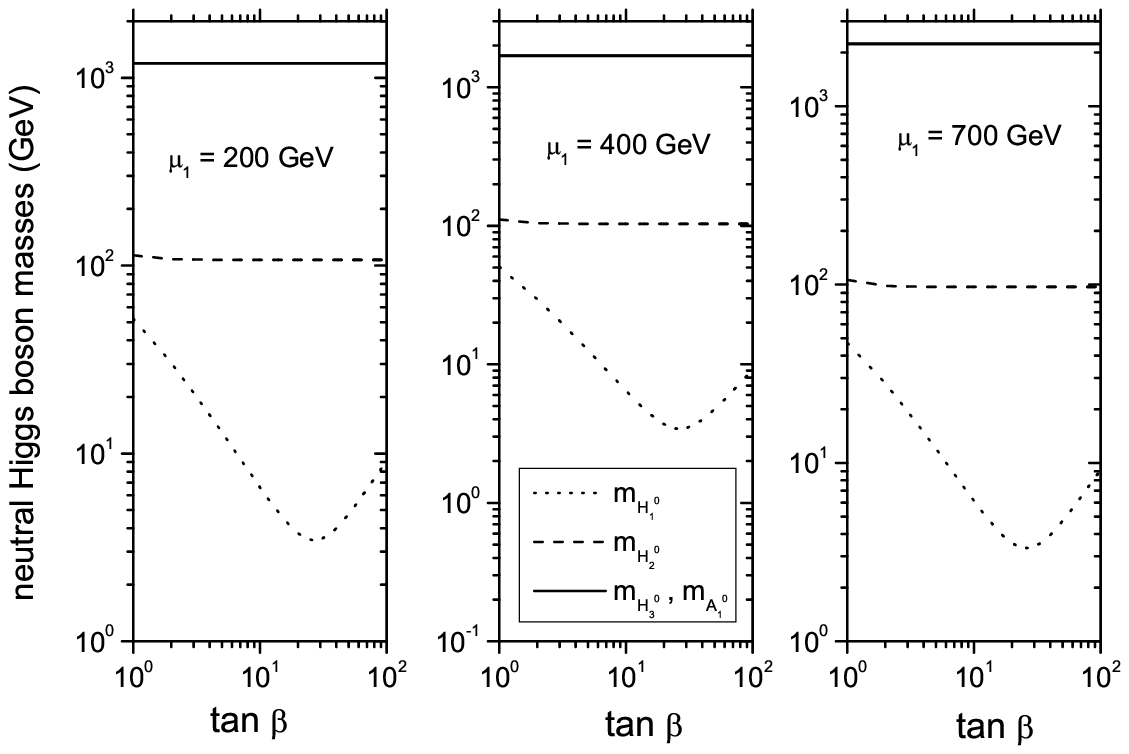}
\caption{ Mass spectrum of the neutral Higgs bosons, taking $A=0$
GeV and $\lambda = 0.5$, for: $\mu_1=200$ GeV  (left), $\mu_1=400$
GeV (center), $ \mu_1=700$ GeV (right). In this scenario (B) only
one massive pseudoscalar appears, as one can see from the limit
$A=0$ (hence $B_1=-A=0$) an extra Goldstone boson should appear, as
discussed in \cite{Espinosa:1991wt}.} \label{fig:en3}
\end{figure}

Finally, in order to consider possible variations with $\mu_1$ and
the behavior in the limit $A\to 0$, we present the charged Higgs
boson masses as function of $\tan\beta$, in Figure \ref{fig:ec3},
for $\lambda = 0.5$, $A=0$ and for $\mu_1=200,\, 400,\, 700$ GeV. We
can see that for all these cases  $ m_{H^{\pm}} \sim 130$ GeV. The
corresponding results for the neutral spectrum are shown in
Figure~\ref{fig:en3} and in Table \ref{tab:4}, which indicate the
presence of two light neutral Higgs states, which lay in the range:
10 GeV $< m_{H^{0}_1} < 50$ GeV, 97 GeV $< m_{H^{0}_2} < 107$ GeV,
for $2 \leq \tan \beta \leq 6$.  In this case we also find
$0.007<R_{H_1^0 Z^0 Z^0}<0.06$, which then gives a small region of $\tan
\beta$ that could be allowed by the LEP2 data. In fact, for $\tan
\beta \geq 6, $ practically all these scenarios are ruled out by
LEP2, because $m_{H^{0}_1} < 10$ GeV \cite{table-LEP}.


%%%%%%%%%%%%%%%%%%%%%%%%%%%%%%%%%%%%%%%%%%%%%%%%%%%%%%%%%%%%%
\section{The vertex $H^{\pm} f f'$ and the decay $t \to H^+_i \, b$}
%%%%%%%%%%%%%%%%%%%%%%%%%%%%%%%%%%%%%%%%%%%%%%%%%%%%%%%%%%

In the two previous papers
\cite{Felix-Beltran:2002tb,Barradas-Guevara:2004qi}, some of us have studied
the bosonic interactions of the charged and neutral Higgs bosons,
now we shall discuss the corresponding fermionic interactions and
their implications for charged Higgs boson production through top
quark decays and via Higgs-strahlung off-top quarks.

%%%%%%%%%%%%%%%%%%%%%%%%%%%%%%%%%%%%%%%%%%%%%%%%%%%%%%%%%%%%%%%%%
\subsection{The Higgs boson coupling to fermions in the MSSM+1CHT}
%%%%%%%%%%%%%%%%%%%%%%%%%%%%%%%%%%%%%%%%%%%%%%%%%%%%%%%%%%%%%%%%%%

As in the MSSM, also in this model only the scalar doublets are coupled
to the fermions, so that the Lagrangian of the Yukawa sector has
the following expression:
\begin{eqnarray}
{\mathcal L}_{Yuk}= -\lambda_u [\bar{u} P_L u \phi_2^0-\bar{u} P_L
d \phi^+_2]-\lambda_d [\bar{d} P_L d \phi_1^0-\bar{d} P_L u
\phi^-_1]+h.c.,
\end{eqnarray}
where the parameters $\lambda_{u,d}$  are related to the fermion
masses via
\begin{eqnarray}
\lambda_u=\frac{\sqrt{2} m_u}{ v_D s_\beta}  \quad ,\quad
\lambda_d=\frac{\sqrt{2} m_d}{ v_D c_\beta} \,.
\end{eqnarray}
The piece of Lagrangian containing the fermion couplings of the charged Higgs
bosons is given by:
%$\bar{u}dH^+_i$, $l\bar{\nu_l}H^+_i$  is:
\begin{eqnarray}
{\mathcal L}_{ffH_i^+}= -\frac{1}{\sqrt{2}v_D } \bar{u} \bigg[
\bigg( \frac{m_d}{c_\beta} (\phi_1^-)^*- \frac{m_u}{s_\beta}
\phi_2^+ \bigg) + \bigg( \frac{m_d}{c_\beta} (\phi_1^-)^*+
\frac{m_u}{s_\beta} \phi^+_2 \bigg) \gamma_5 \bigg]d  +h.c.,
\end{eqnarray}
where $(\phi_1^-)^*$, $\phi_2^+$ are related to the physical
charged Higgs boson states $(H_1^+,H_2^+,H_3^+)$ as follows:
\begin{eqnarray}
(\phi_1^-)^* &=&  \sum_j^3 U_{2,j+1} H_j^+, \quad \phi_2^+=
\sum_j^3 U_{1,j+1} H_j^+, \nonumber \\ H_j^+&=&
(H_1^+,H_2^+,H_3^+).
\end{eqnarray}
%The $U_{ij} ' s$ denote the elements of the rotation matrix for
%the charged sector.

The $U_{jk}$'s denote the elements of the mixing-matrix that
relates the physical charged Higgs bosons ($H^+_1,H^+_2,H^+_3$)
and the Goldstone boson $G^+$ (which gives mass to the $W^+$)
with the fields $\phi_2^+$, $\phi_1^-{}^{*}$, $\xi_2^+$ and
$\xi_1^-{}^{*}$ as follows:
\begin{eqnarray}
\left (
\begin{array}{r}
\phi_2^+ \\
\phi_1^-{}^{*} \\
\xi_2^+ \\
\xi_1^-{}^{*} \\
\end{array} \right ) \ =  \
\left (
\begin{array}{rrrr}
U_{11} & U_{12} & U_{13}  & U_{14} \\
U_{21} & U_{22} & U_{23}  & U_{24} \\
U_{31} & U_{32} & U_{33}  & U_{34} \\
U_{41} & U_{42} & U_{43}  & U_{44} \end{array} \right ) \left (
\begin{array}{r}
G^+ \\
H_1^+ \\
H_2^+ \\
H_3^+ \\
\end{array} \right ) \ .
\end{eqnarray}

Then, the couplings $\bar{u}dH^+_i$, $\bar{\nu_l}lH^+_i$ are given
by:
\begin{eqnarray}
\label{coups1}
g_{H_i^+\bar{u}d} &=& -\frac{i }{v_D \sqrt{2}}  (A_i^{ud}
+V_i^{ud} \gamma_5), \quad g_{H_i^- u \bar{d}}= -\frac{i }{v_D
\sqrt{2}}  (A_i^{ud} -V_i^{ud} \gamma_5), \nonumber\\
g_{H_i^+\bar{\nu_l}l}&=& -\frac{i }{v_D \sqrt{2}} A_i^l (1 +
\gamma_5), \quad g_{H_i^- \nu_l \bar{l}}= -\frac{i }{v_D \sqrt{2}}
A_i^l (1-\gamma_5),
\end{eqnarray}
where $A_i^{ud}$ and $V_i^{ud}$ are defined as:
\begin{eqnarray}
\label{coups2}
A_i^{ud} &=& m_d t_\beta \frac{U_{2,i+1}}{s_\beta}-m_u \cot_\beta
\frac{U_{1,i+1}}{c_\beta}, \label{AV} \nonumber \\ V_i^{ud} &=&
m_d t_\beta \frac{U_{2,i+1}}{s_\beta}+m_u \cot_\beta
\frac{U_{1,i+1}}{c_\beta}, \nonumber \\ A_i^l&=& m_l t_\beta
\frac{U_{2,i+1}}{s_\beta}.
\end{eqnarray}
One can see that the formulae in Eq.~(10) become the couplings
$\bar{u}dH^+_i$, $\bar{\nu_l}lH^+_i$ of the MSSM when we replace
$U_{2,i+1} \to s_\beta$ and $U_{1,i+1} \to
-c_\beta$\cite{kanehunt}. The vertex $\bar{u}dH^+_i$ induces at
tree-level the decay $t\to H^+ \, b$, which will be studied in the
next section.

%%%%%%%%%%%%%%%%%%%%%%%%%%%%%%%%%%%%%%%%%%%%%%%%%%%%%%%%%%%%%
\subsection{The decay $t \to H^+_i \, b$}
%%%%%%%%%%%%%%%%%%%%%%%%%%%%%%%%%%%%%%%%%%%%%%%%%%%%%%%%%%

In order to study this top quark BR we must consider both the
decays $t \to H^+_i \, b$ for $i=1,2$, because both modes could be
kinematically allowed for several parameter configurations within our model. The
decay width of these modes takes the following form:
\begin{eqnarray}
\Gamma(t \to H^+_i b) &=& \frac{g^2}{64 \pi (m_W^2 - 2 g^2 v_T^2)}
m_t^3 \lambda^{1/2}(1,q_{H^+_i},q_b) \nonumber\\
& & \times \, \bigg[ (1-q_{H^+_i}+ q_b) \bigg(
\frac{U^2_{1,i+1}}{s_\beta^2}+ q_b \frac{U^2_{2,i+1}}{c_\beta^2}
\bigg)-4 q_b \frac{U_{1,i+1}U_{2,i+1}}{s_\beta c_\beta} \bigg],
\end{eqnarray}
where $\lambda$ is the usual kinematic factor $\lambda(a,b,c)=
(a-b-c)^2-4bc$ and $q_{b,H^+}=m^2_{b,H^+}/m_t^2$.

Furthermore, we shall neglect the decay width for the light fermion
generations. As we mentioned before, if one replaces $U_{2,i+1} \to
s_\beta$ and $U_{1,i+1} \to -c_\beta$, the formulae of the decay
width also reduce to the MSSM case: see, e.g., \cite{Carena:1999py}. In
general the decay width for $t \to H^+_i \, b$ depends on the
Superpotential parameters through the elements $U_{(1,2),i+1}$. When
we choose an MSSM+1CHT scenario where spontaneous EWSB is
dominated by the effects of the Higgs doublets, the decay width is
practically similar to the one obtained in the MSSM. If we do not, results
can be very different in the two models.
We shall
discuss this in the forthcoming numerical analysis.

%%%%%%%%%%%%%%%%%%%%%%%%%%%%%%%%%%%%%%%%%%%%%%%%%%%%%%%%%%%%%
\subsection{Numerical results for the decay $t \to H^+_i \, b$ in the MSSM+1CHT }
%%%%%%%%%%%%%%%%%%%%%%%%%%%%%%%%%%%%%%%%%%%%%%%%%%%%%%%%%%

We explore several theoretically allowed regions of our MSSM+1CHT
scenario and constrain these by
using experimental bounds on the BR$(t \to H^+ \, b)$. In the so-called
``tauonic Higgs model''
\cite{Abulencia:2005jd}, the decay mode ($H^+ \to \tau^+ \,
{\nu}_{\tau}$) dominates the charged Higgs boson decay width, and
BR$(t \to H^+ \, b)$ is constrained to be less than 0.4 at 95 \%
C.L. \cite{Abulencia:2005jd}. However, if no assumption is
made on the charged Higgs boson decay, BR$(t \to H^+ \, b)$ is
constrained to be less than 0.91 at 95 \% C.L.
\cite{Abulencia:2005jd}. Conversely, the combined LEP data
excluded a charged Higgs boson with mass less than 79.3 GeV at 95 \%
C.L., a limit valid for an arbitrary BR$(H^+ \to \tau^+ \, {\nu_\tau})$
\cite{partdat,LEP}.

Thus, in order to conclude in this regard, we need to discuss all
the charged Higgs boson decays following the steps of  our previous
paper \cite{Barradas-Guevara:2004qi}. In the present work, we shall
evaluate all charged Higgs boson decays relevant masses below that
of the top quark, thus
including the
modes $ \tau^+ {\nu_\tau}, c \bar{s},  c \bar{b}, W^+ H_1^0, W^+
A_1^0$. In what follows, we want to find out whether a light charged
Higgs boson (with $m_{H^{\pm}} \leq m_{W^{\pm}}$) is still allowed
phenomenologically. As usual, we refer to our two benchmark
scenarios.

\begin{figure}
\centering
\includegraphics[width=6in]{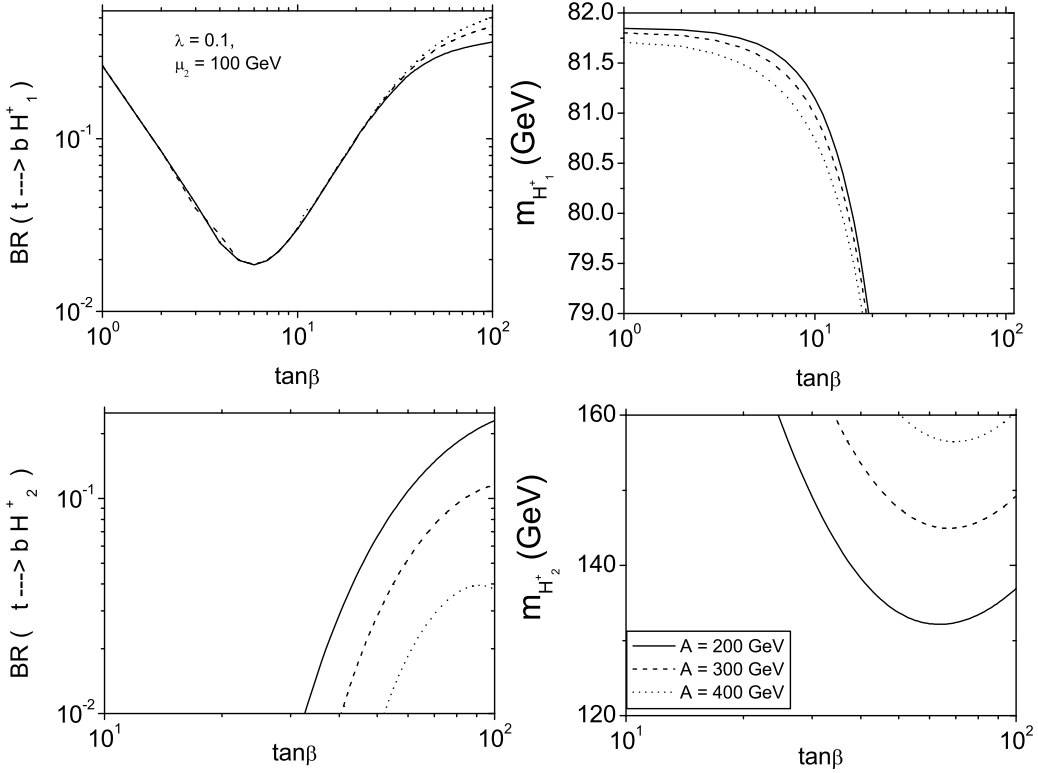} \label{a21}
\caption{ It is plotted:  a) the BR($t \to b \, H^+_1 $) vs. $\tan
\beta$ (top-left), b) the $\tan\beta-m_{H^+_1}$ plane (top-right),
c) the BR($t \to b \, H_2^+ $) vs. $\tan\beta$ (bottom-left), d) the
$\tan\beta-m_{H_2^+}$ plane (bottom-right), in Scenario A by taking
$\lambda = 0.1$, for: $A=200$ GeV (solid), $A=300$ GeV (dashes), $
A=400$ GeV (dots).} \label{fig:dt11}
 \end{figure}

\begin{figure}
\centering
\includegraphics[width=5in]{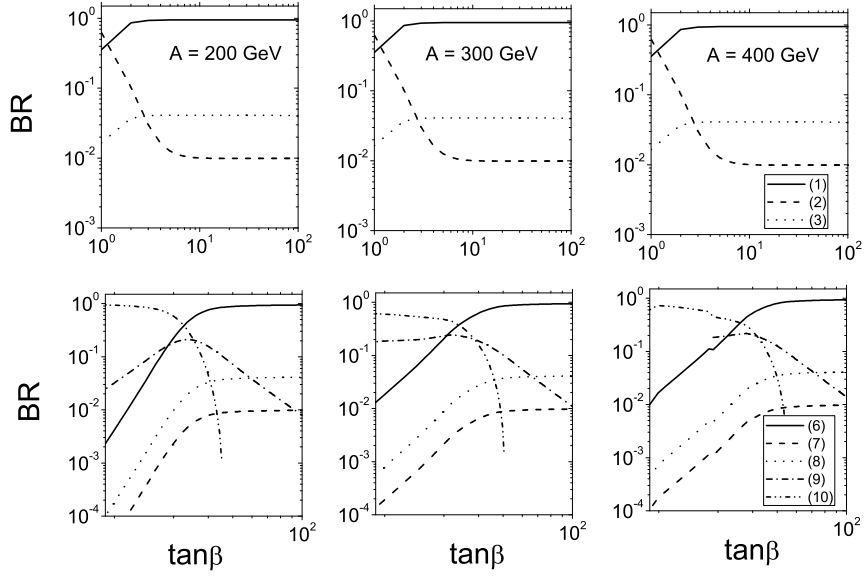}
\caption{ The figure shows the branching ratios of $H^+_1$ (top) and
$H^+_2$ (bottom) decaying into the principal modes in Scenario A,
with $\lambda = 0.1$ and $\mu_2 = 100$ GeV, for: $A=200$ GeV (left),
$A=300$ GeV (center), $ A=400$ GeV (right). The lines correspond to:
(1) BR($H^+_1 \to \tau^+ {\nu_\tau} $), (2) BR($H^+_1 \to c \bar{s}$),
(3) BR($H^+_1 \to c \bar{b}$), (6) BR($H^+_2 \to \tau^+ {\nu_\tau} $),
(7) BR($H^+_2 \to c \bar{s}$), (8) BR($H^+_2 \to c \bar{b}$),(9)
BR($H^+_2 \to W^+ H^0_1$), (10) BR($H^+_2 \to W^+ A^0_1$).}
\label{fig:dh11}
\end{figure}

\vskip0.25cm \noindent {\bf Scenario A.} Remember that this scenario
was defined by taking $B_1=\mu_1=0$, $B_2=-A$, and $\mu_2=100$ GeV
while for $\lambda$ we considered the values
$\lambda=0.1,\,0.5,\,1.0$. In Figure~\ref{fig:dt11} we present plots
of: a) the BR($t \to b \, H_1^+ $) vs. $\tan \beta$; b) the
$\tan\beta-m_{H_1^+}$ plane; c) the BR($t \to b \, H_2^+ $) vs.
$\tan\beta$ and d) the $\tan\beta-m_{H_2^+}$ plane; for the case
$\lambda = 0.1$, taking $A=200,\,300,\,400$ GeV. We can observe that
both modes $t \to b \, H_i^+$ for $i=1,2$ are kinematically allowed.
Also, we see that a charged Higgs boson with mass in the range 80
GeV $< m_{H_1^+} < 82$ GeV and for $1 < \tan\beta < 15$ satisfies the
constraint BR$(t \to b \, H^+) < 0.4$. Furthermore, from the plots
of Figure~\ref{fig:dh11} we can see that in this scenario the dominant
decay mode is into $\tau^+{\nu}_\tau$, therefore they fall within
the scope of the tauonic Higgs model, so that BR$(t \to H^+ \,
b) \leq 0.4$ applies. This also happens for a heavier charged Higgs
boson with mass in the range 125 GeV $< m_{H_2^+} < 160$ GeV and for
$35 < \tan\beta < 100$. However, although this scenario is
consistent with current experimental bounds from Tevatron on BR$(t \to b
\,
H^+)$, it is excluded after one considers the LEP2 limits on the
neutral Higgs sector via $R_{H_1^0 Z^0 Z^0}$ \cite{table-LEP}.

\begin{figure}
\centering
\includegraphics[width=5in]{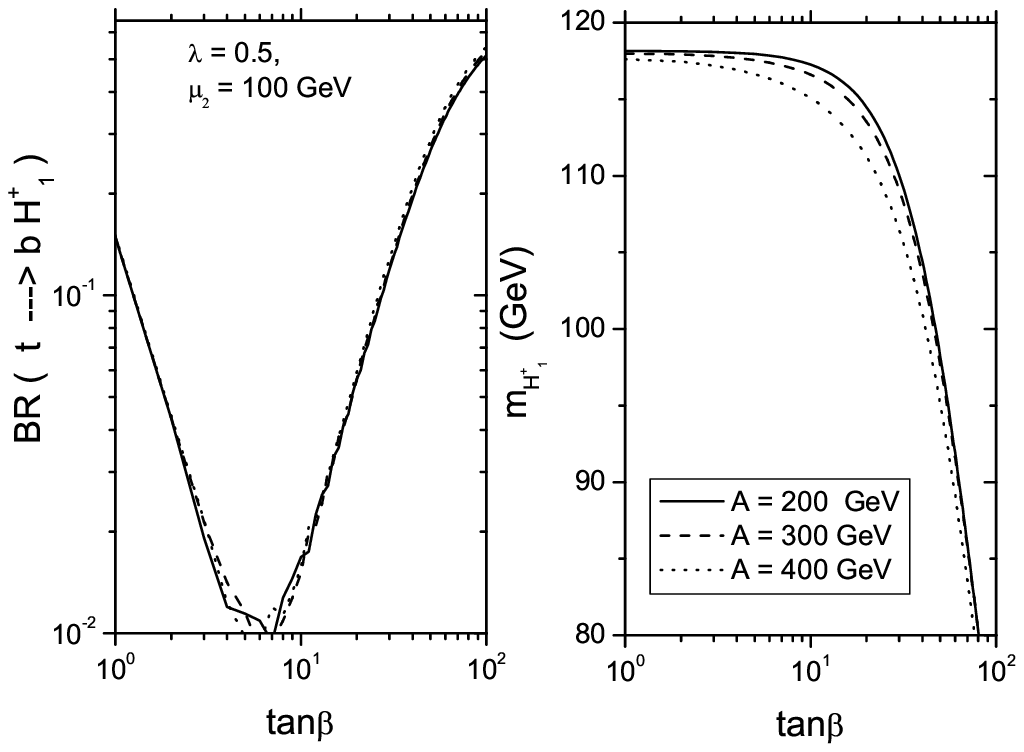}
\caption{ It is plotted: the BR($t \to b \, H_1^+ $) vs. $\tan\beta$
(left), the $\tan\beta-m_{H_1^+}$ plane (right), in Scenario A
taking $\lambda = 0.5$, for: $A=200$ GeV (solid), $A=300$ GeV
(dashes), $ A=400$ GeV (dots).} \label{fig:dt12}
\end{figure}

\begin{figure}
\centering
\includegraphics[width=5in]{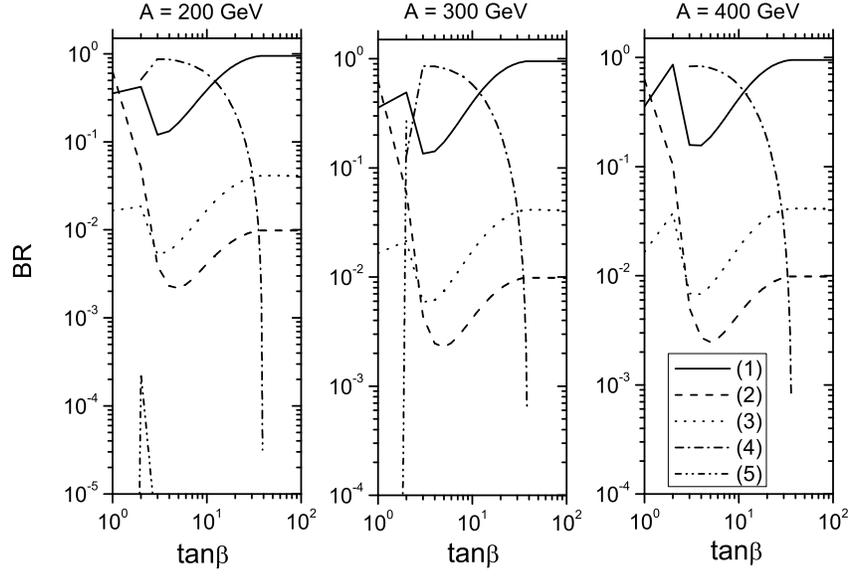}
\caption{ The figure shows the branching ratios of $H^+_1$ decaying
into the principal modes in Scenario A, with $\lambda = 0.5$ and
$\mu_2 = 100$ GeV, for: $A=200$ GeV (left), $A=300$ GeV (center), $
A=400$ GeV (right). The lines correspond to: (1) BR($H^+_1 \to \tau^+
{\nu_\tau} $), (2) BR($H^+_1 \to c \bar{s}$), (3) BR($H^+_1 \to c
\bar{b}$), (4) BR($H^+_1 \to W^+ H^0_1$), (5) BR($H^+_1 \to W^+
A^0_1$).} \label{fig:dh12}
\end{figure}

\begin{figure}
\centering
\includegraphics[width=5in]{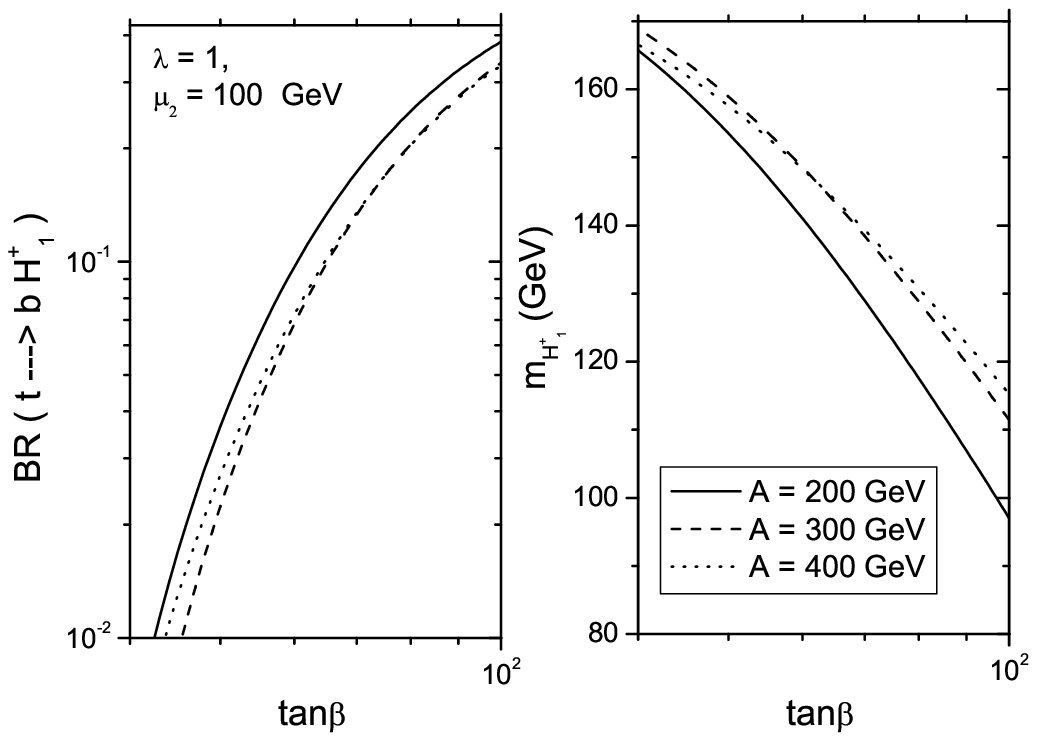}
\caption{ Same as Figure \ref{fig:dt12}, but taking $\lambda = 1$.}
\label{fig:dt13}
\end{figure}

In Figure~\ref{fig:dt12} we present similar plots for the case
$\lambda = 0.5$, taking $A=200,\,300,\,400$ GeV. We can observe that
the mode $t \to b \, H_1^+$ satisfies the constraint BR$(t \to b \,
H^+) < 0.4$ in the ranges 80 GeV $< m_{H_1^+} < 115$ GeV and $30 <
\tan\beta < 100$. Then, from the plots of Figure \ref{fig:dh12}, we see
that in this range the dominant decay mode is into
$\tau^+{\nu}_\tau$, therefore they also fall within the realm of
the tauonic Higgs model, so that BR$(t \to H^+ \, b) \leq 0.4$.
Besides, this scenario is very interesting because we can have
$m_{H^\pm} \sim m_{W^\pm}$  and a light neutral Higgs boson, which
could be consistent with the experimental LEP2 limits, as
discussed in section II and shown in Table \ref{tab:1}.  In
Figure \ref{fig:dt13} we present the corresponding plots for the case
$\lambda = 1.0$, taking again $A=200,\,300,\,400$ GeV. We can
observe that in this case only the mode $t \to b \, H_1^+$ is
kinematically allowed and satisfies the constraint BR$(t \to b \,
H^+) < 0.4$ in the ranges
140 GeV $< m_{H_1^+} < 160$ GeV and $60 < \tan\beta < 100$. \\
%%%%%%%%%%%%%%%%%%%%%%%%%%%%%%%%%
%%%%%%%%%%%%%%%%%%%%%%%%%%%%%%%%%%
%%%%%%%%%%%%%%%%%%%%%%%%%%%%%%%%%%%%
%%%%%%%%%%%%%%%%%%%%%%%%%%%%%%%%%%%%
\begin{figure}
\centering
\includegraphics[width=5in]{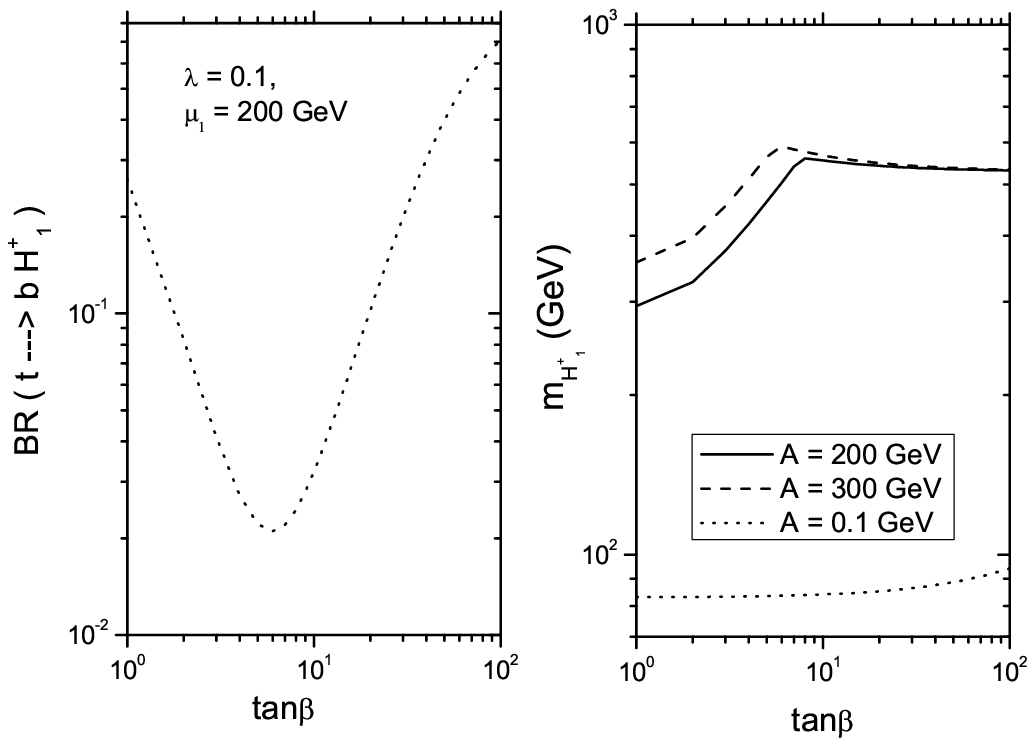}
\caption{It is plotted: BR($t \to b \, H_1^+ $) vs. $\tan\beta$
 (left), the $\tan\beta-m_{H_1^+}$ plane (right), in Scenario B, taking $\lambda = 0.1$
for: $A=200$ GeV  (solid), $A=300$ GeV (dashes), $ A=0.1$ GeV
(dots).} \label{fig:dt21}
\end{figure}
\begin{figure}
\centering
\includegraphics[width=5in]{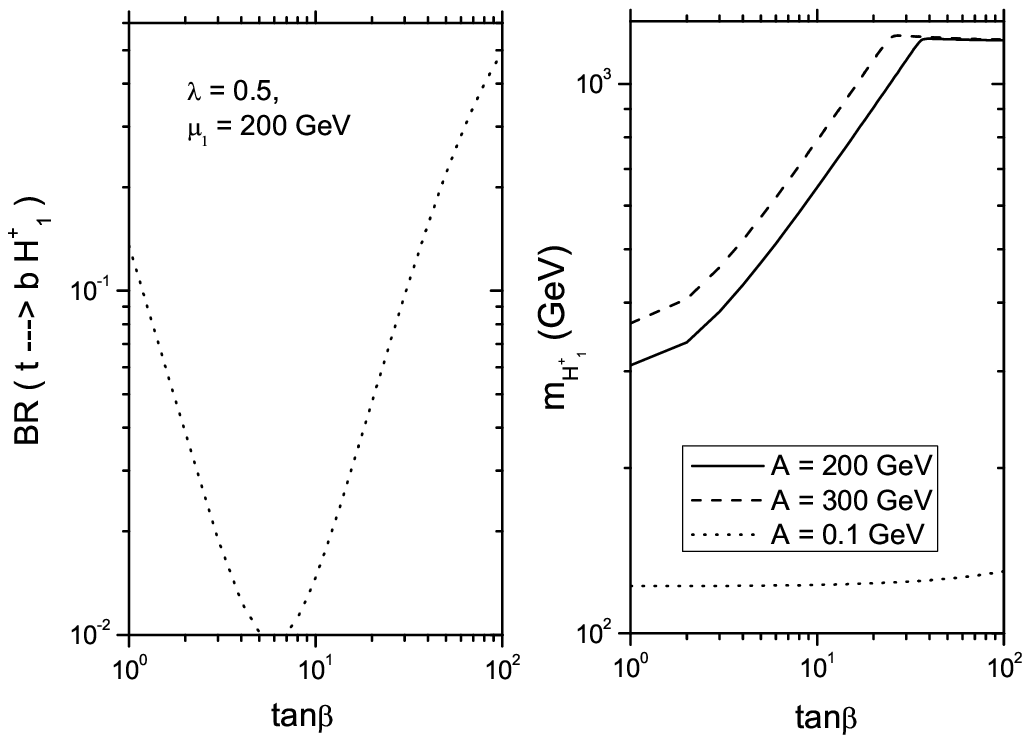}
\caption{ Same as Figure \ref{fig:dt21}, but taking $\lambda = 0.5$.}
\label{fig:dt22}
\end{figure}
\begin{figure}
\centering
\includegraphics[width=5in]{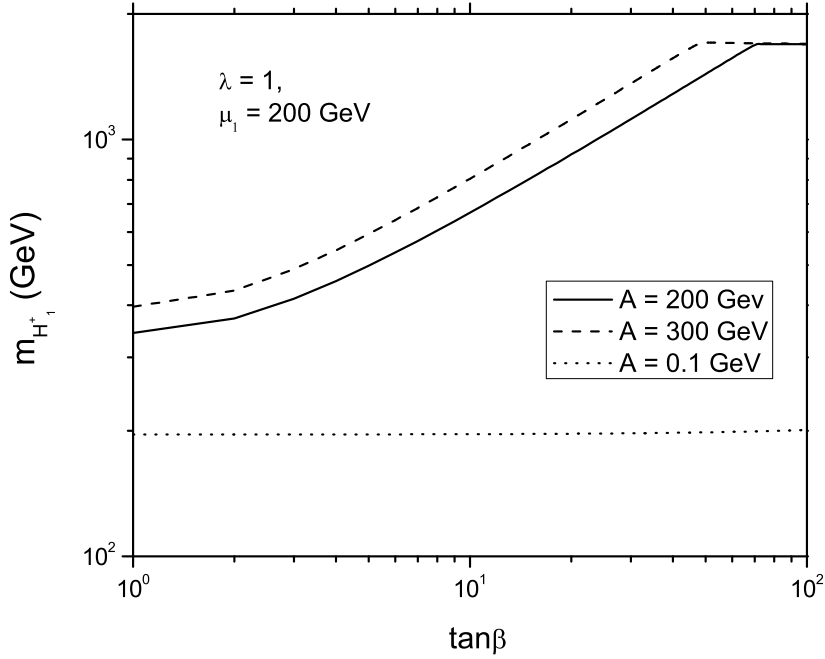}
\caption{The $\tan\beta- m_{H^+_1}$ plane, in  Scenario B, taking
$\lambda =1.0$ for:$A =200$ GeV  (solid), $A=300$ GeV (dashes),
$A=0.1$ GeV (dots).} \label{fig:dt23}
\end{figure}
\begin{figure}
\centering
\includegraphics[width=5in]{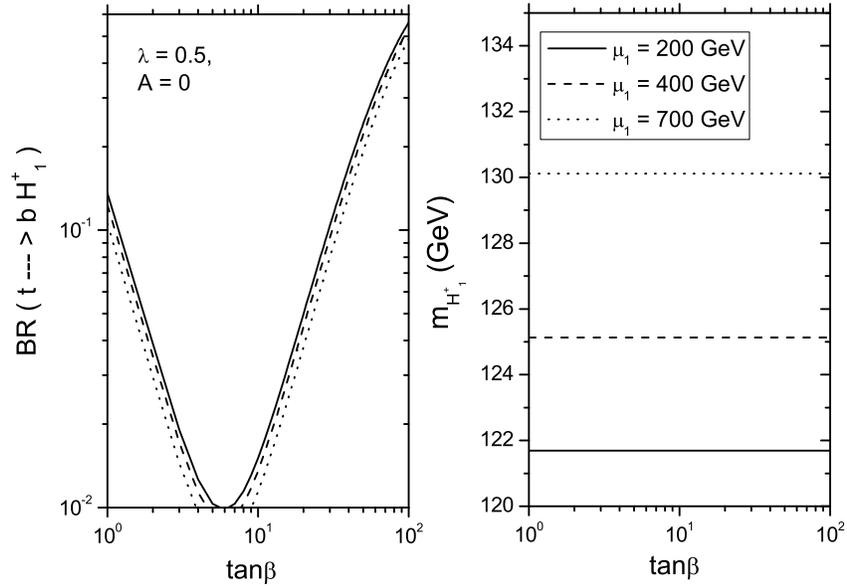}
\caption{ It is plotted: BR($t \to b \, H_1^+ $) vs. $\tan \beta$
(left), the $\tan\beta-m_{H_1^+}$ plane (right), in  Scenario B,
taking $\lambda = 0.5$ and $A=0$ for: $\mu_1=200$ GeV (solid),
$\mu_1=400$ GeV (dashes), $ \mu_1=700$ (dots).} \label{fig:dt24}
\end{figure}
%%%%%%%%%%%%%%%%%%%%%%%%%%%%%%%%%
\begin{figure}
\centering
\includegraphics[width=5in]{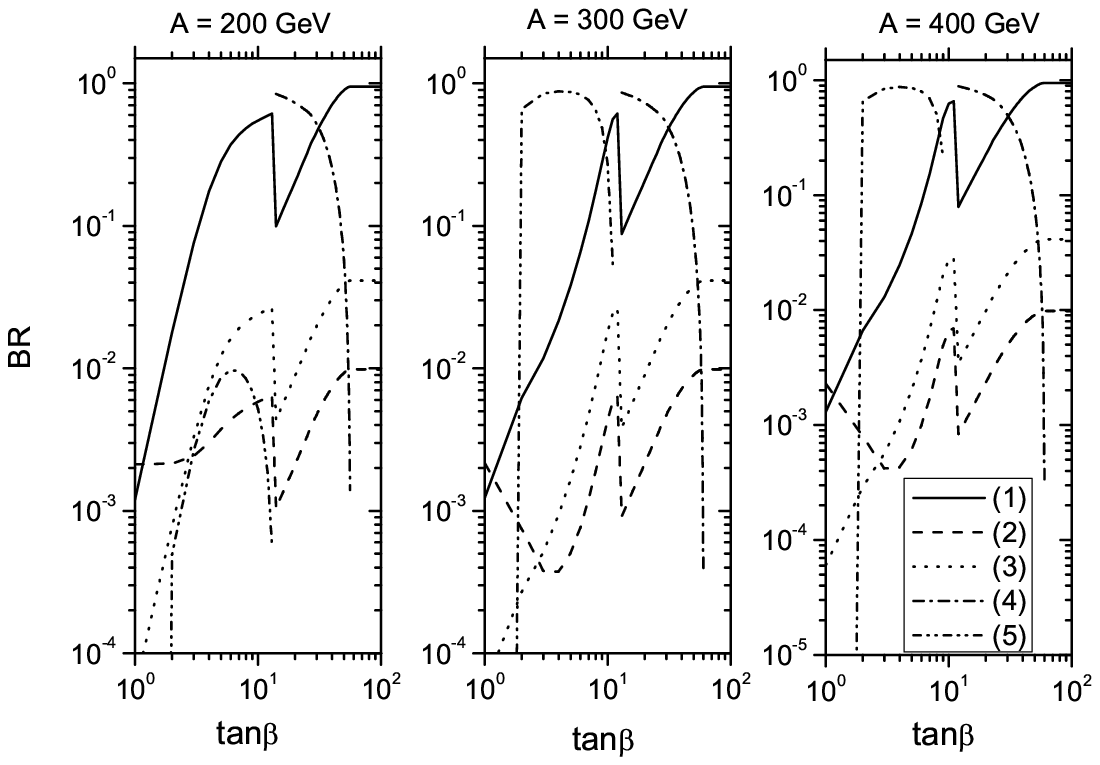}
\caption{ Same as Figure \ref{fig:dh12} but taking $\lambda = 1$.}
\label{fig:dh13}
\end{figure}
%%%%%%%%%%%%%%%%%%%%%%%%%%%%%%%%
\vskip0.25cm \noindent{\bf Scenario B.} Recall that this was defined
by taking $B_2=\mu_2=0$, $B_1=-A$, and $\mu_1=200$ GeV while for
$\lambda$ we considered the values $\lambda=0.1,\,0.5$. In
Figure~\ref{fig:dt21} we present plots of: a) the BR($t \to b \, H_1^+
$) vs. $\tan \beta$; b) the $\tan\beta-m_{H_1^+}$ plane; for the
case $\lambda = 0.1$, taking $A=200,\,300,\,0.1$ GeV. We can observe
that the mode $t \to b \, H_1^+$ is  kinematically allowed for the
case $A=0.1$ GeV and, if we combine the results of Table
\ref{tab:3},  one can see that for charged Higgs boson masses in the
range 84 GeV $< m_{H_1^+} < 89$ GeV and for $12 < \tan\beta < 50$
the model fulfills the constraint BR$(t \to b \, H^+) < 0.4$. Again
we have a charged Higgs mass $\sim m_{W^\pm}$. Similarly, in
Figure~\ref{fig:dt22} we present similar plots but for the case
$\lambda = 0.5$, taking $A=200,\,300,\,0.1$ GeV. We can observe here
that the mode $t \to b \, H_1^+$ is kinematically permitted again
for the case $A=0.1$ and the model is fulfilling the constraint
BR$(t \to b \, H^+) < 0.4$ in the ranges 121 GeV $< m_{H_1^+} < 129$
GeV and $20 < \tan\beta < 80$. In Figure~\ref{fig:dt23} we present
only a plot of the $\tan\beta-m_{H_1^+}$ plane, for the case
$\lambda = 1.0$, taking $A=200,\,300,\,0.1$ GeV. We can deduce from
here that the mode $t \to b \, H_1^+$ is kinematically forbidden,
because $m_{H_1^+} \approx 200$ GeV in the range $1 < \tan\beta <
100$. Finally, in Figure~\ref{fig:dt24} we present plots of: a) the
BR($t \to b \, H_1^+ $) vs. $\tan \beta$; b) the
$\tan\beta-m_{H_1^+}$ plane, for the case $\lambda = 0.5$ and $A=0$,
taking $\mu_1=200,\,400,\,700$ GeV. We note here that the mode $t
\to b \, H_1^+$ is kinematically allowed and we see that a charged
Higgs boson with mass in the range 121 GeV $< m_{H_1^+} < 131$ GeV
and for $1 < \tan\beta < 70$ satisfies the constraint BR$(t \to b \,
H^+) < 0.4$. Combining the results of Table \ref{tab:4}, one can
observe that only the small region $1 < \tan\beta <  6$ is allowed
by the LEP collaborations' results.

%%%%%%%%%%%%%%%%%%%%%%%%%%%%%%%%%%%%%%%%%%%%%%%%%%%%%%%%%%%%%%%%%
\subsection{\bf Decays of charged Higgs bosons in the MSSM+1CHT }
%%%%%%%%%%%%%%%%%%%%%%%%%%%%%%%%%%%%%%%%%%%%%%%%%%%%%%%%%%%%%%%%%%
Let us now discuss the decay modes of the charged Higgs bosons within our
model, which have an interest independently of whether these states
are themselves produced in top decays.
As usual, we refer to our two customary benchmark scenarios.
%%%%%%%%%%%%%%%%%%%%%%%%%%%%%%%%%%%%%%%%%%%%%%
% BR`s for scenario A
%%%%%%%%%%%%%%%%%%%%%%
\vskip0.25cm
\noindent
{\bf Scenario A}. In Figure~\ref{fig:dh11} we present the BR's
of the channels $H_i^+ \to \tau^+
{\nu_\tau}, c \bar{s},  c \bar{b}, W^+ H_1^0, W^+ A_1^0$ for $i =
1, 2$  as a function  of $\tan\beta$ in the range $1< \tan\beta < 100
$ for the case $\lambda = 0.1$, taking $A=200, 300, 400$ GeV. When $t
\to b H_i^+$ is kinematically allowed for both $i=1,2$, the dominant decay
of the charged Higgs bosons is via the mode $\tau^+ {\nu_\tau}$, with
BR($H_i^+ \to \tau^+ {\nu}_\tau$) $\approx 1$. We can observe that for
$H_2^+$ the decay mode $ W^+ A_1^0$ is dominant for $\tan\beta < 30$, although
 the decay $t \to b H_2^+$ is not kinematically allowed.
In Figure~\ref{fig:dh12} we present similar plots for $H_1^+$, but now
with $\lambda = 0.5$ and again in this case the dominant decay mode
is into  $\tau^+ {\nu_\tau}$ for the range $\tan \beta > 15$. Then,
from Figure~\ref{fig:dh12} one gets that BR($H_i^+ \to \tau^+
{\nu_\tau}$) $\approx 1$ when $t \to b H_1^+$ is kinematically
allowed. Similarly, in Figure~\ref{fig:dh13} we present the
corresponding plots for $H_1^+$ in the case $\lambda =1.0$. For $A =
200$ GeV the dominant decay of  the considered charged Higgs boson
is the mode  $\tau^+ {\nu_\tau}$, except in the range $15< \tan \beta
<35$, where the decay channel $W^+ H_1^0$ is also relevant. For $A=300, 400$
GeV the dominant decay of the charged Higgs state is via the mode
$\tau^+ {\nu_\tau}$ when $\tan\beta > 35$, but for $12 < \tan\beta <35$
the decay channel $W^+ H_1^0$ becomes
the leading one
whereas for the range $2 < \tan \beta < 9$ the mode $W^+ A_1^0$ is dominant.\\
\begin{figure}
\centering
\includegraphics[width=5in]{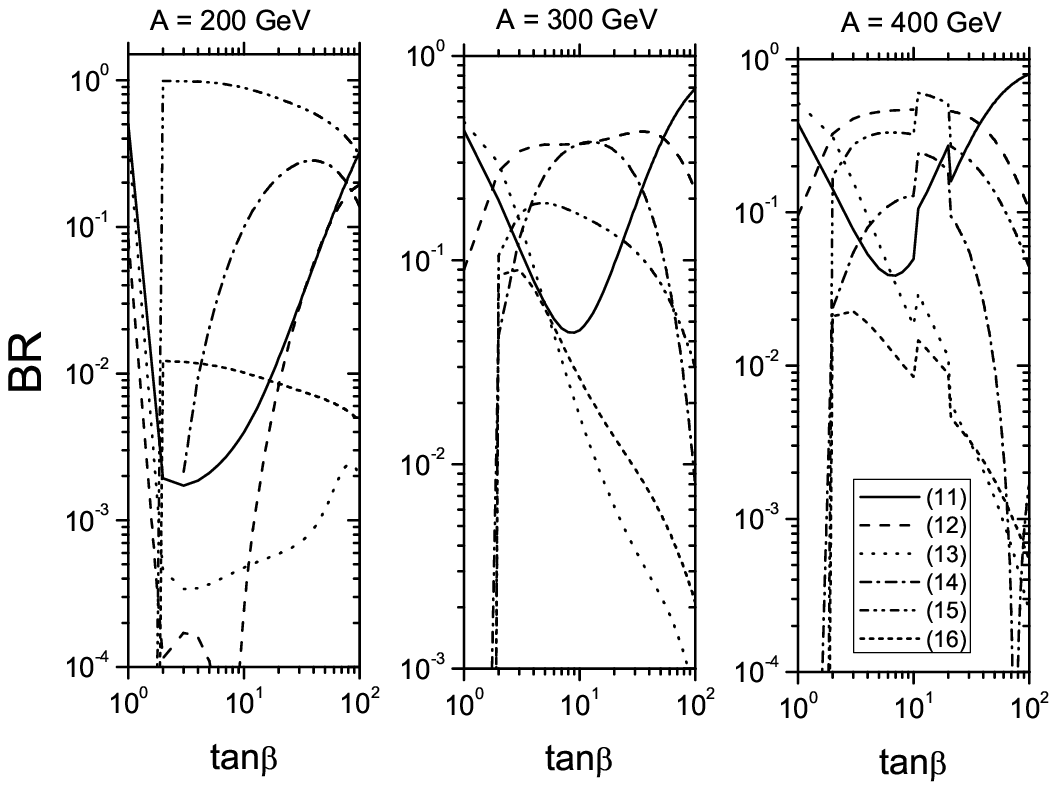}
\caption{ The figure shows the branching ratios of  $H^+_3$ decaying
into the principal modes in Scenario A, taking $\lambda = 0.1$ and
$\mu_2 = 100$ GeV for: $A=200$ GeV (left), $A=300$ GeV (center), $
A=400$ GeV (right). The lines  correspond to: (11) BR($H^+_3 \to t
\bar{b} $), (12) BR($H^+_3 \to W^+ H^0_1$),  (13) BR($H^+_3 \to W^+
H^0_2$), (14) BR($H^+_3 \to W^+ H^0_3$), (15) BR($H^+_3 \to W^+
A^0_1$), (16) BR($H^+_3 \to W^+ Z^0$).} \label{fig:dha13}
\end{figure}
Now we discuss the decay modes of $H^+_3$ for the case $\lambda =
0.1$. We can see in Figure~\ref{fig:dha13} for $A=200$ GeV that the
mode $t\bar{b}$ is dominant when $\tan \beta < 2$, but for $2< \tan
\beta$ the mode $W^+ A_1^0$ is the leading one. For the case $A =
300$ GeV there are three dominant decay modes:  $W^+ H_2^0$
in the range $\tan \beta < 2$, $W^+ H_1^0$ for $2<\tan
\beta<60$ and $t \bar{b}$ when $60< \tan \beta$. For $A = 400$ GeV
the relevant decay channels are: $W^+ H_2^0$ in the range $\tan
\beta < 2$,  $W^+ H_1^0$ for the two ranges $2<\tan \beta<10$ and $20<
\tan \beta< 40$, $W^+ A_1^0 $ for $10<\tan \beta<20$, $t \bar{b}$
when $40< \tan \beta$.
\begin{figure}
\centering
\includegraphics[width=5in]{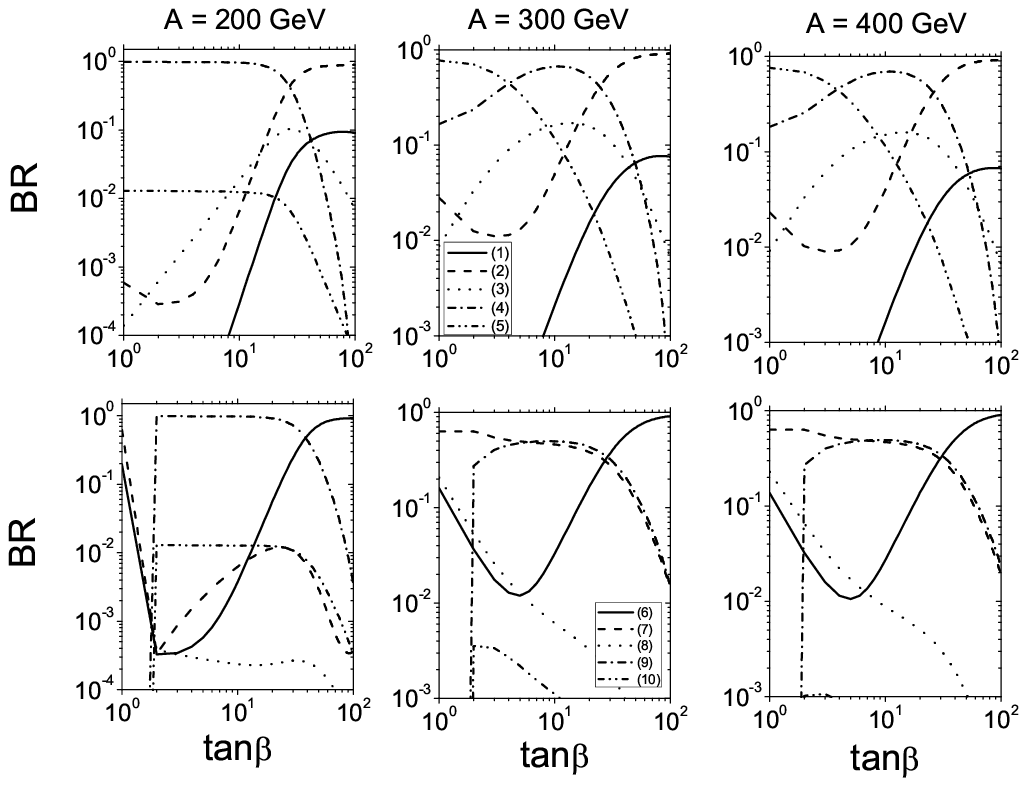}
\caption{ The figure shows the branching ratios of  $H^+_2$ (top)
and $H^+_3$ (bottom) decaying into the principal modes in Scenario
A, taking $\lambda = 0.5$ and $\mu_2 = 100$ GeV for $A=200$ GeV
(left), $A=300$ GeV (center), $ A=400$ GeV (right). The  lines
correspond to: (1) BR($H^+_2 \to \tau^+ {\nu_\tau} $), (2) BR($H^+_2
\to t \bar{b}$), (3) BR($H^+_2 \to W^+ H^0_1$), (4) BR($H^+_2 \to
W^+ A^0_1$), (5) BR($H^+_2 \to W^+ Z^0$), (6) BR($H^+_3 \to t
\bar{b}$),(7) BR($H^+_3 \to W^+ H^0_1$), (8) BR($H^+_3 \to W^+
H^0_2$), (9) BR($H^+_3 \to W^+ A^0_1$), (10) BR($H^+_3 \to W^+ Z^0$).}
\label{fig:dha223}
\end{figure}
In  Figure~\ref{fig:dha223} we present the corresponding plots for the
BR's  of the  channels $H_i^+ \to \tau^+ {\nu_\tau}, t \bar{b},  W^+
H_j^0, W^+ A_1^0, W^+ Z^0$ for $i = 2,3$, $ j= 1, 2$ as a function  of
$\tan\beta$ in the range $1< \tan\beta < 100$ for the case $\lambda
= 0.5$, taking $A=200, 300, 400$ GeV. When $A =200$ GeV the dominant
decay modes for $H_2^+$  are: $W^+ A_1^0$ in the range $\tan
\beta<25 $ and $t\bar{b}$ for $25< \tan \beta$.  We obtain similar
results for $H_3^+$, except for $\tan \beta <2$, where the mode $ W^+
H^0_1$ is dominant. For the cases $A =300$, $400$ GeV the relevant
decay modes of $H_2^+$  are: $W^+ Z^0$ when $\tan \beta<4$, $W^+
A_1^0$ in the range $4< \tan \beta<25 $ and $t\bar{b}$ for $25<
\tan \beta$. As for $H_2^+$, we obtain the same dominant modes as for
$H_3^+$  in the range $4 < \tan \beta $, because the decay channel
$W^+ H_1^0$  is relevant for $\tan \beta < 4$.
\begin{figure}
\centering
\includegraphics[width=5in]{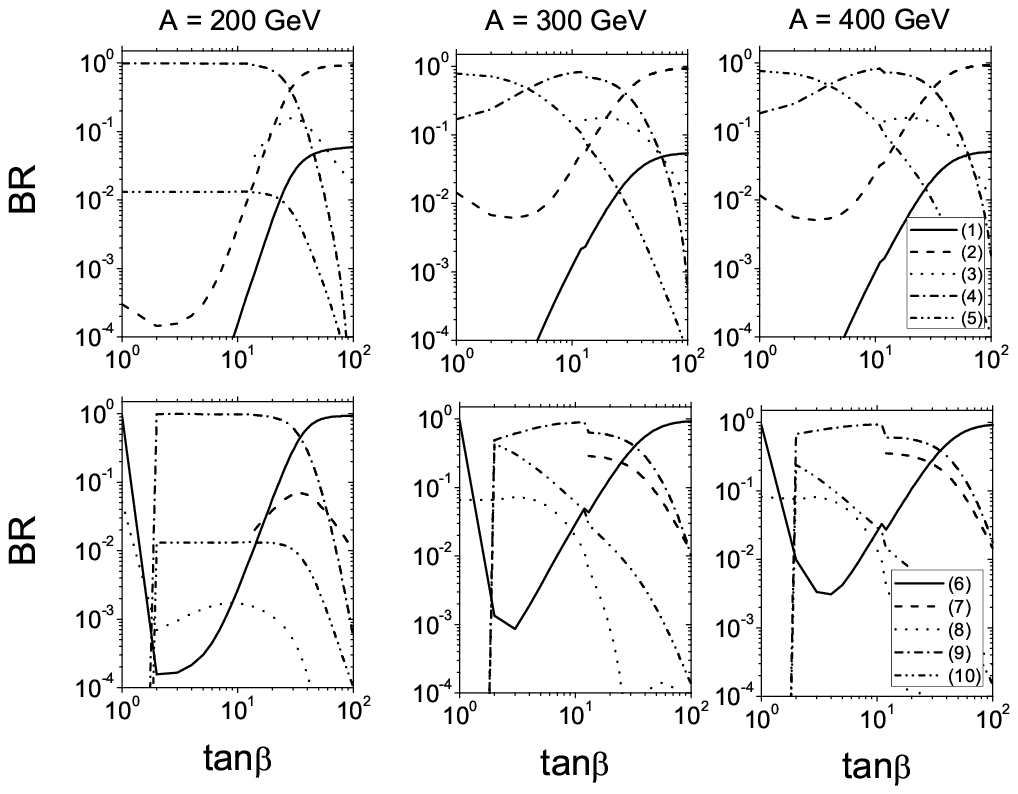}
\caption{ Same as  Figure \ref{fig:dha223}, but taking $\lambda= 1$.}
\label{fig:dha323}
\end{figure}
In  Figure~\ref{fig:dha323} we present  plots for the BR's  of the
channels $H_i^+ \to \tau^+ {\nu_\tau}, t \bar{b},  W^+ H_j^0, W^+
A_1^0, W^+ Z^0$ for $i = 2,3$, $ j=1, 2$ as a function  of $\tan\beta$
in the range $1< \tan\beta < 100$ for the case $\lambda = 1$, taking
$A=200, 300, 400$ GeV. As in the case $\lambda = 0.5$, when $A =200$
GeV the dominant decay modes for $H_2^+$  are: $W^+ A_1^0$ in the
range $\tan \beta<25 $, $t\bar{b}$ for $25< \tan \beta$.  We obtain
similar results for $H_3^+$, except for $\tan \beta <2$ where the
mode $t\bar{b}$ is dominant. For the cases $A =300$, $400$ GeV the
relevant decay modes of $H_2^+$  are: $W^+ Z^0$ when $\tan \beta<4$,
$W^+ A_1^0$ in the range $4< \tan \beta<25 $ and $t\bar{b}$ for
$25< \tan \beta$. We get the following dominant modes for  $H_3^+$:
$W^+ H_2^0$  for $1.5< \tan \beta<2 $,  $W^+ A_1^0$ when $2<\tan
\beta <25$ while in two ranges $\tan \beta <1.5 $ and $25<\tan \beta$
the mode $t \bar{b}$  becomes the most important one.
\begin{figure}
\centering
\includegraphics[width=4in]{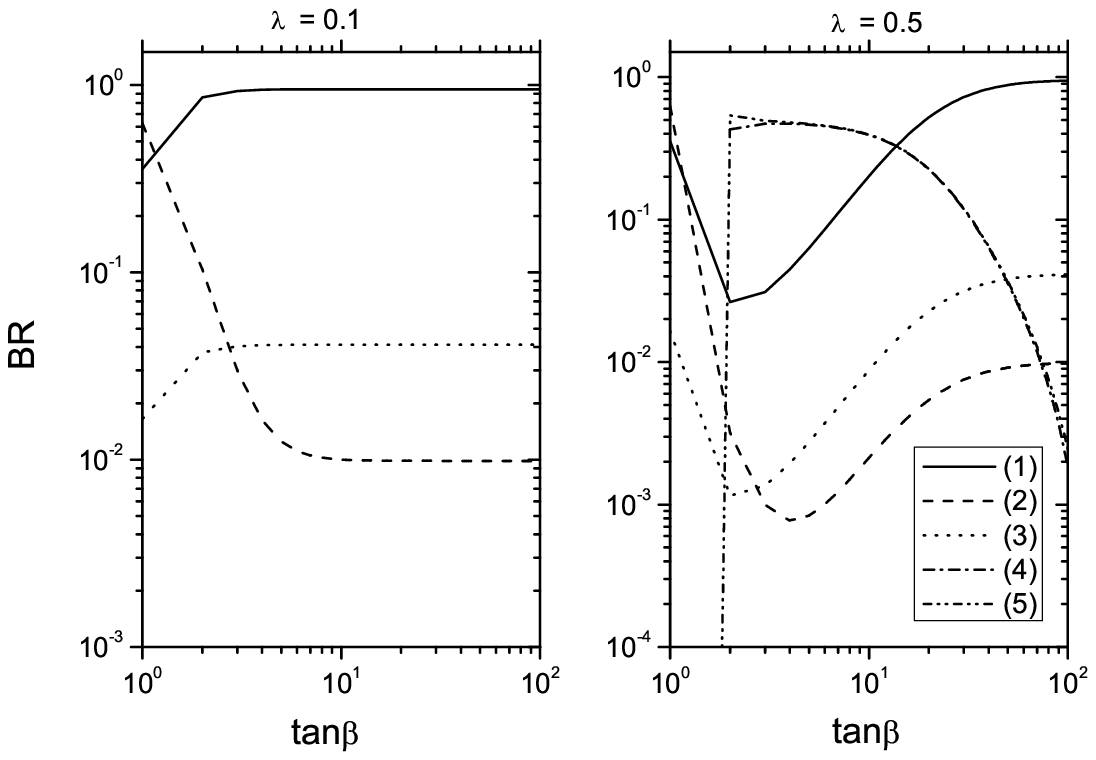}
\caption{ The figure shows the branching ratios of $H^+_1$ decaying
into the principal modes in Scenario B considering the cases:
$\lambda=0.1$  (left), $ \lambda=0.5$ (right), with $A= 0.1$ GeV and
$\mu_1 = 200$ GeV. The lines correspond to: (1) BR($H^+_1 \to \tau^+
{\nu_\tau} $), (2) BR($H^+_1 \to c \bar{s}$), (3) BR($H^+_1 \to c
\bar{b}$), (4) BR($H^+_1 \to W^+ H^0_1$), (5) BR($H^+_1 \to W^+
A^0_1$).} \label{fig:dh212}
\end{figure}
\begin{figure}
\centering
\includegraphics[width=5in]{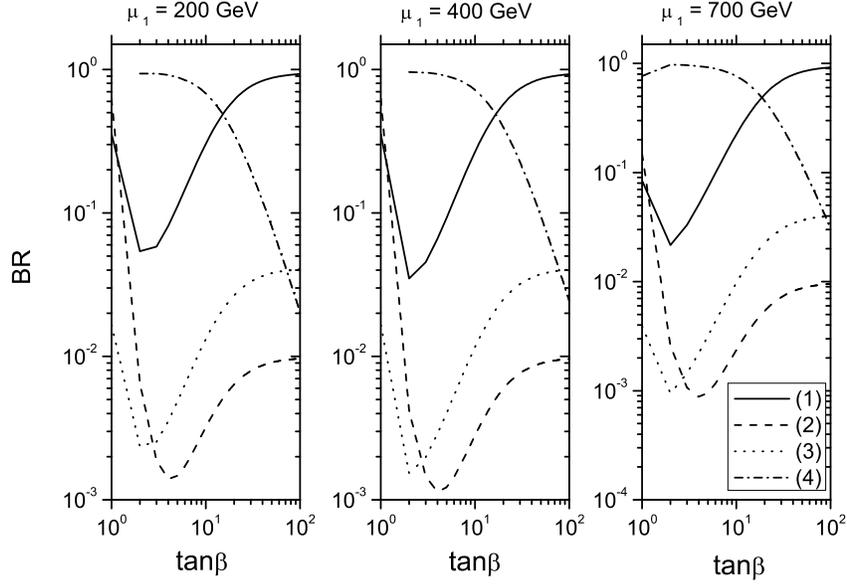}
\caption{ The figure shows the branching ratios of $H^+_1$ decaying
into the principal modes in Scenario B, with $\lambda = 0.5$ and $A
= 0$ GeV for: $\mu_1=200$ GeV (left), $\mu_1=400$ GeV (center), $
\mu_1=700$ GeV (right). The lines correspond to: (1) BR($H^+_1 \to
\tau^+ {\nu_\tau} $), (2) BR($H^+_1 \to c \bar{s}$), (3) BR($H^+_1 \to
c \bar{b}$), (4) BR($H^+_1 \to W^+ H^0_1$).} \label{fig:dh24}
\end{figure}
%%%%%%%%%%%%%%%%%%%%%%%%%%%%%%%%%%%%%%%%%%%%%%
% BR`s for scenario B
%%%%%%%%%%%%%%%%%%%%%%
\vskip0.25cm \noindent {\bf Scenario B}. In  Figure \ref{fig:dh212} we
present the BR($H_i^+ \to \tau^+ {\nu_\tau}, c \bar{s},  c \bar{b}, W^+
H_1^0, W^+ A_1^0$)'s  as a function  of $\tan\beta$ in the range $1<
\tan\beta < 100 $ for the case $A= 0.1$ GeV, taking $\lambda = 0.1,
0.5$. When the mode $t \to b H_1^+$ is kinematically allowed, the
dominant decay of the charged Higgs bosons is via the mode $\tau^+
{\nu_\tau}$ and it is obtained that BR($H_1^+ \to \tau^+ {\nu_\tau}$)
$\approx 1$. We can observe that for $\lambda = 0.5$ the mode $ W^+
A_1^0$ is dominant for $2< \tan\beta < 15$. In Figure \ref{fig:dh24}
we present BR($H_i^+ \to \tau^+ {\nu_\tau}, c \bar{s},  c \bar{b}, W^+
H_1^0, W^+ A_1^0$)'s  as a function  of $\tan\beta$ in the range $1<
\tan\beta < 100 $ for the case $A= 0$ GeV, $\lambda =  0.5$, taking
$\mu_1 = 200, 400, 700$ GeV. In these cases the decay $t \to b
H_1^+$ is kinematically allowed and the dominant decay of the
charged Higgs boson is via $\tau^+ {\nu_\tau}$ for $\tan \beta > 20$,
since BR($H_1^+ \to \tau^+ {\nu_\tau}$) $\approx 1$. In contrast,
when  $\tan \beta < 20$ the mode $W^+ H_1^0$ is dominant.
\begin{figure}
\centering
\includegraphics[width=4in]{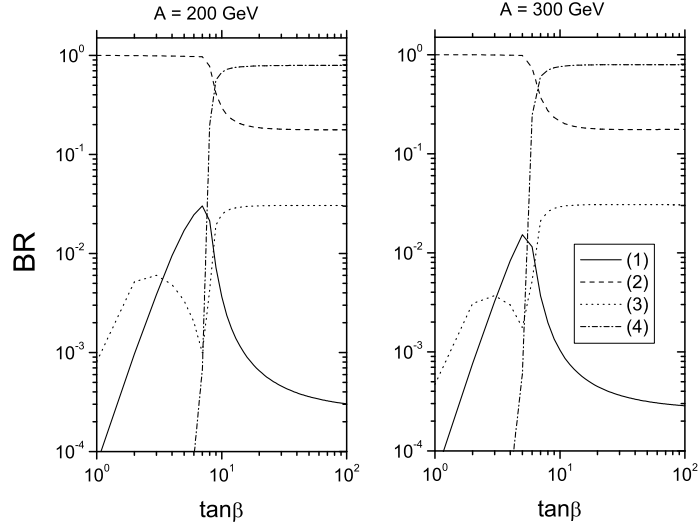}
\caption{ The figure shows the branching ratios of  $H^+_1$ decaying
into the principal modes in Scenario B taking $\lambda = 0.1$ GeV
and $\mu_1 = 200$ GeV, for: $A=200$  GeV (left), $ A=300$ GeV
(right). The  lines correspond to: (1) BR($H^+_1 \to \tau^+ {\nu_\tau}
$), (2) BR($H^+_1 \to t \bar{b}$), (3) BR($H^+_1 \to W^+ H^0_1$),
(4) BR($H^+_1 \to W^+ Z^0$).} \label{fig:dhb11}
\end{figure}
\begin{figure}
\centering
\includegraphics[width=4in]{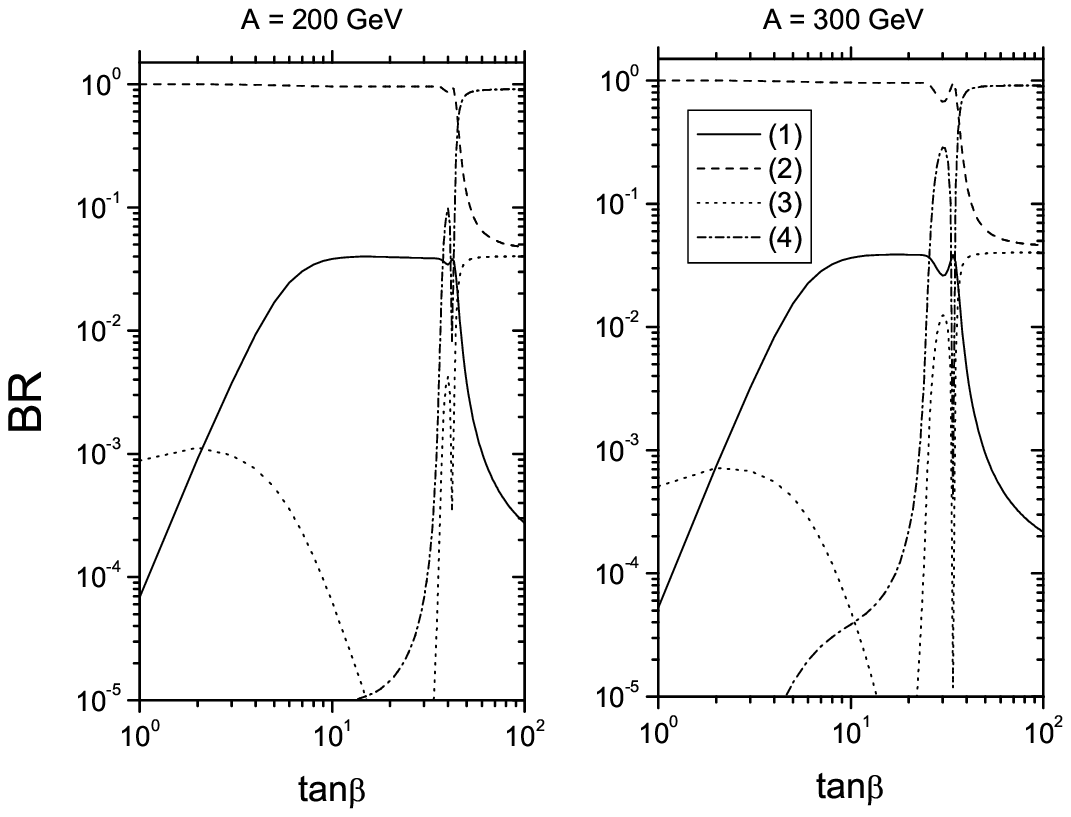}
\caption{ Same as  Figure \ref{fig:dhb11}, but taking $\lambda= 0.5$.}
\label{fig:dhb21}
\end{figure}

In  Figure \ref{fig:dhb11} we present the BR's of the relevant decay
channels of the charged Higgs $H_1^+$ for $A= 200 $, 300 GeV, taking
$\lambda=0.1$. When $\tan \beta <6$ the dominant mode is $t
\bar{b}$. When $6 < \tan \beta$ though, the dominant
mode becomes $W^+ Z^0$. In Figure \ref{fig:dhb21} we observe that
for the same previous values for $A$, but with
$\lambda=0.5$, the dominant decays channels are: $t \bar{b}$ for
$\tan \beta <40$ and $W^+ Z^0$ for  $\tan \beta >40$. \\
%%%%%%%%%%%%%%%%%%%%%%%%
\begin{figure}
\centering
\includegraphics[width=5in]{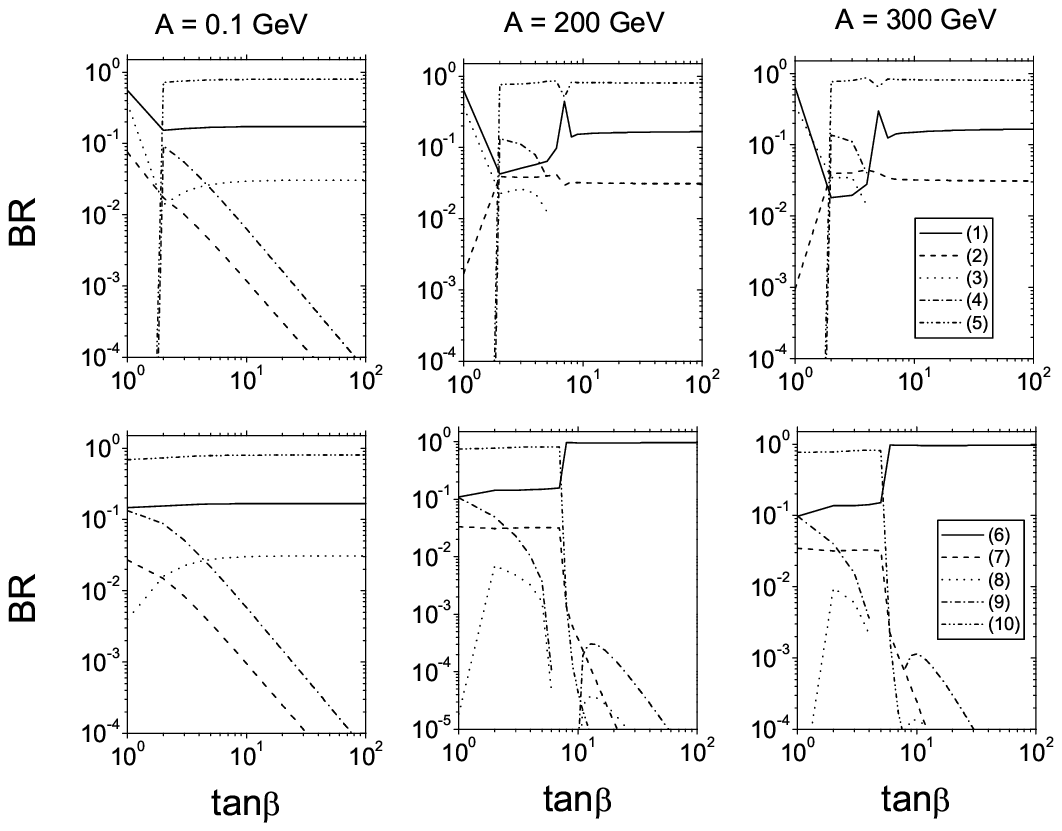}
\caption{ The figure shows the branching ratios of $H^+_2$ (top) and
$H^+_3$ (bottom) decaying into the principal modes in Scenario B,
with $\lambda = 0.1$ and $\mu_1 = 200$ GeV, for: $A=0.1$ GeV (left),
$A=200$ GeV (center), $ A=300$ GeV (right). The lines correspond to:
(1) BR($H^+_2 \to t \bar{b} $), (2) BR($H^+_2 \to W^+ H^0_1$), (3)
BR($H^+_2 \to W^+ H^0_2$), (4) BR($H^+_2 \to W^+ A^0_1$), (5)
BR($H^+_2 \to W^+ Z^0$), (6) BR($H^+_3 \to t \bar{b}$),(7) BR($H^+_3
\to W^+ H^0_1$), (8) BR($H^+_3 \to W^+ H^0_2$), (9) BR($H^+_3 \to
W^+ A^0_1$), (10) BR($H^+_3 \to W^+ Z^0$).} \label{fig:dhb123}
\end{figure}
In  Figure~\ref{fig:dhb123} we present plots for the BR's  of the
channels $H_i^+ \to  t \bar{b},  W^+ H_j^0, W^+ A_1^0, W^+ Z^0$ for $i
= 2,3$, $ j= 1, 2$ as a function  of $\tan\beta$ in the range $1<
\tan\beta < 100$ for the case $\lambda = 0.1$, taking $A=0.1, 200,
300$ GeV.  When $A =0.1$ GeV the dominant decay modes for $H_2^+$
are:  $t\bar{b}$ for $ \tan \beta<2$ and $W^+ Z^0$ when $2<\tan
\beta$.  For $H_3^+$  the mode $ W^+ Z^0$ is the dominant one in the
range $1< \tan\beta < 100$. For $A =200$, $300$ GeV the relevant
decay modes of $H_2^+$  are: $t\bar{b}$ when $\tan \beta<2$ and
$W^+ Z^0$ when $2< \tan \beta $.  We obtain similar results for
$H_3^+$ , but now the mode $W^+ Z^0$ is dominant when $6 < \tan \beta
$, and $t\bar{b}$  is the relevant mode for $\tan \beta < 6$.
\begin{figure}
\centering
\includegraphics[width=5in]{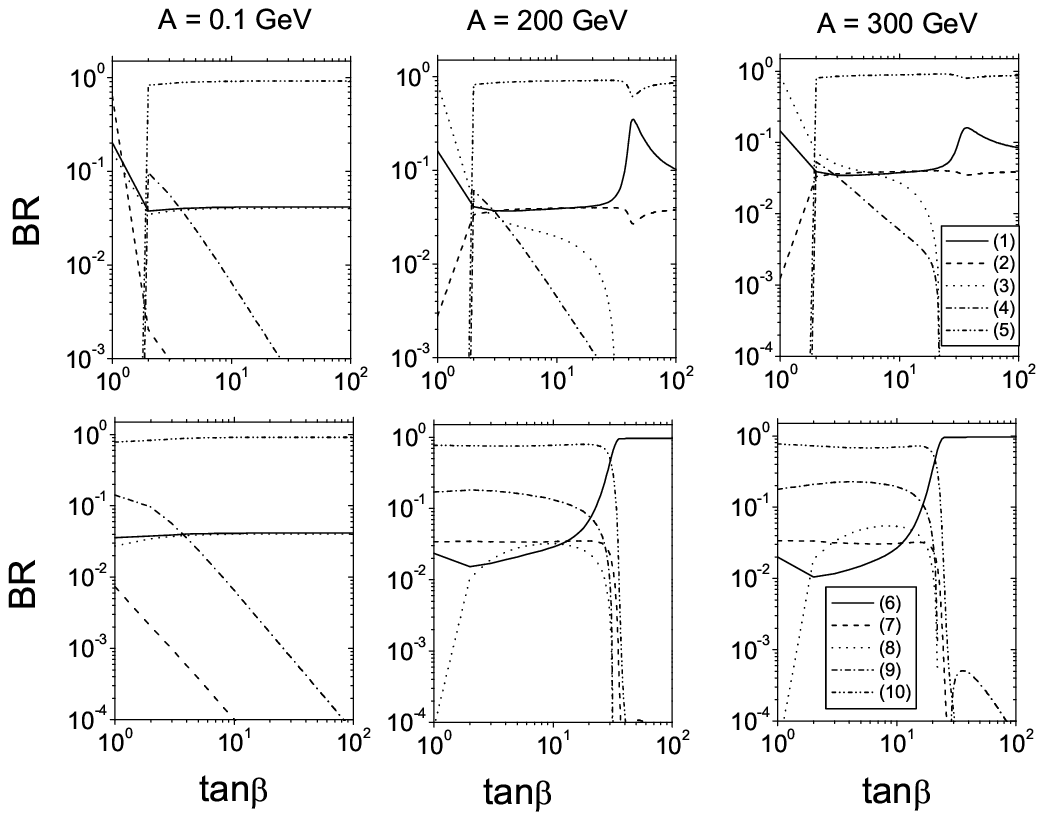}
\caption{ Same as  Figure \ref{fig:dhb123}, but taking $\lambda= 0.5$.}
\label{fig:dhb223}
\end{figure}
In  Figure~\ref{fig:dhb223} we present  plots for the BR's  of
channels $H_i^+ \to t \bar{b},  W^+ H_j^0, W^+ A_1^0, W^+ Z^0$ for $i
= 2,3$, $ j= 1, 2$ as a function  of $\tan\beta$ in the range $1<
\tan\beta < 100$, taking $\lambda = 0.5$ for $A=0.1, 200, 300$ GeV.
When $A =0.1$ GeV the dominant decay modes for $H_2^+$  become: $W^+
H_1^0$ in the range $\tan \beta<1.5 $,  $t\bar{b}$ for $1.5< \tan
\beta$ and $W^+Z^0$ when $2<\tan \beta$.  For $H_3^+$  the mode $W^+
Z^0$ is dominant for all $\tan \beta$. For  $A =200$, $300$ GeV the
relevant decay modes of $H_2^+$  are:  $W^+ H_2^0$ when $2< \tan
\beta $ and $W^+Z^0$ for $2< \tan \beta$. Similar dominant modes are
obtained for  $H_3^+$,  but now  when $\tan \beta <30$ the mode $W^+
Z^0$  becomes the principal one, and for $30<\tan \beta$ the mode $t
\bar{b}$ is the relevant one.
%%%%%%%%%%%%%
\begin{figure}
\centering
\includegraphics[width=5in]{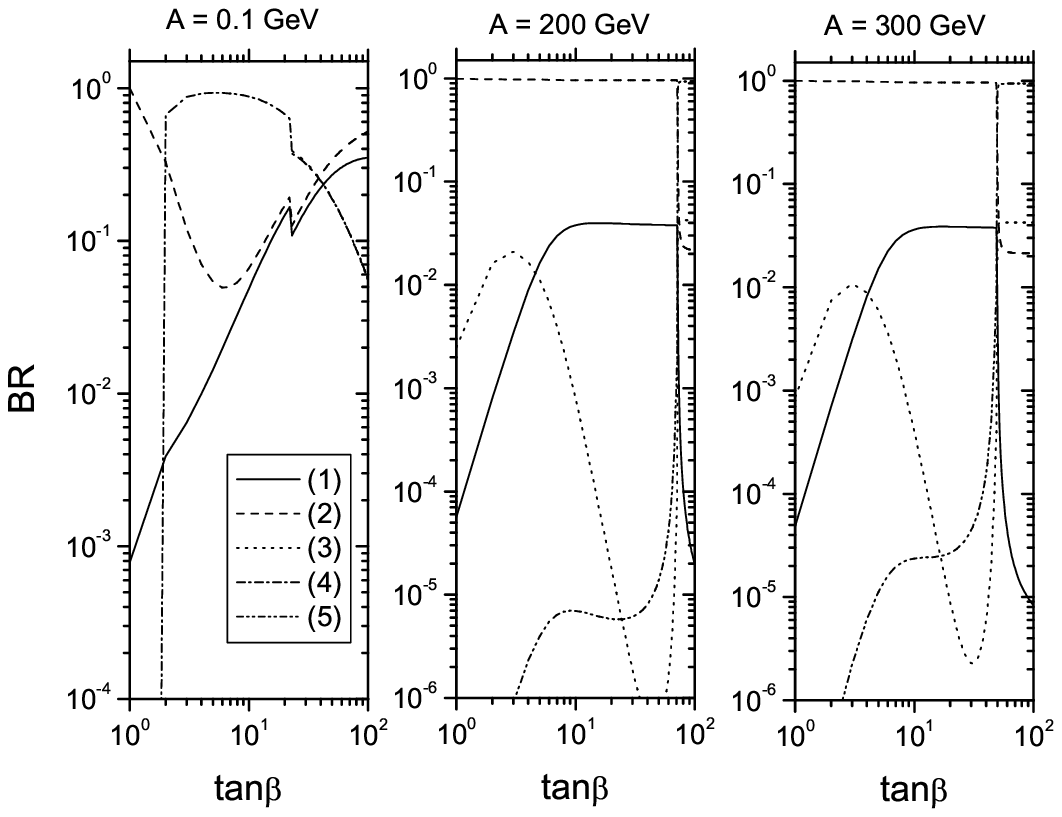}
\caption{ The figure shows the branching ratios of  $H^+_1$ decaying
into the principal modes in Scenario B, taking $\lambda= 1$ and
$\mu_1 = 200$ GeV for: $A=0.1$ GeV (left), $A=200$ GeV (center), $
A=300$ GeV (right). The lines  correspond to the modes: (1)
BR($H^+_1 \to \tau^+ {\nu_\tau} $), (2) BR($H^+_1 \to t \bar{b}$), (3)
BR($H^+_1 \to W^+ H^0_1$), (4) BR($H^+_1 \to W^+ A^0_1$), (5)
BR($H^+_1 \to W^+ Z^0$).} \label{fig:dhb31}
\end{figure}
In Figure~\ref{fig:dhb31} we present  plots for the BR's  of the
channels $H_1^+ \to \tau^+ {\nu_\tau}, t \bar{b},  W^+ H_1^0, W^+
A_1^0, W^+ Z^0$  as a function  of $\tan\beta$ in the range $1<
\tan\beta < 100$ for  $\lambda = 1$, taking $A=0.1, 200, 300$ GeV.
For $A =0.1$ GeV the dominant decay modes for $H_1^+$  are: $W^+
A_1^0$ in the range $2<\tan \beta<40 $, $t\bar{b}$ for $\tan \beta
<2 $ and $40< \tan \beta$.   For  $A =200$ ($300$) GeV the relevant
decay modes of $H_2^+$  are: $t\bar{b}$ when $\tan \beta<70 (50)$
and $W^+ A_1^0$ when $70 (50)< \tan \beta$.
\begin{figure}
\centering
\includegraphics[width=5in]{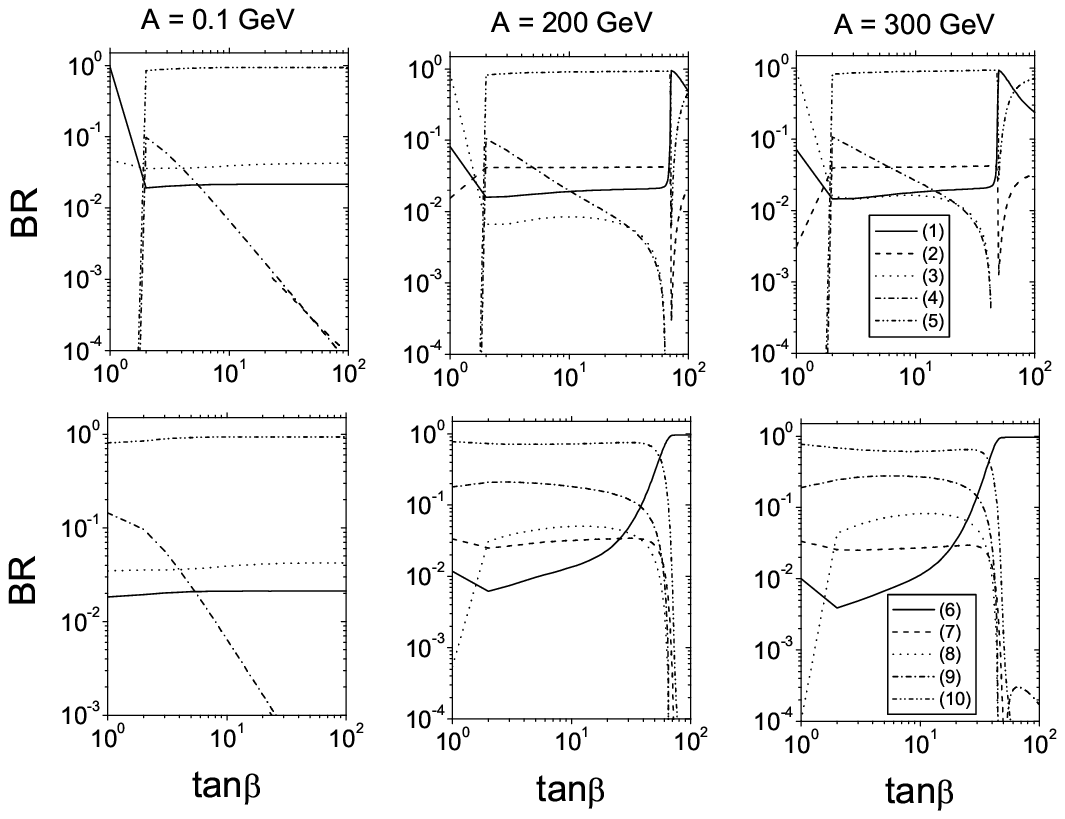}
\caption{ Same as  Figure \ref{fig:dhb123}, but taking $\lambda= 1$.}
\label{fig:dhb323}
\end{figure}
In  Figure~\ref{fig:dhb323} we present  plots for the BR's  of
channels $H_i^+ \to t \bar{b},  W^+ H_j^0, W^+ A_1^0, W^+ Z^0$ for $i
= 2,3$, $ j= 1, 2$ as a function  of $\tan\beta$ in the range $1<
\tan\beta < 100$, taking $\lambda = 1$ for $A=0.1, 200, 300$ GeV. As
in the case $\lambda = 0.5$, when $A =0.1$ GeV the dominant decay
modes for $H_2^+$  are:  $t\bar{b}$ for $\tan \beta<2$ and $W^+Z^0$
when $2<\tan \beta$.  For $H_3^+$  the mode $W^+ Z^0$ is dominant for
the entire $\tan \beta$ range. For the case $A =200$ ($300)$ GeV the
relevant decay modes of $H_2^+$  are:  $W^+ H_2^0$ when $2< \tan
\beta $, $W^+Z^0$ for $2< \tan \beta<60 (50)$, $t \bar{b}$ when $60
(50)<\tan \beta <100 (70)$. The  dominant modes for the $H_3^+$ are:
$W^+ Z^0$ in the range $\tan \beta <60$, and for $60<\tan \beta$ the
mode $t \bar{b}$ is relevant.
\begin{figure}
\centering
\includegraphics[width=5in]{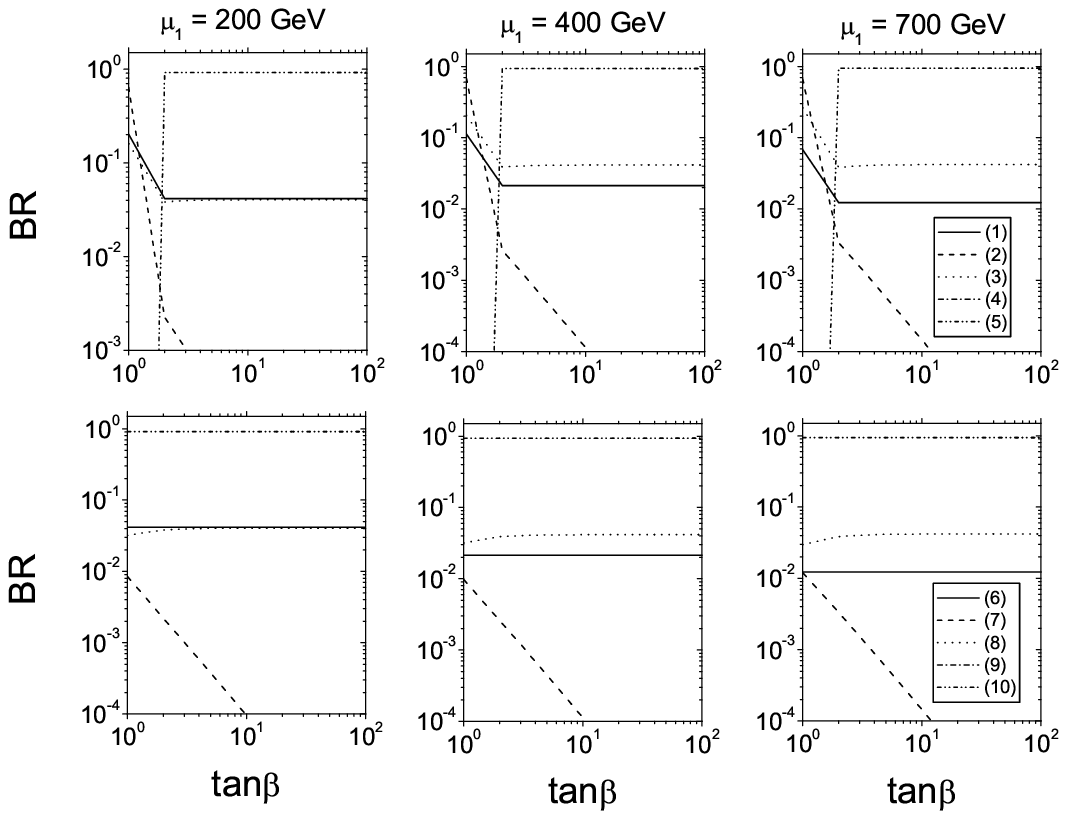}
\caption{ The figure shows the branching ratios of  $H^+_2$ (top)
and $H^+_3$ (bottom) decaying into the principal modes in Scenario
B, taking $\lambda = 0.1$ and $A = 0$ GeV for: $\mu_1=200$ GeV
(left), $\mu_1=400$ GeV (center), $ \mu_1=700$ GeV (right). The
lines correspond to: (1) BR($H^+_2 \to t \bar{b} $), (2) BR($H^+_2
\to W^+ H^0_1$), (3) BR($H^+_2 \to W^+ H^0_2$), (4) BR($H^+_2 \to
W^+ A^0_1$), (5) BR($H^+_2 \to W^+ Z^0$), (6) BR($H^+_3 \to t
\bar{b}$),(7) BR($H^+_3 \to W^+ H^0_1$), (8) BR($H^+_3 \to W^+
H^0_2$), (9) BR($H^+_3 \to W^+ A^0_1$), (10) BR($H^+_3 \to W^+ Z^0$).}
\label{fig:dhb423}
\end{figure}
In Figure~\ref{fig:dhb423} we present the corresponding plots for
the BR's  of the  channels $H_i^+ \to t \bar{b},  W^+ H_j^0, W^+
A_1^0, W^+ Z^0$ for $i = 2,3$, $ j= 1, 2$ as a function  of
$\tan\beta$ in the range $1< \tan\beta < 100$, taking $\lambda = 0.5$
for $\mu_1= 200, 400, 700$ GeV with $A = 0$ GeV.  The dominant decay
modes for $H_2^+$  are: $W^+ H_2^0$ in the range $\tan \beta<1.5$,
$t\bar{b}$ for $1.5<\tan \beta<2$ and $W^+Z^0$ when $2<\tan \beta$.
For $H_3^+$  the mode $W^+ Z^0$ is dominant for all values of $\tan
\beta$. For the case $\mu_1 =400$, $700$ GeV the relevant decay
modes of $H_2^+$  are:  $W^+ H_1^0$ in the range $ \tan \beta<1.5 $,
$W^+ H_2^0$  when $1.5< \tan \beta<2 $, $W^+Z^0$ for $2< \tan
\beta<60$. Finally, the  dominant mode for the $H_3^+$ is
$W^+ Z^0$.

%%%%%%%%%%%%%%%%%%%%%%%%%%%%%%%%%%%%%%%%%%%%%%%%%%%%%%%%%%%%%%%%%%%%%%%
\section{Direct charged Higgs production at the LHC in the MSSM+1CHT}
%%%%%%%%%%%%%%%%%%%%%%%%%%%%%%%%%%%%%%%%%%%%%%%%%%%%%%%%%%%%%%%%%%%%%%%

We have found that, in some of the MSSM+1CHT scenarios
envisaged here, light charged Higgs bosons could exist that have
not been excluded by current experimental bounds, chiefly from LEP2
and Tevatron. Their discovery potential should therefore be studied
in view of the upcoming LHC and we shall then turn our attention now to
presenting the corresponding hadro-production cross sections via
direct channels, i.e., other than as secondary products in (anti)top
quark decays.

As dealt with so far, if the charged Higgs boson mass $m_{H^\pm_i}$
satisfies $m_{H^\pm_i} < m_{t} - m_{b}$, where $ m_{t}$ is the top
quark mass and $ m_{b}$ the bottom quark mass, $H^\pm_i$ particles
could be produced in the decay of on-shell (i.e., $\Gamma_t\to0$)
top (anti-)quarks $ t \rightarrow b H^+$, and the charge conjugated
(c.c.) process, the latter being in turn produced in pairs via
$q\bar q$ annihilation and $gg$ fusion. We denote such a $H^\pm_i$
production channel as $q\bar q$, $ gg \rightarrow t\bar t
\rightarrow t\bar bH^-_i$ + c.c.  (i.e., if due to (anti-)top
decays) whilst we use the notation $ q\bar q$, $ gg \rightarrow
t\bar bH^-_i$ + c.c.  to signify when further production diagrams
are included\footnote{Altogether, they represent the full gauge
invariant set of Feynman graphs pertaining to the $2\to3$ body
process with a $t\bar bH^-_i$ + c.c. final state: two for the case of
$q\bar q$
annihilation and eight for gluon-gluon fusion, see, e.g., Eq.~(1.1)
of \cite{Guchait:2001pi}.}. In fact, owing to the large top decay
width ($ \Gamma_{t} \simeq 1.5$~GeV) and due to  the additional
diagrams which do not proceed via direct $ t\bar t$ production but
yield the same final state $t\bar bH^-_i$ + c.c.
\cite{Borzumati:1999th,Miller:1999bm,Moretti:1999bw}, charged Higgs
bosons could also be produced at and beyond the kinematic top decay
threshold. The importance of these effects in the so-called
`threshold' or `transition'  region ($m_{H^\pm_i}\approx m_t$) was
emphasized in various Les Houches
proceedings~\cite{Cavalli:2002vs,Assamagan:2004mu} as well as in
Refs.~\cite{Alwall:2003tc,Guchait:2001pi,Moretti:2002ht,Assamagan:2004gv},
so that the calculations of
Refs.~\cite{Borzumati:1999th,Miller:1999bm} (based on the
appropriate $q\bar q,gg\to tb H^\pm_i$ description) are now
implemented in
HERWIG\,\cite{herwig,Corcella:2000bw,Corcella:2002jc,Moretti:2002eu}\,and
PYTHIA\,\cite{pythia,Alwall:2004xw}. A comparison between the two
generators was carried out in Ref.~\cite{Alwall:2003tc}. For any
realistic simulation of $H^\pm_i$ production with $m_{H^\pm_i}\gsim
m_t$, as can well be the case here, the use of either of these two
implementations is of paramount importance.

\begin{figure}
\centering
\includegraphics[width=5in]{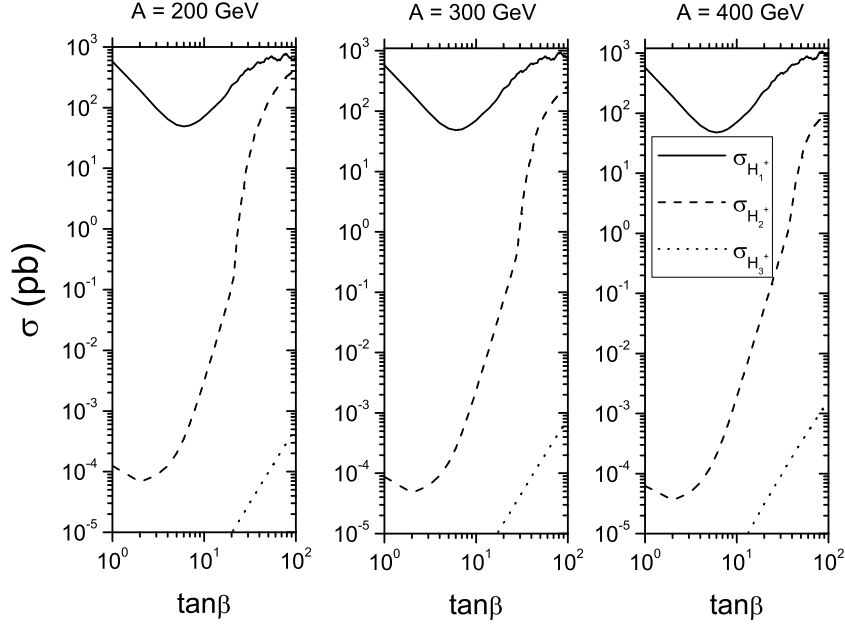}
\caption{ The figure shows the cross sections of $H^+_{1,2,3}$ at
the LHC through the channel $q\bar q,gg\to t\bar b H^-$ + c.c. in
Scenario A with $\lambda=0.1$ and for: $A= 200, 300, 400 $ GeV,
respectively.} \label{fig:Xsec-a1}
\end{figure}

\begin{figure}
\centering
\includegraphics[width=5in]{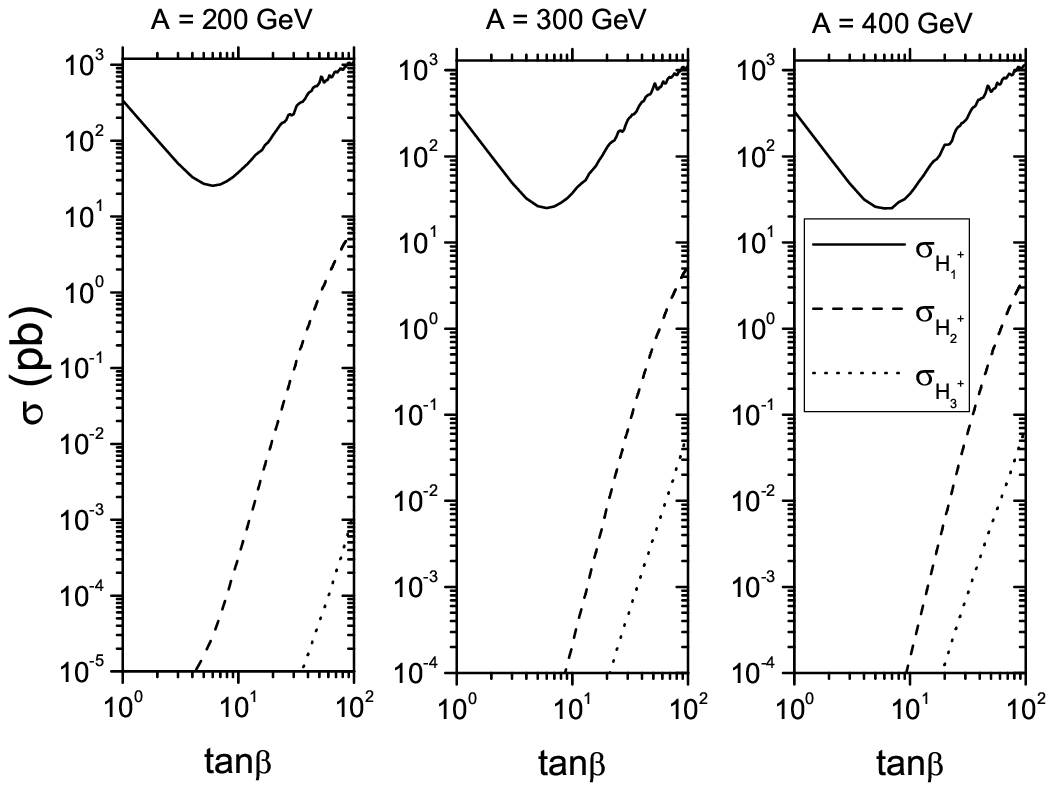}
\caption{ The figure shows the cross sections of $H^+_{1,2,3}$ at
the LHC through the channel $q\bar q,gg\to t\bar b H^-$ + c.c. in
Scenario A with $\lambda=0.5$ and for: $A= 200, 300, 400 $ GeV,
respectively.} \label{fig:Xsec-a2}
\end{figure}

\begin{figure}
\centering
\includegraphics[width=5in]{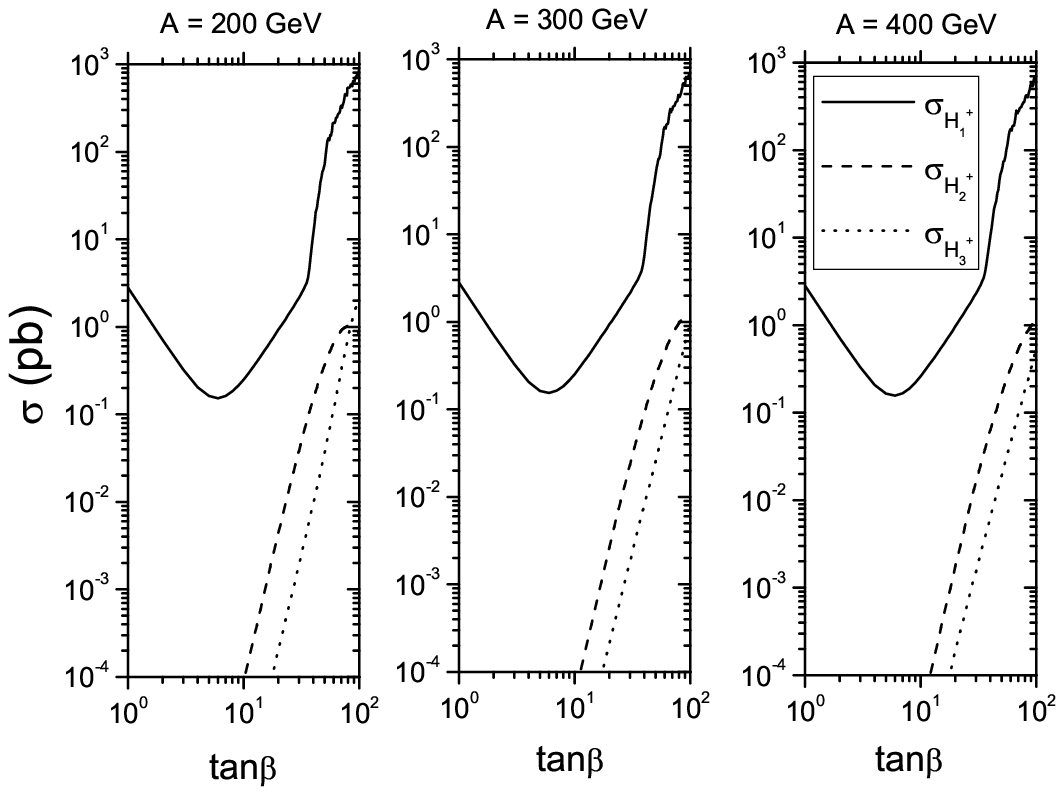}
\caption{ The figure shows the cross sections of $H^+_{1,2,3}$ at
the LHC through the channel $q\bar q,gg\to t\bar b H^-$ + c.c. in
Scenario A with $\lambda=1.0$ and for: $A= 200, 300, 400 $ GeV,
respectively.} \label{fig:Xsec-a3}
\end{figure}

\begin{figure}
\centering
\includegraphics[width=5in]{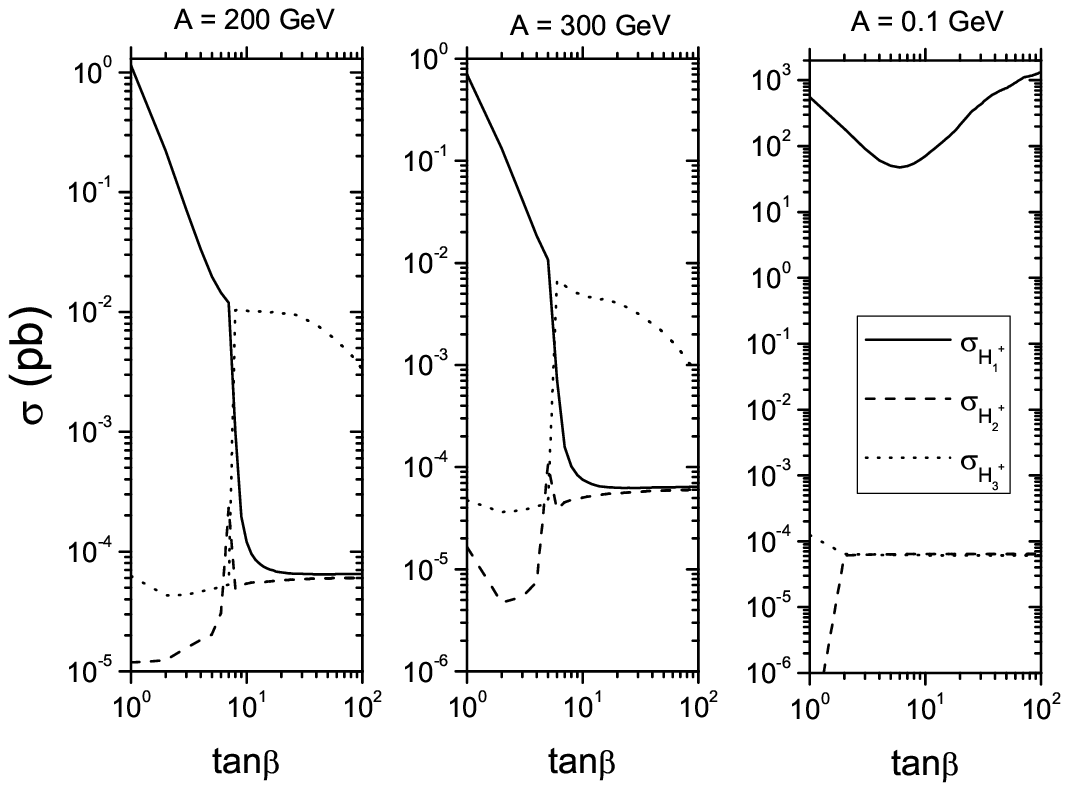}
\caption{ The figure shows the cross sections of $H^+_{1,2,3}$ at
the LHC through the channel $q\bar q,gg\to t\bar b H^-$ + c.c. in
Scenario B with $\lambda=0.1$ and for: $A= 200, 300, 0.1 $ GeV,
respectively.} \label{fig:Xsec-b1}
\end{figure}

\begin{figure}
\centering
\includegraphics[width=5in]{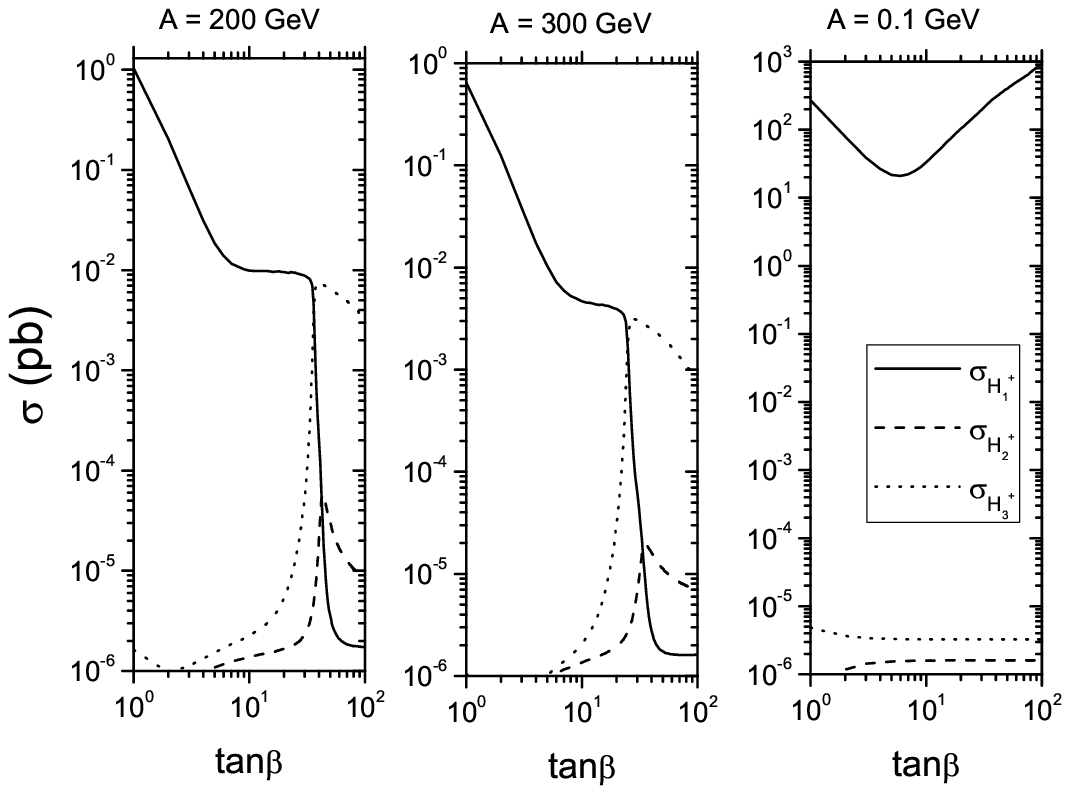}
\caption{ The figure shows the cross sections of $H^+_{1,2,3}$ at
the LHC through the channel $q\bar q,gg\to t\bar b H^-$ + c.c. in
Scenario B with $\lambda=0.5$ and for: $A= 200, 300, 0.1 $ GeV,
respectively.} \label{fig:Xsec-b2}
\end{figure}

\begin{figure}
\centering
\includegraphics[width=5in]{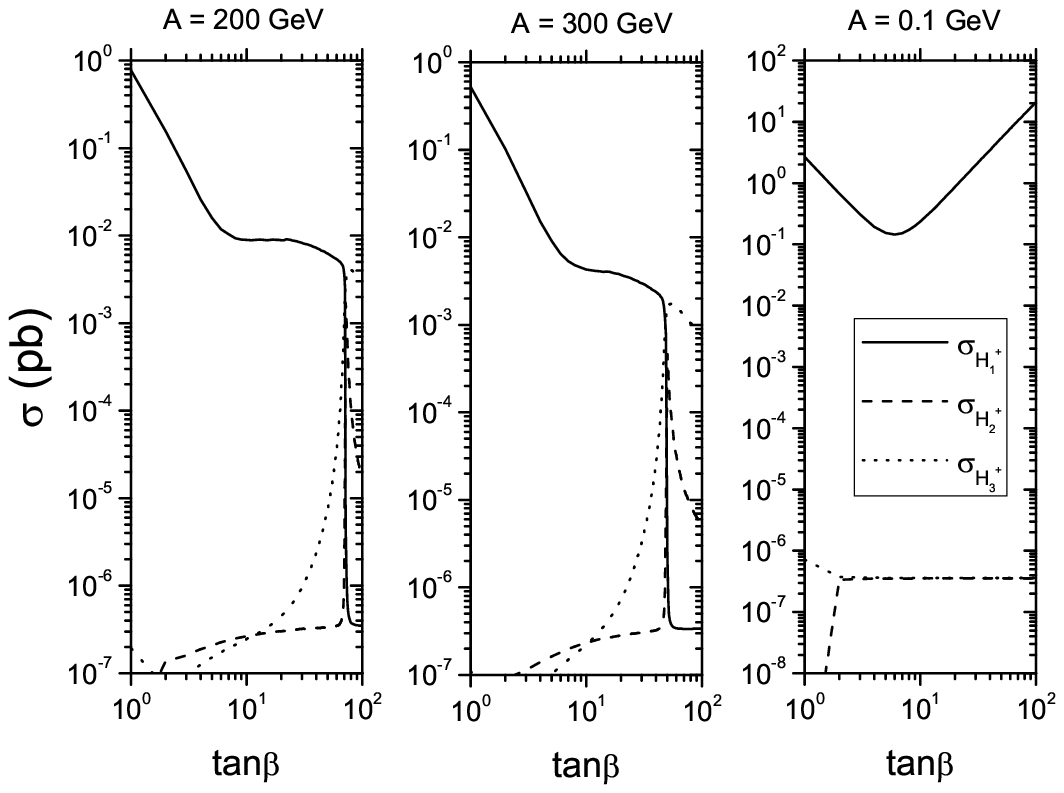}
\caption{ The figure shows the cross sections of $H^+_{1,2,3}$ at
the LHC through the channel $q\bar q,gg\to t\bar b H^-$ + c.c. in
Scenario B with $\lambda=1.0$ and for: $A= 200, 300, 0.1$ GeV,
respectively.} \label{fig:Xsec-b3}
\end{figure}

\begin{figure}
\centering
\includegraphics[width=5in]{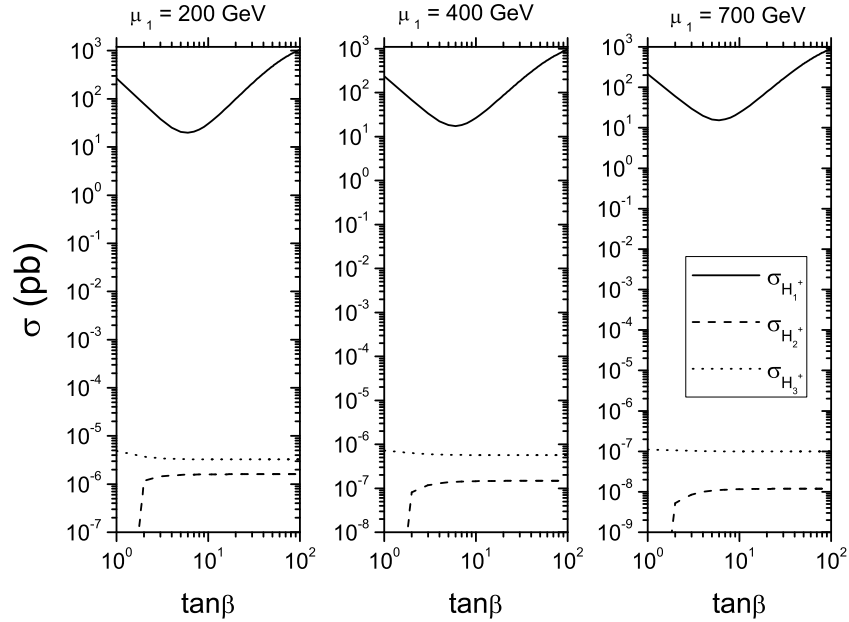}
\caption{ The figure shows the cross sections of $H^+_{1,2,3}$ at
the LHC through the channel $q\bar q,gg\to t\bar b H^-$ + c.c. in
Scenario B with $A=0$ GeV, $\lambda=0.5$ and for: $\mu_1=200, 400,
700 $ GeV, respectively.} \label{fig:Xsec-b4}
\end{figure}

Here, we use HERWIG version 6.510 in default configuration, by
onsetting the subprocess {\tt IPROC~=~3839}, wherein we have
overwritten the default MSSM/2HDM couplings and masses with those
pertaining to the MSSM+1CHT: see
Eqs.~(\ref{coups1})--(\ref{coups2}). The production cross sections
are found in Figures~\ref{fig:Xsec-a1}--\ref{fig:Xsec-a3} and
Figures~\ref{fig:Xsec-b1}--\ref{fig:Xsec-b3} for our usual A
($\mu_1=0$) and B ($\mu_2=0$) Scenarios, respectively, for  various
different choices of $\lambda$ and $A$. (See also
Figure~\ref{fig:Xsec-b4}, illustrating the cross section dependence on
$\tan \beta$ for $\lambda=0.5$, $A=0$ and  $\mu_1= 200$, 400, 700
GeV.)

The pattern of the $H^\pm_1$ cross sections reflects the usual
dependence of the $H^\pm$ state of the MSSM/2HDM,
$\sim(m_t^2\cot\beta^2+m_b^2\tan\beta^2)/(m_t^2+m_b^2)$,  induced by
the Yukawa couplings inside the $t\bar b H^-$ vertex for the case of the
$H^\pm_1$, with a minimum at $\tan\beta\approx6$, as seen in
Figures~\ref{fig:Xsec-a1}--\ref{fig:Xsec-a3}, our Scenario A. (In the
last plot, one may appreciate also some peculiar interference
effects between the $q\bar q$, $ gg \rightarrow t\bar t \rightarrow
t\bar bH^-_i$ + c.c. diagrams and the remaining ones, modulated by the top
width.) The same can be said for the two heavier MSSM+1CHT states,
$H^\pm_2$ and  $H^\pm_3$, with the minima shifted to lower
$\tan\beta$ values, the more so the heavier the Higgs boson,
signalling that are the Yukawa couplings of these last two particles
those differing most from the MSSM/2HDM limit. In the case of
Scenario B, Figures~\ref{fig:Xsec-b1}--\ref{fig:Xsec-b4}, the
aforementioned coupling induced dependence upon $\tan\beta$ is only
seen for the $H^\pm_1$ state for very small $A$'s ($A=0.1$ and
$\mu_1=200$ GeV or even for $A=0$ GeV and $\mu_1$ arbitrary). In
all other parameter configurations, the $H^\pm_1$ trends differ
dramatically from the usual (nearly parabolic) dependence typical of
the MSSM/2HDM limit. The same can be said for all setups chosen in
the case of the $H^\pm_2$ and  $H^\pm_3$. These peculiar patterns
can be understood by using the analytic expressions given in
Ref.~\cite{Espinosa:1991wt} for the  mass spectrum of the charged
Higgs bosons in the approximation $R\ll 1$. In this approximation,
we can obtain in Scenario B that:
$$m_{H^\pm_3} \sim 2 \mu_1 A/ \sin
2\beta,$$
\begin{equation}
m_{H^\pm_{1,2}} \sim \lambda v/ (2 \sqrt{2} R) ( A \sin 2\beta+ 2
\mu_1) \pm 1/2 (g^2- \lambda^2)v^2.\end{equation}

\noindent In the range $5 \lsim \tan \beta \lsim 100$, $\sin 2
\beta$ decreases from $\sim 0.1$ to $\sim 0.01$. This explains the
shapes of the curves showed in the
Figures~\ref{fig:Xsec-b1}--\ref{fig:Xsec-b3}. The parameter $\lambda$
just determines where  the curves for $m_{H^\pm_{1,2}}$ begin to fall.
On the other hand, the pattern
$(m_t^2\cot\beta^2+m_b^2\tan\beta^2)/(m_t^2+m_b^2)$ is suppressed by
a factor of $10^{-2}$ for $m_{H^\pm_{1,2}}$ when $A$ and $\tan \beta$ are
large.

Altogether, by comparing the  $q\bar q,gg\to t\bar b H^-_i$ + c.c.
cross sections herein with, e.g., those of the MSSM in
\cite{Djouadi:2005gj} or the 2HDM in
\cite{Moretti:2002ht,Moretti:2001pp}, it is clear that the MSSM+1CHT
rates can be very large and thus the discovery potential in ATLAS
and CMS can be substantial, particularly for a very light
$H_1^\pm$, which may pertain to our MSSM+1CHT but not the MSSM or
general 2HDM. However, it is only by combining the production rates of
this section with the decay ones of the previous ones that actual
event numbers at the LHC can be predicted.

%%%%%%%%%%%%%%%%%%%%%%%%%%%%%%%%%%%%%%%%%%%%%%%%%%%%%%%%%%%%%%%%%%%%%%%
\section{Charged Higgs boson event rates at the LHC in the MSSM+1CHT}
%%%%%%%%%%%%%%%%%%%%%%%%%%%%%%%%%%%%%%%%%%%%%%%%%%%%%%%%%%%%%%%%%%%%%%%

To illustrate the type of charged Higgs signatures that have the
potential to be detectable at the LHC, we show in Tables \ref{tab:5}
and \ref{tab:6} a summary of results for masses, LHC cross sections
($\sigma$'s),
BR's and event rates. We focus on those cases where the
charged Higgs boson mass is above the threshold for $t \to H^+ b$.
Thus, for Scenario A, all the entries for $\sigma$'s and BR's in Table
\ref{tab:5} correspond to the second charged Higgs boson
$H_2^{\pm}$, while for Scenario B, Table \ref{tab:6} shows the
corresponding results for $H_1^{\pm}$. All of these rates
correspond to $\mu_2=100$ GeV for Scenario A and $\mu_1=200$ GeV for
Scenario B. The cases where $t \to H^+_i b$ is allowed have been
discussed previously in section III. We shall also assume an
integrated luminosity of $10^{-5}$ pb$^{-1}$.

To illustrate these results, let us comment one case within each
scenario. From Table \ref{tab:5}, we can see that for Scenario A,
with $\lambda=0.5$, $A=200$ GeV and $\tan\beta=20$, we have
that
${H_2^{+}}$ is heavier than $m_t -m_b$, with a mass
$m_{H^+_2}=390$
GeV, this precluding top decay contributions, so that
 in this case $\sigma(pp \to t \bar{b} H^+) \approx 1.2
\times 10^{-2}$ pb, while the dominant decays are $H_2^+ \to t
\bar{b}, \, W^+ A_1^0, \, W^+ Z^0, \, W^+ H_1^0$ which give a number
of events of 180, 900, 12, 92, respectively. In this case the most
promising signal is $H_2^+ \to W^+ A_1^0$. However, when $\lambda=1.0$
we have that all event rates get decreased, partially because the
masses get enhanced. For instance $m_{H_2^+}$ becaomes 545 GeV, however
the number of events at the LHC for $H_2^+ \to W^+ A_1^0$ can still be
substantial, as it is about 349.

Then, for Scenario B, we have that $H_1^+$ is already above the
threshold for $t \to H^+_i b$. So, for the declared values of the relevant
parameters, we obtain that the lightest charged Higgs boson mass is
$m_{H_1^+}=906$ and 921 GeV for $\lambda=0.5$ and $\lambda=1.0$,
respectively. In such a case only the decay $H_1^+ \to t \bar{b}$ can
reach significant numbers for the LHC. We obtain a number of events of
912 and 854, respectively. The other decay that has a large BR is
$H_1^+ \to \tau^+ \nu_{\tau}$, but in all cases the number of events
is at most of order 20--30, which seems quite difficult to be
detectable at the CERN machine.

Thus, we conclude that signatures in Scenario A are more diverse
than for Scenario B. However, in order to reach quantitative
conclusions we need to perform a study of signal {\it versus}
background, but this seems more tractable if we concentrate in each
mode individually, rather than while taking the general view that was
attempted in this paper.
%%%%%%%%%%%%%%%%%%%%%%%%%%%%%%%%
%%%%%%%%%%%%%%%%%%%%%%%%%%%%%%%%%%%%%%%%%%%%%%
%%%%%%%%%%%%%%%%%%%%%%%%%%%%%%%%
%%%%%%%%%%%%%%%%%%%%%%%%%%%%%%%%%%%%%%%%%%%%%%
%%%%%%%%%%%%%%%%%%%%%%%%%%%%%%%%
%%%%%%%%%%%%%%%%%%%%%%%%%%%%%%%%%%%%%%%%%%%%%%
%%%%%%%%%%%%%%%%%%%%%%%%%%%%%%%%
%%%%%%%%%%%%%%%%%%%%%%%%%%%%%%%%%%%%%%%%%%%%%%
%%%%%%%%%%%%%%%%%%%%%%%%%%%%%%%%
%%%%%%%%%%%%%%%%%%%%%%%%%%%%%%%%%%%%%%%%%%%%%%
%%%%%%%%%%%%%%%%%%%%%%%%%%%%%%%%
%%%%%%%%%%%%%%%%%%%%%%%%%%%%%%%%%%%%%%%%%%%%%%
%%%%%%%%%%%%%%%%%%%%%%%%%%%%%%%%
%%%%%%%%%%%%%%%%%%%%%%%%%%%%%%%%%%%%%%%%%%%%%%

%%%%%%%%%%%%%%%%%%%%%%%%%%%%%%%%%%%%%%%%%%%%%%%%%%%%%%%%%%%%%%%%%%%%%%%
\section{Conclusions}
%%%%%%%%%%%%%%%%%%%%%%%%%%%%%%%%%%%%%%%%%%%%%%%%%%%%%%%%%%%%%%%%%%%%%%%

After describing the entire Higgs mass and coupling spectrum,
we have studied the fermion-charged Higgs bosons vertices within the
context of an extension of the Minimal Supersymmetric Standard Model
with an additional Complex Higgs Triplet (MSSM+1CHT). Then we have
analyzed the decay $t \to b \, H_i^+$ $(i=1,2,3.)$ in the framework
of this model. We have found that in some plausible scenarios of the
MSSM+1CHT the experimental data allow regions in the plane
$M_{H_i^{\pm}}-\tan\beta$ for the lightest charged Higgs boson
$H_1^{\pm}$ (even for $H_2^{\pm}$) that are forbidden in the case of
the MSSM. For instance, in this model it is possible to have a
charged Higgs boson with mass $m_{H^{\pm}} \approx m_{W^{\pm}}$
which is not excluded yet by any of the current data. Furthermore,
there are other regions of parameters where the top quark can decay
into two charged Higgs states, so that plenty of these states could
be produced at the LHC in (anti)top quark decays. Whenever the mass
of the charged Higgs boson is larger than than $m_t$, there is scope
to exploit direct $H_i^\pm$ hadro-production through the associate
mode $q\bar q,gg\to t\bar b H^-_i$ + c.c.

On the one hand, the detection at the LHC of charged Higgs bosons in the regions
of parameter space accessible in the MSSM (or a 2HDM) would not contradict
the
MSSM+1CHT hypothesis. On the other hand, the observation at such a machine
of several charged Higgs bosons would correspond
to a model with a more elaborate Higgs
sector than the MSSM (or a 2HDM), such as the MSSM+1CHT.

\bigskip

%%%%%%%%%%%%%%%%%%%%%%%%%%%%%%%%%%%%%%%%%%%%%%%%%%%%%%%%%%%%%%%%%%%%%%%%%%
\noindent
{\bf Acknowledgements}:
This work was supported by CONACyT and SNI (M\'exico).
J.H.S. thanks in particular CONACyT (M\'exico) for the grant J50027-F.
S.M. acknowledges the FP6 RTN
MRTN-CT-2006-035505 (HEPTOOLS, Tools and Precision Calculations
for Physics Discoveries at Colliders) for partial support and The Royal
Society (London, UK) for a Conference Grant to attend the meeting
`Charged2006', where this work was started.
%%%%%%%%%%%%%%%%%%%%%%%%%%%%%%%%%%%%%%%%%%%%%%%%%%%%%%%%%%%%%%%%%%%%%%%%%%%

\newpage

\squeezetable
\begin{table*}[htdp]
\caption{\label{tab:5} Summary of LHC event rates for Scenario A with
$\mu_2=100$ GeV for an integrated luminosity of $10^{5}$
pb$^{-1}$, for several different signatures.}

\begin{tabular}{|c|c|c|c|c|c|}
\hline $\lambda$ & $(A,\tan\beta)$ & $m_{H_i^+}$ in GeV &
$\sigma(pp \to H^+_2 tb)$ in pb & Relevant BR's & Nr. Events\\

\hline \multicolumn{1}{|c|}{ 0.5 }& (200,5) &(118,740,790) &
$1.6\times 10^{-5}$ &\begin{tabular}{l} BR$\left( H^{+}_{2} \to
tb\right)\approx 5.8 \times 10^{-4} $\\
BR$\left( H^{+}_{2} \to
W^{+}A^{0}_1\right) \approx 9.8 \times 10^{-1} $\\
BR$\left( H^{+}_{2} \to
W^{+}Z^{0}\right) \approx 1.2 \times 10^{-2} $\\
BR$\left( H^{+}_{2} \to
W^{+}H^{0}_{1}\right) \approx 4.2 \times 10^{-3} $%
\end{tabular}
& \multicolumn{1}{|c|}{$
\begin{tabular}{r}
0\\
1\\
0\\
0%
\end{tabular}
$} \\ \hline

\hline \multicolumn{1}{|c|}{ 0.5 }& (200,20) &(114,390,470) &
$1.2\times 10^{-2}$ &\begin{tabular}{l} BR$\left( H^{+}_{2} \to
tb\right)\approx 1.5 \times 10^{-1} $\\
BR$\left( H^{+}_{2} \to
W^{+}A^{0}_1\right) \approx 7.5 \times 10^{-1} $\\
BR$\left( H^{+}_{2} \to
W^{+}Z^{0}\right) \approx 1.0 \times 10^{-2} $\\
BR$\left( H^{+}_{2} \to
W^{+}H^{0}_{1}\right) \approx 7.7 \times 10^{-2} $%
\end{tabular}
& \multicolumn{1}{|c|}{$
\begin{tabular}{r}
180\\
900\\
12\\
92%
\end{tabular}
$} \\ \hline

\hline \multicolumn{1}{|c|}{ 0.5 }& (200,50) &(98,290,370) &
$8.7\times 10^{-1}$ &\begin{tabular}{l} BR$\left( H^{+}_{2} \to
\tau^+ \nu_\tau \right) \approx 8.2 \times 10^{-2} $\\
BR$\left( H^{+}_{2} \to
tb\right)\approx 8.4 \times 10^{-1} $\\
BR$\left( H^{+}_{2} \to
W^{+}A^{0}_1\right) \approx 2.4 \times 10^{-2} $\\
BR$\left( H^{+}_{2} \to
W^{+}H^{0}_{1}\right) \approx 4.8 \times 10^{-2} $%
\end{tabular}
& \multicolumn{1}{|c|}{$
\begin{tabular}{r}
7134\\
73080\\
2088\\
4176%
\end{tabular}
$} \\ \hline

\hline \multicolumn{1}{|c|}{ 1.0 }& (200,5) &(191,1047,1087) &
$3.4\times 10^{-6}$ &\begin{tabular}{l} BR$\left( H^{+}_{2} \to
\tau^+\nu_\tau\right)\approx 4.6 \times 10^{-6} $\\
BR$\left( H^{+}_{2} \to
tb \right)\approx 3.0 \times 10^{-4} $\\
BR$\left( H^{+}_{2} \to
W^{+}A^{0}_{1}\right) \approx 9.8 \times 10^{-1} $\\
BR$\left( H^{+}_{2} \to W^{+}Z^{0}\right) \approx 1.3 \times
10^{-2} $%
\end{tabular}
& \multicolumn{1}{|c|}{$
\begin{tabular}{r}
0\\
0\\
0\\
0%
\end{tabular}
$} \\ \hline

\hline \multicolumn{1}{|c|}{ 1.0 }& (200,20) &(185,545,610) &
$4.5\times 10^{-3}$ &\begin{tabular}{l} BR$\left( H^{+}_{2} \to
tb\right)\approx 1.1 \times 10^{-1} $\\
BR$\left( H^{+}_{2} \to
W^{+}A^{0}_1\right) \approx 7.7 \times 10^{-1} $\\
BR$\left( H^{+}_{2} \to
W^{+}Z^{0}\right) \approx 1.1 \times 10^{-2} $\\
BR$\left( H^{+}_{2} \to
W^{+}H^{0}_{1}\right) \approx 1.0 \times 10^{-2} $%
\end{tabular}
& \multicolumn{1}{|c|}{$
\begin{tabular}{r}
50\\
349\\
5\\
4%
\end{tabular}
$} \\ \hline

\hline \multicolumn{1}{|c|}{ 1.0 }& (200,50) &(153,400,450) &
$3.6\times 10^{-1}$ &\begin{tabular}{l} BR$\left( H^{+}_{2} \to
\tau^+\nu_\tau\right)\approx 5.0 \times 10^{-2} $\\
BR$\left( H^{+}_{2} \to
tb\right)\approx 8.4 \times 10^{-1} $\\
BR$\left( H^{+}_{2} \to
W^{+}A^{0}_1\right) \approx 2.6 \times 10^{-2} $\\
BR$\left( H^{+}_{2} \to
W^{+}H^{0}_{1}\right) \approx 7.9 \times 10^{-2} $%
\end{tabular}
& \multicolumn{1}{|c|}{$
\begin{tabular}{r}
1800\\
30240\\
936\\
2844%
\end{tabular}
$} \\ \hline

\end{tabular}
\label{default5}
\end{table*}

\newpage

\squeezetable
\begin{table*}[htdp]
\caption{\label{tab:6} Summary of LHC event rates for Scenario B with
$\mu_1=200$ GeV for an integrated luminosity of $10^{5}$
pb$^{-1}$, for several different signatures. }

\begin{tabular}{|c|c|c|c|c|c|}
\hline $\lambda$ & $(A,\tan\beta)$ & $m_{H_i^+}$ in GeV &
$\sigma(pp \to
H^+_1tb)$ in pb & Relevant BR's & Nr. Events\\

\hline \multicolumn{1}{|c|}{ 0.5 }& (200,5) &(473,1304,1305) &
$1.8 \times 10^{-2}$ &\begin{tabular}{l}
BR$\left( H^{+}_{1} \to
\tau^+ \nu_\tau \right)\approx 1.7 \times 10^{-2} $\\
BR$\left( H^{+}_{1}\to
tb\right)\approx 9.8 \times 10^{-1} $\\
BR$\left( H^{+}_{1} \to
W^{+}Z^{0}\right) \approx 4.2 \times 10^{-6} $\\
BR$\left( H^{+}_{1} \to
W^{+}H^{0}_{1}\right) \approx 5.4 \times 10^{-4} $%
\end{tabular}
& \multicolumn{1}{|c|}{$
\begin{tabular}{r}
31\\
1764\\
0\\
1%
\end{tabular}
$} \\ \hline

\hline \multicolumn{1}{|c|}{ 0.5 }& (200,20) &(906,1223,1225) &
$9.6 \times 10^{-3}$ &\begin{tabular}{l} BR$\left( H^{+}_{1} \to
\tau^+ \nu_\tau \right)\approx 3.9 \times 10^{-2} $\\
BR$\left( H^{+}_{1}\to
tb\right)\approx 9.5 \times 10^{-1} $\\
BR$\left( H^{+}_{1} \to
W^{+}Z^{0}\right) \approx 1.4 \times 10^{-5} $\\
BR$\left( H^{+}_{1} \to
W^{+}H^{0}_{1}\right) \approx 2.1 \times 10^{-6} $%
\end{tabular}
& \multicolumn{1}{|c|}{$
\begin{tabular}{r}
37\\
912\\
0\\
0%
\end{tabular}
$} \\ \hline

\hline \multicolumn{1}{|c|}{ 0.5 }& (200,50) &(1206,1207,1424) &
$2.0\times 10^{-6}$ &\begin{tabular}{l} BR$\left( H^{+}_{1} \to
\tau^+ \nu_\tau \right)\approx 3.8 \times 10^{-3} $\\
BR$\left( H^{+}_{1}\to
tb\right)\approx 1.3 \times 10^{-1} $\\
BR$\left( H^{+}_{1} \to
W^{+}Z^{0}\right) \approx 8.3 \times 10^{-1} $\\
BR$\left( H^{+}_{1} \to
W^{+}H^{0}_{1}\right) \approx 3.6 \times 10^{-2} $%
\end{tabular}
& \multicolumn{1}{|c|}{$
\begin{tabular}{r}
0\\
0\\
0\\
0%
\end{tabular}
$} \\ \hline

\hline \multicolumn{1}{|c|}{ 1.0 }& (200,5) &(497,1838,1850) &
$1.6 \times 10^{-2}$ &\begin{tabular}{l} BR$\left( H^{+}_{1} \to
\tau^+ \nu_\tau \right)\approx 1.6 \times 10^{-2} $\\
BR$\left( H^{+}_{1}\to
tb\right)\approx 9.7 \times 10^{-1} $\\
BR$\left( H^{+}_{1} \to
W^{+}Z^{0}\right) \approx 4.0 \times 10^{-6} $\\
BR$\left( H^{+}_{1} \to
W^{+}H^{0}_{1}\right) \approx 1.1 \times 10^{-2} $%
\end{tabular}
& \multicolumn{1}{|c|}{$
\begin{tabular}{r}
26\\
1552\\
0\\
18%
\end{tabular}
$} \\ \hline

\hline \multicolumn{1}{|c|}{ 1.0 }& (200,20) &(921,1724,1738) &
$8.9 \times 10^{-3}$ &\begin{tabular}{l} BR$\left( H^{+}_{1} \to
\tau^+ \nu_\tau \right)\approx 3.9 \times 10^{-2} $\\
BR$\left( H^{+}_{1}\to
tb\right)\approx 9.6 \times 10^{-1} $\\
BR$\left( H^{+}_{1} \to
W^{+}Z^{0}\right) \approx 5.8 \times 10^{-6} $\\
BR$\left( H^{+}_{1} \to
W^{+}H^{0}_{1}\right) \approx 1.8 \times 10^{-5} $%
\end{tabular}
& \multicolumn{1}{|c|}{$
\begin{tabular}{r}
35\\
854\\
0\\
0%
\end{tabular}
$} \\ \hline

\hline \multicolumn{1}{|c|}{ 1.0 }& (200,50) &(1434,1699,1713) &
$6.2 \times 10^{-3}$ &\begin{tabular}{l} BR$\left( H^{+}_{1} \to
\tau^+ \nu_\tau \right)\approx 3.8 \times 10^{-2} $\\
BR$\left( H^{+}_{1}\to
tb\right)\approx 9.6 \times 10^{-1} $\\
BR$\left( H^{+}_{1} \to
W^{+}Z^{0}\right) \approx 1.4 \times 10^{-5} $\\
BR$\left( H^{+}_{1} \to
W^{+}H^{0}_{1}\right) \approx 5.8 \times 10^{-7} $%
\end{tabular}
& \multicolumn{1}{|c|}{$
\begin{tabular}{r}
24\\
595\\
0\\
0%
\end{tabular}
$} \\ \hline

\end{tabular}
\label{default6}
\end{table*}


\newpage


\begin{thebibliography}{99}

\bibitem{stanmod} S. L. Glashow, Nucl. Phys. {\bf 22}, 579 (1961); S.
Weinberg, Phys. Rev. Lett. {\bf 19}, 1264 (1967); A. Salam, Proc.
8th NOBEL Symposium, ed. N. Svartholm (Almqvist and Wiksell,
Stockholm, 1968), p. 367.

\bibitem{kanehunt} S. Dawson et al., {\it The Higgs Hunter's Guide},
2nd ed., Frontiers in Physics Vol. {\bf 80} (Addison-Wesley,
Reading MA, 1990).

\bibitem{LHCrev} For a review of SM and MSSM Higgs Physics at the LHC see, e.g.:
A. Djouadi, hep-ph/0503172 and hep-ph/0503173.

\bibitem{susyhix}
M. Carena {\it et al.}, hep-ph/0010338; C. Balazs
{\it et al.}, Phys. Rev. {\bf D59}, 055016 (1999); J. L.
D\'{\i}az-Cruz {\it et al.}, Phys. Rev. Lett. {\bf 80}, 4641
(1998).

\bibitem{stronghix} See, e.g., recent work on Little Higgs models:
  N.~Arkani-Hamed, A.~G.~Cohen, E.~Katz and A.~E.~Nelson,
  %``The littlest Higgs,''
  JHEP {\bf 0207}, 034 (2002).
  %%CITATION = JHEPA,0207,034;%%
And on AdS/CFT Higgs models:
  R.~Contino, Y.~Nomura and A.~Pomarol,
  %``Higgs as a holographic pseudo-Goldstone boson,''
  Nucl.\ Phys.\  B {\bf 671}, 148 (2003);
%  [hep-ph/0306259];
  %%CITATION = NUPHA,B671,148;%%
%\cite{Aranda:2007tg}
%\bibitem{Aranda:2007tg}
  A.~Aranda, J.~L.~Diaz-Cruz, J.~Hernandez-Sanchez and R.~Noriega-Papaqui,
  %``Fundamental and composite scalars from extra dimensions,''
  0708.3821 [hep-ph].
  %%CITATION = 0708.3821;%%

%\cite{Rizzo:1990uu}
\bibitem{Rizzo:1990uu}
  T.~G.~Rizzo,
  %``Updated Bounds On Higgs Triplet Vacuum Expectation Values And The Tree
  %Level Rho Parameter From Radiative Corrections,''
  Mod.\ Phys.\ Lett.\ A {\bf 6}, 1961 (1991).
  %%CITATION = MPLAE,A6,1961;%%


\bibitem{cppaper}
  J.~W.~F.~Valle,
  %``Neutrino physics overview,''
  hep-ph/0608101;
  %%CITATION = HEP-PH 0608101;%%
 S.~Kanemura, T.~Kasai, G.~L.~Lin, Y.~Okada, J.~J.~Tseng and C.~P.~Yuan,
  %``Phenomenology of Higgs bosons in the Zee-model,''
  Phys.\ Rev.\ D {\bf 64}, 053007 (2001).
%  [hep-ph/0011357].
  %%CITATION = HEP-PH 0011357;%%

\bibitem{ourexdims}
  A.~Aranda, C.~Balazs and J.~L.~Diaz-Cruz,
  %``Where is the Higgs boson? ((V)),''
  Nucl.\ Phys.\ B {\bf 670}, 90 (2003);
 % [hep-ph/0212133];
  %%CITATION = HEP-PH 0212133;%%
%\cite{Aranda:2005ze}
%\bibitem{Aranda:2005ze}
  A.~Aranda and J.~L.~Diaz-Cruz,
  %``Gauge-Higgs unification with brane kinetic terms,''
  Phys.\ Lett.\  B {\bf 633}, 591 (2006).
%  [hep-ph/0510138].
  %%CITATION = PHLTA,B633,591;%%

\bibitem{vegunwud} J.~F.~ Gunion, R.~ Vega and J.~ Wudka,
%(UC, Davis) . UCD-89-13, Dec 1989. 40pp.
Phys.\ Rev.\ D {\bf42}, 1673 (1990).
%\cite{Espinosa:1991wt}

\bibitem{Espinosa:1991wt}
  J.~R.~Espinosa and M.~Quiros,
  %``Higgs triplets in the Supersymmetric standard model,''
  Nucl.\ Phys.\ B {\bf 384}, 113 (1992).
  %%CITATION = NUPHA,B384,113;%%



%\cite{Felix-Beltran:2002tb}
\bibitem{Felix-Beltran:2002tb}
  O.~Felix-Beltran,
  %``Higgs masses and coupling within an extension of the MSSM with Higgs
  %triplets,''
  Int.\ J.\ Mod.\ Phys.\ A {\bf 17}, 465 (2002).
  %%CITATION = IMPAE,A17,465;%%

\bibitem{Hdecays}
S.~Moretti and W.~J.~Stirling,
  %``Contributions of below threshold decays to MSSM Higgs branching ratios,''
  Phys.\ Lett.\  B {\bf 347}, 291 (1995)
  [Erratum-ibid.\  B {\bf 366}, 451 (1996)];
%  [hep-ph/9412209];
  %%CITATION = PHLTA,B347,291;%%
A.~Djouadi, J.~Kalinowski and P.~M.~Zerwas,
  %``Two- and Three-Body Decay Modes of SUSY Higgs Particles,''
  Z.\ Phys.\  C {\bf 70}, 435 (1996).
%  [hep-ph/9511342].
  %%CITATION = ZEPYA,C70,435;%%

\bibitem{hcdecay}
J. L. D\'{\i}az-Cruz and M.A. P\' erez, Phys. Rev. {\bf D33}, 273
(1986); J.F. Gunion, G.L. Kane and J. Wudka, Nucl. Phys. {\bf B299},
231 (1988); A. Mendez and A. Pomarol, Nucl. Phys. {\bf B349}, 369
(1991); E. Barradas {\it et al.}, Phys. Rev. {\bf D53}, 1678 (1996); M.
Capdequi-Peyranere, H. E. Haber, and P. Irulegui, Phys. Rev. {\bf
D44}, 191 (1991); S. Kanemura, Phys. Rev. {\bf D61}, 095001
(2000); K.A. Assamagan {\it et al.},
hep-ph/0406152; J. Hern\'andez-S\'anchez {\it et
al.}, Phys. Rev. {\bf D69}, 095008 (2004).

\bibitem{hctoSUSY}
%\cite{Bisset:2000ud}
%\bibitem{Bisset:2000ud}
  M.~Bisset, M.~Guchait and S.~Moretti,
  %``Signatures of MSSM charged Higgs bosons via chargino neutralino decay
  %channels at the LHC,''
  Eur.\ Phys.\ J.\  C {\bf 19}, 143 (2001);
%  [hep-ph/0010253];
  %%CITATION = EPHJA,C19,143;%%
%\cite{Bisset:2003ix}
%\bibitem{Bisset:2003ix}
  M.~Bisset, F.~Moortgat and S.~Moretti,
  %``Trilepton + top signal from chargino neutralino decays of MSSM charged
  %Higgs bosons at the LHC,''
  Eur.\ Phys.\ J.\  C {\bf 30}, 419 (2003).
%  [hep-ph/0303093].
  %%CITATION = EPHJA,C30,419;%%

\bibitem{ourpaper}
J. L. D\'{\i}az-Cruz, J. Hern\'andez-S\'anchez and
 J.J. Toscano, Phys. Lett. {\bf B512}, 339 (2001).
%\cite{Barradas-Guevara:2004qi}

\bibitem{Barradas-Guevara:2004qi}
  E.~Barradas-Guevara, O.~F\'elix-Beltr\'an, J.~Hern\'andez-S\'anchez and
A.~Rosado,
  %``The decay H+- $\to$ W-+ h0 within an extension of the MSSM with one
  %complex Higgs triplet,''
  Phys.\ Rev.\ D {\bf 71}, 073004 (2005).
%  [hep-ph/0408196].
  %%CITATION = HEP-PH 0408196;%%

\bibitem{Ketevi}
%\cite{Assamagan:2004wq}
%\bibitem{Assamagan:2004wq}
  K.~A.~Assamagan, J.~Guasch, S.~Moretti and S.~Penaranda,
  %``Determining the ratio of the H+ --> tau nu to H+ --> t anti-b decay  rates
  %for large tan(beta) at the Large Hadron Collider,''
  hep-ph/0402212;
  %%CITATION = HEP-PH/0402212;%%
%\cite{Assamagan:2004ji}
%\bibitem{Assamagan:2004ji}
  K.~A.~Assamagan, J.~Guasch, S.~Moretti and S.~Penaranda,
  %``Distinguishing Higgs models in H+ --> tau+ nu / t anti-b at large tan
  %beta,''
  Czech.\ J.\ Phys.\  {\bf 55}, B787 (2005).
%  [hep-ph/0409189].
  %%CITATION = CZYPA,55,B787;%%

\bibitem{ldcysampay}
J. L. D\'{\i}az-Cruz and O. A. Sampayo, Phys. Rev. {\bf D50}, 6820
(1994); J.F. Gunion {\it et al.}, Nucl. Phys. {\bf B294}, 621 (1987); M.
A. P\' erez and A. Rosado, Phys. Rev. {\bf D30}, 228 (1984).

\bibitem{newhcprod}
S.~Moretti and K.~Odagiri,  Phys.\ Rev.\  D {\bf 55}, 5627 (1997);
{F.~Borzumati, J.-L.~Kneur and N.~Polonsky}, {Phys. Rev.} D {\bf 60}, 115011 (1999);
D.J.~Miller, S.~Moretti, D.P.~Roy and W.J.~Stirling, Phys. Rev. {\bf D61}, 055011 (2000);
S.~Moretti and D.~P.~Roy, Phys.\ Lett.\ B {\bf 470}, 209 (1999);
A.~A.~Barrientos Bendezu and B.~A.~Kniehl, Phys.\ Rev.\  D {\bf 59}, 015009 (1999),
Phys.\ Rev.\  D {\bf 61}, 097701 (2000) and   Phys.\ Rev.\  D {\bf 63}, 015009 (2001);
%Z. Fei et al., Phys. Rev. {\bf D63}, 015002 (2001);
  O.~Brein, W.~Hollik and S.~Kanemura,
  %`The MSSM prediction for W+- H-+ production by gluon fusion,''
  Phys.\ Rev.\  D {\bf 63}, 095001 (2001)
  [arXiv:hep-ph/0008308];
  %%CITATION = PHRVA,D63,095001;%%
  A.~Belyaev, D.~Garcia, J.~Guasch and J.~Sola,
  JHEP {\bf 0206},  059 (2002) and
  Phys.\ Rev.\  D {\bf 65},  031701 (2002).


\bibitem{newhcprod-HO}
S.~H.~Zhu,
  Phys.\ Rev.\  D {\bf 67},  075006 (2003);
T.~Plehn,
  Phys.\ Rev.\  D {\bf 67}, 014018 (2003);
E.~L.~Berger, T.~Han, J.~Jiang and T.~Plehn,
  Phys.\ Rev.\  D {\bf 71}, 115012 (2005).


%\cite{Abulencia:2005jd}
\bibitem{Abulencia:2005jd}
  A.~Abulencia {\it et al.}  [CDF Collaboration],
  %``Search for charged Higgs bosons from top quark decays in p anti-p
  %collisions at s**(1/2) = 1.96-TeV,''
  Phys.\ Rev.\ Lett.\  {\bf 96}, 042003 (2006).
%  [hep-ex/0510065].
  %%CITATION = HEP-EX 0510065;%%

\bibitem{lepbounds} For a review see:
F. Borzumati and A. Djouadi, Phys. Lett. {\bf B549}, 170 (2002);
D. P. Roy, Mod. Phys. Lett. {\bf A19}, 1813 (2004).

%\bibitem{hcwhdetect} S. Moretti, Phys. Lett. {\bf B481}, 49 (2000);
%K. A. Assamagan, Y. Coadou and A. Deandrea, Eur. Phys. J. {\bf
%C4}, 9 (2002).

%\bibitem{jfg}
%J. F. Gunion and H. E. Haber,
%Phys. Rev. {\bf D67}, 075019 (2003).

%\bibitem{npb380}
%J. L. Diaz Cruz and A. Mendez,
%Nucl. Phys.{\bf B380}, 39 (1992).

%%\cite{Eidelman:2004wy}
\bibitem{partdat}
S.~Eidelman {\it et al.}  [Particle Data Group],
%``Review of particle physics,''
Phys.\ Lett.\ B {\bf 592}, 1 (2004).
%%CITATION = PHLTA,B592,1;%%

%
%\bibitem{cotaneu}
%Review of Particle Physics,
%Phys. Rev. {\bf D66}, 309 (2002).

%%\cite{Djouadi:1997yw}
%\bibitem{Djouadi:1997yw}
%  A.~Djouadi, J.~Kalinowski and M.~Spira,
%  %``HDECAY: A program for Higgs boson decays in the standard model and its
%  %Supersymmetric extension,''
%  Comput.\ Phys.\ Commun.\  {\bf 108}, 56 (1998)
%  [hep-ph/9704448].
%  %%CITATION = HEP-PH 9704448;%%


%\cite{Ellis:1991}
\bibitem{Ellis-Ridolfi}
  J.~R.~Ellis, G.~Ridolfi and F.~Zwirner,
  %``On radiative corrections to Supersymetric Higgs boson masses and their implications for
     %LEP searches,''
  Phys.\ Lett.\ B {\bf 257}, 83 (1991) and Phys.\ Lett.\ B {\bf 262}, 477 (1991).
%  [hep-ph/9912516].
  %%CITATION = HEP-PH 9912516;%%

\bibitem{espqui2}
  J.~R.~Espinosa and M.~Quir\'{o}s,
  Phys. Rev. Lett. {\bf 81}, 516 (1998),
  Phys.\ Lett.\ B {\bf 302}, 51 (1993) and Phys.\ Lett.\ B {\bf 279}, 92 (1992).


\bibitem{future-work}
J. L. D\'{\i}az-Cruz, J. Hern\' andez--S\' anchez,  S. Moretti  and A. Rosado,
in preparation.


% Search for neutral MSSM Higgs bosons at LEP
\bibitem{table-LEP}
S. Schael {\it et al.}  [ALEPH Collaboration, DELPHI Collaboration, L3 Collaboration, OPAL Collaborations
and LEP Working Group for Higgs Boson Searches], Eur.\ Phys.\ J.{\bf C47},
547 (2006).% [ hep-ex/0602042]

%\cite{Carena:1999py}
\bibitem{Carena:1999py}
  M.~Carena, D.~Garcia, U.~Nierste and C.~E.~M.~Wagner,
  %``Effective Lagrangian for the anti-t b H+ interaction in the MSSM and
  %charged Higgs phenomenology,''
  Nucl.\ Phys.\ B {\bf 577}, 88 (2000).
%  [hep-ph/9912516].
  %%CITATION = HEP-PH 9912516;%%


\bibitem{LEP}
%\cite{:2001xy}
%\bibitem{:2001xy}
    [LEP Higgs Working Group for Higgs boson searches],
  %``Search for charged Higgs bosons: Preliminary combined results using LEP
  %data collected at energies up to 209-GeV,''
  hep-ex/0107031;
  %%CITATION = HEP-EX 0107031;%%
%\cite{Heister:2002ev}
%\bibitem{Heister:2002ev}
  A.~Heister {\it et al.}  [ALEPH Collaboration],
  %``Search for charged Higgs bosons in e+ e- collisions at energies up to
  %s**(1/2) = 209-GeV,''
  Phys.\ Lett.\ B {\bf 543}, 1 (2002).
%  [hep-ex/0207054].
  %%CITATION = HEP-EX 0207054;%%



%\bibitem{akeroyd1} A.G. Akeroyd, Nucl. Phys. {\bf B544}, 557 (1999).

%\bibitem{akeroyd2} A.G. Akeroyd, A. Arhrib and E. Naimi, Eur. Phys. J.
%{\bf C12}, 451 (2000).

%\cite{Guchait:2001pi}
\bibitem{Guchait:2001pi}
M.~Guchait and S.~Moretti,
%``Improving the discovery potential of charged Higgs bosons at Tevatron  run
%2,''
JHEP {\bf 0201}, 001 (2002).
%[hep-ph/0110020].
%%CITATION = HEP-PH 0110020;%%

\bibitem{Borzumati:1999th}
\mbox{F.~Borzumati, J.-L.~Kneur and N.~Polonsky},  in Ref.~\cite{newhcprod}.

\bibitem{Miller:1999bm}
\mbox{D.J.~Miller, S.~Moretti, D.P.~Roy and W.J.~Stirling}, in
Ref.~\cite{newhcprod}.

\bibitem{Moretti:1999bw}
S.~Moretti and D.~P.~Roy,  in Ref.~\cite{newhcprod}.

\bibitem{Cavalli:2002vs}
D.~Cavalli {\it et al.},
%[The {Higgs} working group: Summary report],
{hep-ph/0203056}.

%\cite{Assamagan:2004mu}
\bibitem{Assamagan:2004mu}
K.~A.~Assamagan {\it et al.},  %[Higgs Working Group Collaboration],
%``The Higgs working group: Summary report 2003,''
%hep-ph/0406152.
%%CITATION = HEP-PH 0406152;%%
in Ref.~\cite{hcdecay}.

\bibitem{Alwall:2003tc} J.~Alwall, C.~Biscarat, S.~Moretti,
J.~Rathsman and A.~Sopczak,
%``The p anti-p $\to$ t b H+- process at the Tevatron in HERWIG and PYTHIA
%simulations,''
Eur.\ Phys.\ J.\ C {\bf 39S1},  37 (2005).
%[hep-ph/0312301].
%%CITATION = HEP-PH 0312301;%%

%\cite{Moretti:2002ht}
\bibitem{Moretti:2002ht}
S.~Moretti,
%``Improving the discovery potential of charged Higgs bosons at the  Tevatron
%and Large Hadron Collider,''
Pramana {\bf 60}, 369 (2003).
%[hep-ph/0205104].
%%CITATION = HEP-PH 0205104;%%

%\cite{Assamagan:2004gv}
\bibitem{Assamagan:2004gv}
K.~A.~Assamagan, M.~Guchait and S.~Moretti,
%``Charged Higgs bosons in the transition region M(H+-) approx. m(t) at  the
%LHC,''
hep-ph/0402057.
%%CITATION = HEP-PH 0402057;%%

\bibitem{herwig}
G. Abbiendi, I. G. Knowles,
G. Marchesini,
M. H. Seymour, L. Stanco
and B. R. Webber, Comp. Phys. Commun. {\bf 67},  465 (1992).

\bibitem{Corcella:2000bw}
G.~Corcella {\it et al.},
{JHEP} {\bf 0101}, 010 (2001).

\bibitem{Corcella:2002jc}
G.~Corcella {\it et al.},
{hep-ph/0210213}.

\bibitem{Moretti:2002eu}
{S.~Moretti, K.~Odagiri, P.~Richardson, M. H.~Seymour and B. R.~Webber},
{JHEP} {\bf 0204}, 028 (2002).

\bibitem{pythia}
T.~Sj\"ostrand, Comp. Phys. Comm. {\bf 82},  {74} ({1994});
T. Sj\"ostrand, P. Ed\'en, C. Friberg, L. L\"onnblad, G. Miu, S. Mrenna and E.
Norrbin, Comp. Phys. Commun. {\bf 135}, 238 (2001);
T. Sj\"ostrand, L. L\"onnblad and S. Mrenna, {hep-ph/0108264};
T.~Sj\"ostrand, L.~L\"onnblad, S.~Mrenna and P.~Skands,
%``PYTHIA 6.3: Physics and manual,''
hep-ph/0308153.

%\cite{Alwall:2004xw}
\bibitem{Alwall:2004xw}
J.~Alwall and J.~Rathsman,
%``Improved description of charged Higgs boson production at hadron colliders,''
JHEP {\bf 0412}, 050 (2004).
%[hep-ph/0409094].
%%CITATION = HEP-PH 0409094;%%

%\cite{Djouadi:2005gj}
\bibitem{Djouadi:2005gj}
  A.~Djouadi,
  %``The anatomy of electro-weak symmetry breaking. II: The Higgs bosons in  the
  %minimal supersymmetric model,''
  second paper in \cite{LHCrev}.
  %%CITATION = HEP-PH/0503173;%%

%\cite{Moretti:2001pp}
\bibitem{Moretti:2001pp}
  S.~Moretti,
  %``Pair production of charged Higgs scalars from electroweak gauge boson
  %fusion,''
  J.\ Phys.\ G {\bf 28}, 2567 (2002).
  %[hep-ph/0102116].
  %%CITATION = JPHGB,G28,2567;%%

\end{thebibliography}
\end{document}